\documentclass[11pt]{article}
\pdfoutput=1 

\usepackage{jcappub}
                     
\usepackage{amsmath}
\usepackage{bm}
\usepackage[T1]{fontenc}
\usepackage{caption}
\usepackage{subcaption}
\usepackage{color}
\usepackage[export]{adjustbox}
\usepackage[normalem]{ulem}
\usepackage{xcolor}
\usepackage{tocvsec2}

\newcommand{\svc}{\langle \sigma v^2\rangle_{\rm can}}

\newcommand{\pushright}[1]{\ifmeasuring@#1\else\omit\hfill$\displaystyle#1$\fi\ignorespaces}
\newcommand{\pushleft}[1]{\ifmeasuring@#1\else\omit$\displaystyle#1$\hfill\fi\ignorespaces}

\newcommand{\beq}{\begin{equation}}
\newcommand{\eeq}{\end{equation}}

\newcommand{\intp}{\int \frac{d^3p}{(2\pi)^3}}

\definecolor{darkgreen}{cmyk}{0.85,0.2,1.00,0.2}

\title{\boldmath Cannibalism's lingering imprint on the matter power spectrum}

\makeatother

\author[a]{Adrienne L. Erickcek,}

\author[b]{Pranjal Ralegankar,}
\author[b]{and Jessie Shelton}
\affiliation[a]{Department of Physics and Astronomy, University of North Carolina at Chapel Hill, Phillips Hall CB 3255, Chapel Hill, NC 27599, USA}
\affiliation[b]{Illinois Center for Advanced Studies of the Universe, Department of Physics, University of Illinois at Urbana-Champaign, Urbana, IL 61801, USA }

\emailAdd{erickcek@physics.unc.edu}
\emailAdd{pranjal6@illinois.edu}
\emailAdd{sheltonj@illinois.edu}

\abstract{The early universe may have contained internally thermalized dark sectors that were decoupled from the Standard Model.  In such scenarios, the relic dark thermal bath, composed of the lightest particle in the dark sector, can give rise to an epoch of early matter domination prior to Big Bang Nucleosynthesis, which has a potentially observable impact on the smallest dark matter structures.   This lightest dark particle can easily and generically have number-changing self-interactions that give rise to ``cannibal'' behavior.  We consider cosmologies where an initially sub-dominant cannibal species comes to temporarily drive the expansion of the universe, and we provide a simple map between the particle properties of the cannibal species and the key features of the enhanced dark matter perturbation growth in such cosmologies.  We further demonstrate that cannibal self-interactions can determine the small-scale cutoff in the matter power spectrum even when the cannibal self-interactions freeze out prior to cannibal domination.}

\begin{document}
\maketitle
\flushbottom

\section{Introduction}
While standard cosmology posits that post-inflationary reheating is followed by uninterrupted radiation domination prior to matter-radiation equality, a variety of well-motivated particle physics scenarios predict departures from radiation domination in the poorly constrained epoch between the end of inflation and Big Bang nucleosynthesis (BBN) \cite{Allahverdi_2021}.   For instance, supersymmetric theories often predict moduli whose energy density can come to dominate the universe as they coherently oscillate in a quadratic potential, giving rise to an early matter-dominated era (EMDE) that ends when the modulus decays \cite{Kane:2015jia}.  The semi-classical evolution of light spin-zero fields can also give rise to epochs of kination when the scalar field's kinetic energy dominates over its potential energy \cite{Spokoiny:1993kt,Joyce:1996cp,Ferreira:1997hj}.
 
Early departures from radiation domination are also generic consequences of theories that contain an internally thermalized hidden sector that is thermally decoupled from the Standard Model (SM).   Such decoupled self-interacting hidden sectors are readily obtained from straight-forward inflationary scenarios \cite{Hodges:1993yb,Berezhiani:1995am,Adshead:2016xxj,Adshead:2019uwj,Halverson:2019kna}, and can naturally provide a cosmological origin for the dark matter (DM) of our universe \cite{Kolb:1985bf,Hodges:1993yb,Chen:2006ni,Feng:2008mu}, a possibility that becomes ever more compelling with the continued absence of direct detection signals to date.
If the lightest state in the hidden sector is massive, then it can easily come to dominate the energy density of the universe after it becomes non-relativistic.  If this particle is effectively pressureless when it dominates, it produces an EMDE \cite{Zhang:2015era,Berlin:2016vnh,Berlin:2016gtr,Dror:2016rxc,Dror:2017gjq}.  However, in many familiar theories, ranging from the simple and minimal example of a single scalar field to the exceptionally well-motivated scenario of a confining Yang-Mills sector, the lightest particle in the dark sector has number-changing ``cannibal'' self-interactions that remain in equilibrium even after the particle becomes non-relativistic \cite{Carlson:1992fn,Boddy:2014yra,Boddy:2014qxa,Bernal:2015ova,Soni:2016gzf,Forestell:2016qhc,Pappadopulo:2016pkp,Farina:2016llk,Buen-Abad2018,Heeba:2018wtf,Heimersheim:2020aoc}.  While this particle dominates the expansion of the universe, these self-interactions sustain appreciable pressure in the cannibal fluid, giving rise to an early cannibal-dominated era (ECDE) \cite{Erickcek:2020wzd}.  In either case, radiation domination is restored when the lightest hidden sector particle decays into SM particles; this must occur prior to neutrino decoupling to avoid altering the abundance of light elements \cite{Kawasaki:1999na,Kawasaki:2000en,Hannestad:2004px,Ichikawa:2005vw} and the anisotropies in the cosmic microwave background \cite{deSalas:2015glj,Hasegawa:2019jsa}.
 
 Altered expansion histories prior to BBN can leave potentially observable footprints in dark matter perturbations on scales that experienced altered growth \cite{ES11, BR14, FOW14, Erickcek:2015jza,Redmond:2018xty}.  Since subhorizon dark matter density perturbations grow linearly with the scale factor during matter domination, an EMDE generates a significantly enhanced population of sub-earth-mass dark matter halos if the dark matter particles are cold enough to form such structures \cite{ES11, Erickcek:2015jza}.  The masses and central densities of the smallest microhalos are determined by the small-scale cutoff in the matter power spectrum.  The rapid growth of perturbations during the EMDE implies that the observational signatures of these microhalos, such as the dark matter annihilation rates within their dense cores, are extremely sensitive to the scale of this cutoff \cite{Erickcek:2015jza, Erickcek:2015bda, Delos:2019dyh}.  If dark matter does not interact with SM particles, the small-scale cutoff is most often determined by the microphysics of the species that produces the altered cosmic evolution, making the microhalo population a probe of the particle physics of the early universe as well as its expansion history.  In the case of ECDEs, the small-scale cutoff in the matter power spectrum results from the thermal pressure of the cannibal particles.  In Ref. \cite{Erickcek:2020wzd}, we determined how the cutoff scale is set by the strength of the cannibal self-interactions and the mass of the cannibal field in scenarios in which the cannibal density exceeds the SM density up until the decay of the cannibal particles.
 
Here we extend our study of ECDEs to models with an arbitrary initial temperature ratio between the relativistic cannibal fluid and the SM plasma, which determines when the cannibal density exceeds the SM density.  We demonstrate that cannibal interactions continue to control the small-scale cutoff even when they freeze out while the cannibal is still subdominant to SM radiation.  We show that DM perturbations that experience the most growth are those with wavelengths on the same scale as the cannibal sound horizon, which is controlled by the strength of the cannibal self-interactions.  This enables us to extend the map between cannibal particle properties and the properties of the resulting microhalo population to the fully general case and establish its dependence on the initial temperature ratio between the hidden sector and the SM.

The organization of this paper is as follows. Section~\ref{sec:bkgd} discusses the homogeneous evolution of cosmologies with a period of early cannibal domination. In section~\ref{sec:pert}, we study perturbation growth in these cosmologies and highlight important length scales, showing that both the magnitude and scale of maximum DM perturbation growth are directly connected to cannibal particle properties. 
We discuss possible breakdowns of the perfect-fluid approximation in section~\ref{sec:can_not_perfect}. Implications of early cannibal-dominated eras for the earliest-forming microhalos are discussed in section~\ref{sec:params}, and we conclude in section~\ref{sec:conclusions}.
Several technical results are given in the appendices:  quantitative detail about our homogeneous background modeling is given in appendix~\ref{sec:A}, while appendix~\ref{sec:steady-state} contains derivations governing the evolution of cannibal density perturbations deep inside the cannibal sound horizon. In appendix~\ref{sec:can_kd} we compute the cannibal 2-to-2 scattering rate. Finally, in appendix~\ref{sec:der_pert} we derive the leading corrections to the cosmological perturbation equations from cannibal decays.

\section{Homogeneous background evolution}
\label{sec:bkgd}

We are interested in a universe comprised of three components: the cannibal species, DM, and the thermal SM radiation bath.  For simplicity, we consider the DM relic abundance to already be in place at the beginning of our analysis.  In the natural and minimal scenario where DM and the cannibal species are part of the same thermal bath in the early universe, we expect DM to be heavier than the cannibal: to experience cannibalism the cannibal species cannot be in equilibrium with any relativistic species while it is non-relativistic, and thus is generically the lightest state in that sector. 
The cannibal species must be thermally decoupled from the SM radiation bath, making the initial temperature ratio between the two sectors  a free parameter. 
Once the cannibal particle becomes non-relativistic, its energy density dilutes more slowly than that of the SM radiation and will eventually come to dominate the universe provided the cannibal is sufficiently long-lived. We focus on the parameter space where the universe undergoes  such an {\em early cannibal-dominated era} (ECDE) and caution that the cannibal may or may not be actively undergoing cannibalism during the ECDE.
The cannibal eventually decays into SM particles, which must occur before neutrino decoupling to avoid spoiling the successful predictions of BBN \cite{Kawasaki:1999na,Kawasaki:2000en,Hannestad:2004px,Ichikawa:2005vw} and altering the features of the CMB \cite{deSalas:2015glj,Hasegawa:2019jsa}.

We consider a simple model of a cannibal species given by a scalar field with cubic and quartic interactions,
\begin{align}\label{eq:L_can}
	\mathcal{L}_{\rm can}=\frac{1}{2}\partial^{\mu}\varphi\partial_{\mu}\varphi-\frac{1}{2}m^2\varphi^2-\frac{g}{3!}\varphi^3-\frac{\lambda}{4!}\varphi^4.
\end{align}
In this model three-to-two scattering processes maintain chemical equilibrium in the cannibal plasma even after the cannibal particles have become non-relativistic. In the non-relativistic limit, the $s$-wave component of the thermally-averaged $3\rightarrow2$ cross-section dominates and is given in terms of the leading $s$-wave piece of the matrix element, $|\mathcal{M}_s|^2$, by \cite{Kuflik:2017iqs}
\begin{align}\label{eq:general}
	\svc=\frac{\sqrt{5}}{2304\pi m^3}|\mathcal{M}_s|^2.
\end{align}
Using the matrix element calculated from the Lagrangian given in eq.~\eqref{eq:L_can}, we obtain
\footnote{In contrast to the expression in Refs.~\cite{Pappadopulo:2016pkp,Farina:2016llk}, we find a negative sign between the cubic and quartic couplings in the square brackets as well as an extra factor of $9/4$ in the normalization. When the cubic interaction arises from spontaneous breaking of the $Z_2$ symmetry taking $\phi\to -\phi$, this expression for the three-to-two cross-section vanishes at threshold, in accordance with the general result \cite{Smith:1992rq}.} 
\begin{align}\label{eq:alpha_can}
	\svc=&\frac{25\sqrt{5}(g/m)^2[(g/m)^2-3\lambda]^2}{147456\pi m^5}\equiv \frac{25\sqrt{5}\pi^2\alpha_c^3}{5184 m^5}.
\end{align} 
Here $\alpha_c$ parametrizes the combination of cannibal couplings that determines the strength of $3\rightarrow 2$ reactions, 
\begin{align}\label{eq:alpha_c}
	(4\pi\alpha_c)^3\equiv \frac{9}{4}(g/m)^2[(g/m)^2-3\lambda]^2.
\end{align}
While eq.~\eqref{eq:L_can} describes a specific cannibal model, it also provides a useful toy model for a broad class of theories with cannibal interactions. For instance, the lightest glueballs in a pure SU($N$) sector have cannibal interactions that can be described with an effective Lagrangian of the form in eq.~\eqref{eq:L_can} \cite{Boddy:2014yra,Boddy:2014qxa,Soni:2016gzf,Forestell:2016qhc,Buen-Abad2018}.

The Boltzmann equations that describe the homogeneous evolution of the cannibal fluid in the early universe, together with DM and SM radiation, are
\begin{align}
    \frac{d\rho_{\rm can}}{dt}+3H\rho_{\rm can}(1 +w_c(a))&=-\Gamma m n_{\rm can}\label{eq:num_can_density}\\
    \frac{d\rho_r}{dt}+4H\rho_r&=\Gamma m n_{\rm can}\label{eq:num_r_density}\\
    \frac{d\rho_{\rm DM}}{dt}+3H\rho_{\rm DM}&=0\label{eq:num_DM_density}\\
    \frac{dn_{\rm can}}{dt}+3Hn_{\rm can}&=\svc  n_{\rm can}^2(n_{\rm can,eq}-n_{\rm can})-\Gamma n_{\rm can}\label{eq:num_can_ndensity},
\end{align}
where the Hubble rate  is given by
\begin{align}\label{eq:Hubble_def}
H=\frac{1}{\sqrt{3}M_{\rm Pl}}\sqrt{\rho_{\rm can}+\rho_r+\rho_{\rm DM}},
\end{align}
$\rho_r$, $\rho_{\rm DM}$ and $\rho_{\rm can}$ are the energy densities of SM radiation, DM and the cannibals, respectively, $w_c$ is the cannibal equation of state, $n_{\rm can}$ is the cannibal number density and $n_{\rm can,eq}$ its equilibrium value, $\Gamma$ is the zero-temperature decay width of the cannibal particle to the SM, and $M_{\rm Pl}=2.435\times 10^{18}$ GeV is the reduced Planck mass. The collision operator describing cannibal decays that appears on the right-hand side of these equations is derived in appendix~\ref{sec:decayderivation}.

 Chemical equilibrium in the cannibal fluid is maintained as long as the $3\to 2$ scattering rate is rapid compared to $H$; the freeze-out of this cannibal interaction is described by
eq.~\eqref{eq:num_can_ndensity}.  We assume that two-to-two cannibal scatterings are fast enough to maintain internal kinetic equilibrium. Thus all the thermal quantities for the cannibal fluid can be expressed in terms of its chemical potential, $\mu$, and its temperature, $T_c$:
\begin{align}\label{eq:can_density_eq}
	\rho_{\rm can}=\int \frac{d^3p}{(2\pi)^3}E f\left(\frac{E-\mu}{T_c}\right), &&
	w_c(a)=\frac{\int \frac{d^3p}{(2\pi)^3}\frac{p^2}{3E} f\left(\frac{E-\mu}{T_c}\right)}{\int \frac{d^3p}{(2\pi)^3}E f\left(\frac{E-\mu}{T_c}\right)}, && 
	n_{\rm can}=\int \frac{d^3p}{(2\pi)^3}f\left(\frac{E-\mu}{T_c}\right),
\end{align}
where $f(x)=(e^{x}-1)^{-1}$ is the Bose-Einstein distribution. Consequently the system of eqs.~\eqref{eq:num_can_density}-\eqref{eq:num_can_ndensity} can be solved for the four unknowns $\rho_r, \rho_{\rm DM}, \mu$, and $T_c$.
We are interested in models that realize an epoch of cannibalization, which requires that the cannibal species becomes non-relativistic before the cannibal interactions freeze out.  Thus in many numerical integrals of interest we can approximate the cannibal phase space distribution with a Maxwell-Boltzmann distribution.  Further details concerning our numerical methods are given in appendix~\ref{sec:can_fz}.

We set our initial conditions at an initial scale factor $a_i$, defined such that 
\begin{align}\label{eq:ai_def}
T_c(a_i)=10m.
\end{align}
The cannibal fluid is in chemical equilibrium initially, so that $\mu(a_i)=0$. We find the initial DM density by scaling the observed relic density back in time.  Since the cannibal fluid and the SM radiation bath are necessarily thermally decoupled, the initial SM temperature $T_r(a_i)$ must be separately specified; we parameterize it with the initial temperature ratio
\begin{align}\label{eq:xi_def}
\xi_i\equiv\frac{T_c(a_i)}{T_r(a_i)}=\frac{10m}{T_r(a_i)}.
\end{align}

\begin{figure}
\centering
\includegraphics[width=0.7\textwidth]{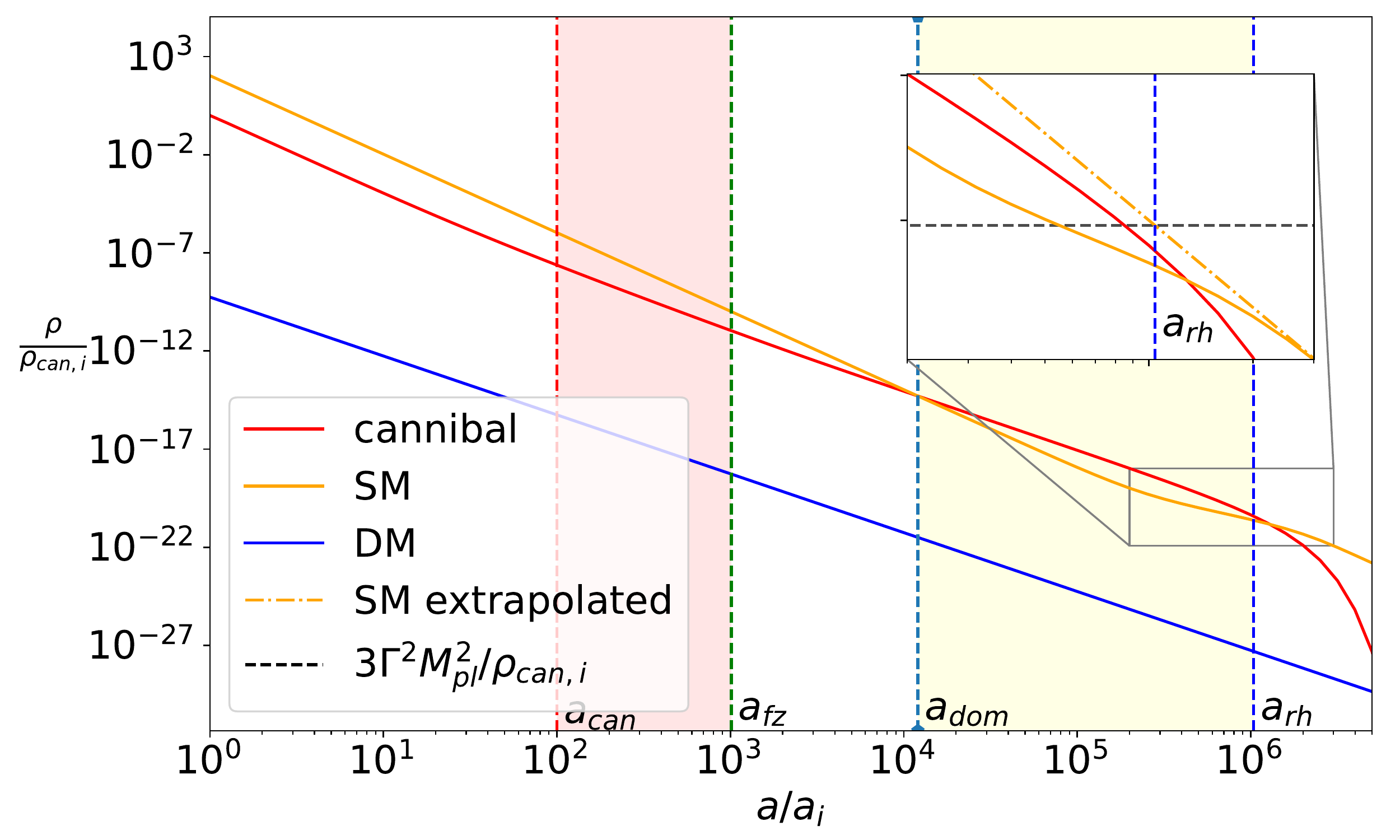}
\caption{Evolution of the cannibal (red), SM radiation (solid orange) and DM (solid blue) energy densities as a function of scale factor. The inset figure highlights our definition of $a_{\rm rh}$ (see eq.~\eqref{eq:Trh_def}) as the scale factor when the SM energy density extrapolated adiabatically back in time (orange dot-dashed) becomes equal to the total energy density required for the Hubble rate to equal decay rate of the cannibal particles (black dashed). The shaded red region highlights the period of cannibalism. The yellow shaded region highlights the period of early matter domination produced by the cannibal particles. This figure takes $m= 190$~GeV, $T_{\rm rh}=10$~MeV, $\alpha_c= 0.06$, and $\xi_i=1$.}
\label{fig:density_evolve}
\end{figure}

Figure~\ref{fig:density_evolve} shows the evolution of energy densities beginning from these initial conditions until the cannibal particles decay.  For $\xi_i=1$ as shown here, the cannibal is subdominant to SM radiation at $a_i$ but eventually comes to dominate. This fluid system goes through four important transitions, which we will discuss in turn: 1) the onset of cannibalism within the hidden sector, 2) the freeze-out of the 3-to-2 cannibal reactions, 3) the transition from SM radiation domination to cannibal domination, and 4) the decay of the cannibal particles into the SM.

At temperatures $T_c\gg m$, the cannibal behaves as radiation.   As the cannibal particles become non-relativistic ($T_c<m$), $2\rightarrow 3$ scattering processes become Boltzmann-suppressed while $3\rightarrow 2$ processes remain active. Thus the comoving number density depletes, which self-heats the cannibal particles  by converting rest mass to kinetic energy. In this ``cannibal'' phase of evolution, the cannibal temperature decreases as $T_c\propto 1/\log(a)$  \cite {Carlson:1992fn}.   More quantitatively, using the equilibrium Maxwell-Boltzmann distribution in eq.~\eqref{eq:can_density_eq} to find the equilibrium cannibal energy density and equation of state as a function of temperature,
then using the resulting expressions for $\rho_{\rm can,eq}$ and $w_{c,\textrm{eq}}$ in eq.~\eqref{eq:num_can_density}, we find while neglecting cannibal decay
\begin{align}
\label{eq:can_eq_evol}
	\rho_{\rm can,eq}\approx \frac{c_1m^4}{(a/a_i)^3\ln(a/(c_2a_i))}\\
	T_{c,\textrm{eq}}\approx\frac{m}{3\ln(a/(c_2a_i))}.
\label{eq:can_eq_evol2}
\end{align}
Here $c_1=148$ and $c_2=25.6$ are numerical factors obtained after numerical integration in the semi-relativistic regime, $10<m/T_c<0.1$; see
appendix~\ref{sec:anl_cannibalism}. As shown in Fig.~\ref{fig:can_intro}, these equations are an accurate description of equilibrium cannibal evolution  for $T_c\lesssim m/5$. Consequently, we define the scale factor $a_{\rm can}$ that marks the onset of cannibalism through 
\begin{align}
	T_c(a_{\rm can})\equiv \frac{m}{5}  ,
	\end{align}
which gives
\begin{align}
\label{eq:acan_def} 
a_{\rm can}=101 a_i.
\end{align}

As long as the cannibal fluid is in chemical equilibrium, the evolution of $\rho_{\rm can}(a)$ with $a$ is independent of the Hubble rate. However, the scale at which the cannibal fluid can no longer maintain chemical equilibrium depends on the Hubble rate and thus on the presence of other species. We define $a_{fz}$, the scale factor when the cannibal $3\rightarrow 2$ reactions freeze out, through
\begin{align}\label{eq:afz_def}
	\svc  n_{\rm can}^2(a_{fz})=H(a_{fz}).
\end{align}
After cannibal freeze-out, the temperature of the cannibal cools as $T_{c}\propto 1/a^2$, as expected for massive non-interacting particles. The residual pressure does not affect the evolution of the homogeneous cannibal density, it does affect the evolution of cannibal perturbations \cite{Erickcek:2020wzd}.  Here, in contrast to Ref.~\cite{Erickcek:2020wzd}, we consider scenarios where the SM radiation bath is important in determining $a_{fz}$, including cases  where the cannibal only comes to dominate after freeze-out (such as the scenario shown in figure~\ref{fig:density_evolve}).

The universe is initially SM radiation-dominated when $\rho_r(a_i)>\rho_{\rm can}(a_i)$ or
\begin{align}\label{eq:SM_dom_ai}
	g_{*}[T_r(a_i)]T_r^4(a_i)>(10m)^4,
\end{align}
where $g_*(T_r)$ 
is the effective number of degrees of freedom in the SM. For $g_*[T_r(a_i)]\sim 100$, SM radiation domination at $a_i$ then requires 
\begin{align}
	\xi_i\lesssim& 3.2.
\end{align}
A universe that is SM radiation dominated at $a_i$ will transition to cannibal domination at the scale factor $a_{\rm dom}$ where
\begin{align}\label{eq:adom_def}
	\rho_{\rm can}(a_{\rm dom})=\rho_r(a_{\rm dom}),
\end{align}
where we have implicitly assumed that the cannibal lifetime is long enough that it will come to dominate before it decays.
 
When $\Gamma$ exceeds the Hubble rate, the cannibal particles decay into the SM radiation bath and the universe then evolves as in the standard $\Lambda$CDM cosmology. We define the reheat temperature, $T_{\rm rh}$, by equating the Hubble rate in a SM radiation-dominated universe with the cannibal decay rate,
\begin{align}\label{eq:Trh_def}
\sqrt{\frac{\pi^2g_{*}(T_{\rm rh})}{30}}\frac{T_{\rm rh}^2}{\sqrt{3}M_{\rm Pl}}\equiv\Gamma.
\end{align}
We define the scale factor at reheating, $a_{\rm rh}$, by isentropically extrapolating the temperature of the SM from $T_{\rm rh}$ to the present-day temperature $T_0$,
\begin{align}\label{eq:entropy_cons}
g_{*s}(T_{\rm rh})(a_{\rm rh}T_{\rm rh})^3=g_{*s}(T_{0})(a_{0}T_{0})^3.
\end{align} 
Here $g_{*s}$ is the effective number of entropic degrees of freedom in the SM and $a_0$ is the present-day scale factor. Note that with the above definition of $a_{\rm rh}$, the temperature of the SM at $a_{\rm rh}$, $T_{r}(a_{\rm rh})$, is not equal to $T_{\rm rh}$. This can be seen in the inset panel of figure~\ref{fig:density_evolve} where the SM energy density (solid orange line) at $a_{\rm rh}$ is smaller than the radiation density when $T=T_{\rm rh}$ (horizontal black dashed line). Figure~\ref{fig:density_evolve} also shows that the SM radiation density evolves adiabatically until the energy injection rate from the cannibal fluid into the radiation becomes of order the Hubble rate ($\rho_{\rm can}\Gamma/\rho_{r}\sim H $).   After this time, the radiation density is proportional to $\Gamma m n_{\rm can}/H$ until the cannibal  energy density becomes subdominant.

\begin{figure}
\centering
\includegraphics[width=0.7\textwidth]{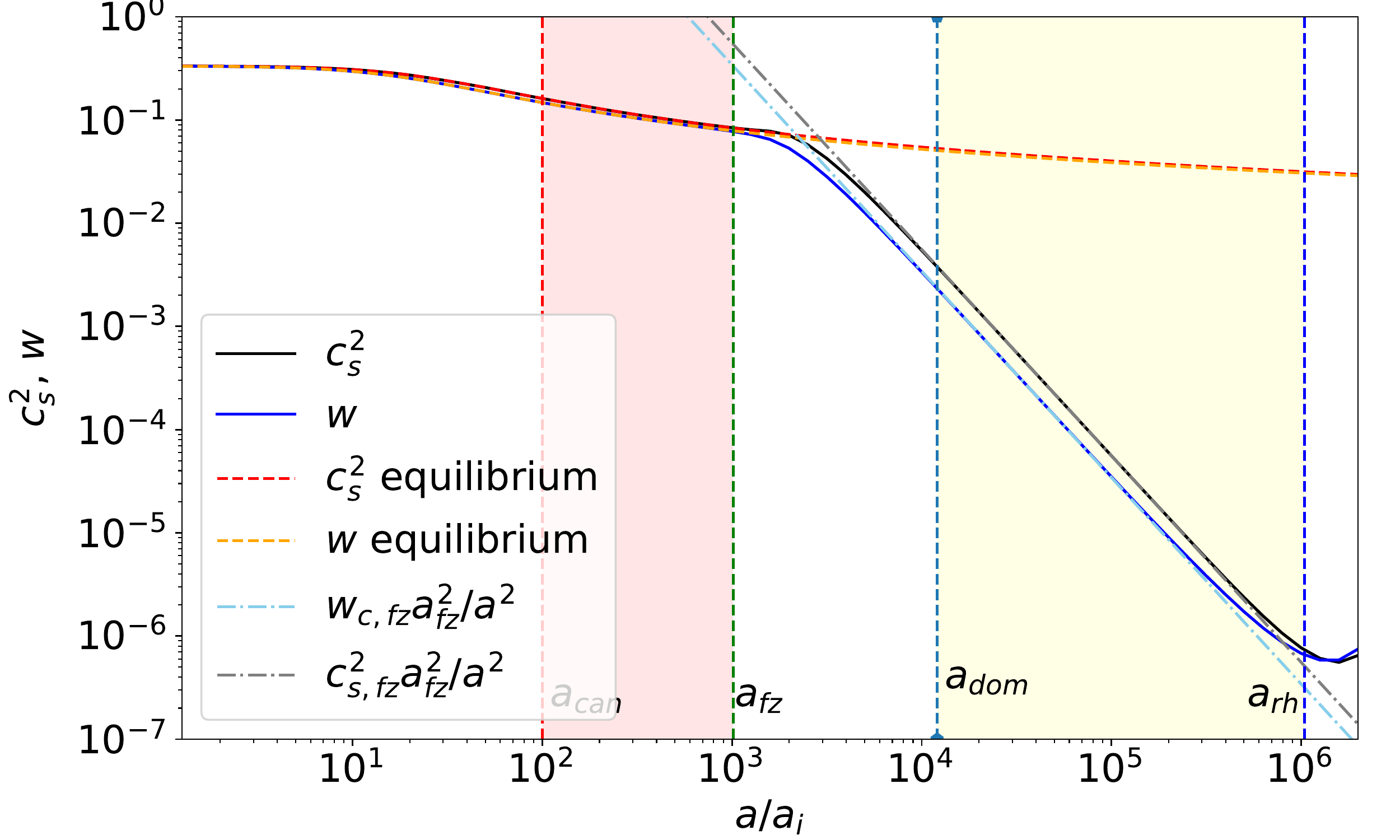}
\caption{Evolution of the cannibal sound speed $c^2_s$ and equation of state $w$ from initial cannibal temperature $T_c(a_i)=10 m$. The red (yellow) dashed line shows the evolution of $c_s^2$ ($w$) in thermal equilibrium, the gray (cyan) dot-dashed line when completely decoupled, and the solid black (blue) line shows the full numerical result. This figure uses the same parameter values as figure~\ref{fig:density_evolve}.}
\label{fig:c2s_fz}
\end{figure}
In solving perturbation equations we need the cannibal equation of state $w_c$ and the cannibal sound speed
\begin{align}\label{eq:c2s_def}
c^2_s=\frac{\partial \mathcal{P}_{\rm can}}{\partial \mathcal{\rho}_{\rm can}}= w_c-\frac{aw_c'(a)}{3(1+w)},
\end{align}
where $\mathcal{P}_{\rm can}=w_c\rho_{\rm can}$ is the pressure of the cannibal fluid. In figure~\ref{fig:c2s_fz} we plot the evolution of both $w_c$ and $c^2_s$ as a function of scale factor. Both quantities begin to deviate from their equilibrium values around $2a_{\rm fz}$. For $a\gg a_{\rm fz}$ both quantities evolve as 
\begin{align}\label{eq:c2s_fz}
c^2_s= c^2_{s,\textrm{fz}}\frac{a_{\rm fz}^2}{a^2} && w_c= w_{c,\textrm{fz}}\frac{a_{\rm fz}^2}{a^2},
\end{align}
where $c^2_{s,\textrm{fz}}$ and $w_{c,\textrm{fz}}$ are constants that give the correct asymptotic evolution (see dot-dashed line in figure~\ref{fig:c2s_fz}). We find that $c_{s,\textrm{fz}}$, to a good approximation, is given by $c_{s, \textrm{eq}}(3a_{\rm fz})$ while $w_{c,\textrm{fz}}$ is given by $w_{c,\textrm{eq}}(2a_{\rm fz})$, where the subscript $eq$ denotes that the variables are computed assuming the cannibal fluid to be in thermal equilibrium. Then using the fact that in the non-relativistic limit $c_{s, \textrm{eq}}^2\approx w_{c,\textrm{eq}}\approx T_{c,\textrm{eq}}/m$, we obtain
\begin{align}
	c_{s,\textrm{fz}}^2\approx\frac{1}{3\ln(3a_{\rm fz}/(c_2a_i))} && w_{c,\textrm{fz}}\approx\frac{1}{3\ln(2a_{\rm fz}/(c_2a_i))}.
\end{align}

In figure~\ref{fig:c2s_fz}, both $w_c$ and $c_s^2$ increase near $a_{\rm rh}$ because the cannibal particles with larger velocities decay later due to time dilation. Consequently, the temperature of the cannibal fluid increases as cannibal particles with slower speeds are removed first. However, the heating near $a_{\rm rh}$ is unimportant for the evolution of dark matter perturbations because we are interested in scenarios with $a_{\rm rh}\gg a_{\rm fz}$. Thus, the cannibal fluid is too cold at $a_{\rm rh}$ for the heating due to time dilation to have any impact.

\subsection{Mapping between cannibal parameters and cosmological scales}\label{sec:param_map}

Our early cannibal-dominated cosmology is governed by four free parameters: the initial temperature ratio $\xi_i$ and the cannibal particle properties
$m,T_{\rm rh},$ and $\alpha_c $. These four parameters determine the three important scales $a_{\rm fz}/a_i,\  a_{\rm dom}/a_i$, and  $a_{\rm rh}/a_i$ that will ultimately control the major features of the matter power spectrum.  Due to the non-trivial evolution of the cannibal density, the exact relations between these cosmological scales and the underlying cannibal parameters are complicated, but useful approximate relations can be obtained by fixing $a/a_i=10^3$ in the logarithm of the expression for $\rho_{\rm can}(a)$ given in eq.~\eqref{eq:can_eq_evol}:
\begin{align}\label{eq:can_eq_evol_sim}
\rho_{\rm can}\sim \frac{40m^4}{(a/a_i)^3}.
\end{align}
This approximation is accurate to $\mathcal{O}(1)$ for scale factors between $a_{\rm can}$ and $a_{\rm rh}/10$ and will enable us to provide simple expressions for key quantities, albeit at the cost of obscuring logarithmic dependence on $a_{\rm fz}/a_i$.

We can  express $a_{\rm rh}/a_i$ in terms of $m$ and $T_{\rm rh}$ by setting the cannibal density at reheating equal to the radiation density. Using eq.~\eqref{eq:can_eq_evol_sim} for the cannibal density then gives
\begin{align}\label{eq:arh_est}
\frac{a_{\rm rh}}{a_i}\sim 2.3\left(\frac{m}{T_{\rm rh}}\right)^{4/3}\left(\frac{g_*(T_{\rm rh})}{10}\right)^{-1/3}.
\end{align}
Similarly, we can find $a_{\rm dom}$ in terms of $\xi_i$ using eq.~\eqref{eq:can_eq_evol_sim} for the cannibal density  in the definition of $a_{\rm dom}$, eq.~\eqref{eq:adom_def}.
With $\rho_r(a_{\rm dom})=\rho_r(a_i)(a_i/a_{\rm dom})^4$, we then find
\begin{align}\label{eq:adom_est}
\frac{a_{\rm dom}}{a_i}\sim 80g_{*}[T_r(a_{i})]\frac{1}{\xi_i^4}.
\end{align}

To express $a_{\rm fz}$ in terms of cannibal parameters, we start with its definition in eq.~\eqref{eq:afz_def}. We then approximate $n_{\rm can}(a_{\rm fz})\approx \rho_{\rm can}(a_{\rm fz})/m$ and express $\rho_{\rm can}$ and $\svc$ using eq.~\eqref{eq:can_eq_evol_sim} and eq.~\eqref{eq:alpha_can} respectively.
In the case where the Hubble rate is dominated by the cannibal density during freeze-out, i.e. $a_{\rm dom}<a_{\rm fz}$, we obtain
\begin{align}\label{eq:asfz_est_can}
	\frac{a_{\rm fz}}{a_i}\sim 3\times 10^4\alpha_c^{2/3}\Big(\frac{\rm GeV}{m}\Big)^{2/9}.
\end{align}
Similarly, in the case where the Hubble rate is dominated by the SM radiation during freeze-out, i.e. $a_{\rm dom}>a_{\rm fz}$, we use $\rho_r(a_{\rm fz})=\rho_r(a_i)(a_i/a_{\rm fz})^4$ to obtain
\begin{align}\label{eq:asfz_est_sm}
	\frac{a_{\rm fz}}{a_i}\sim 3.3\times 10^4 \alpha_c^{3/4}\xi_i^{1/2}\left(\frac{\rm GeV}{m}\right)^{1/4}\left(\frac{g_*(10m/\xi)}{100}\right)^{-1/8}.
\end{align}
We see that $a_{\rm fz}/a_i$ decreases slowly as $m$ increases because increasing $m$ reduces  the $3\rightarrow 2$ cross-section for fixed $\alpha_c$. When the universe is SM dominated at $a_{\rm fz}$, $a_{\rm fz}/a_i$ decreases as we decrease $\xi_i$ because decreasing $\xi_i$ increases $\rho_r$, which in turn increases the Hubble rate, causing freeze-out to occur earlier.

\section{Evolution of perturbations}
\label{sec:pert}

In this section we describe the evolution of cosmological perturbations during an ECDE with particular focus on the physics underlying the growth in DM density perturbations. 
We follow the conventions used in Ma and Bertschinger \cite{Ma:1995ey}. We work in conformal Newtonian gauge with metric given by
\begin{align}
ds^2=-(1+2\psi)dt^2+a^2(t)(1-2\phi)dx^2,
\end{align}
where $\phi$ and $\psi$ are spatial and temporal metric perturbations respectively. We consider all fluids to be perfect fluids. 
Thus there is no anisotropic stress, which implies that
\begin{align}
\psi=\phi.
\end{align}
In section~\ref{sec:can_not_perfect} we revisit the perfect-fluid assumption for the DM and cannibal fluids.

Perturbations of perfect fluids can be described by two quantities: $\delta=[\rho(t,x^i)-\bar{\rho}(t)]/\bar{\rho}$, which is the density perturbation over the uniform background density $\bar{\rho}$, and $\theta=\partial_{j}v^j$, which is the comoving divergence of the physical fluid velocity, $v^j=a\,dx^j/dt$. Our cannibal perturbation equations are similar to those in \cite{Buen-Abad2018, Heimersheim:2020aoc, deLaix:1995vi} but we also include leading-order contributions from cannibal decays into radiation, which are derived in appendix~\ref{sec:der_pert}. Our suite of perturbation equations is then
\begin{gather}
    \delta_c'(a)+(1+w_c)(\frac{ {\theta_c}}{a^2 {H}}-3\phi')+\frac{3}{a}\left(1-\frac{\Gamma}{2H}\right)(c_s^2-w_c)\delta_c=-\frac{ {\Gamma}}{a {H}}\phi\left(1-\frac{3}{2}w_c\right),\label{eq:delta_can_eq}\\
     {\theta_c}'(a)+\frac{1}{a}(1-3w_c)\theta_c+\frac{w_c'}{1+w_c} {\theta_c}-\frac{c_s^2}{1+w_c}\frac{ {k}^2}{a^2 {H}}\delta_c-\frac{ {k}^2}{a^2 {H}}\phi=\frac{\Gamma}{aH}\theta_c c_s^2,\label{eq:theta_can_eq}\\
    \delta_{\rm DM}'(a)+\frac{ {\theta}_{\rm DM}}{a^2 {H}}-3\phi'=0,\label{eq:delta_dm_eq}\\
     {\theta}_{\rm DM}'(a)+\frac{1}{a} {\theta}_{\rm DM}-\frac{ {k}^2}{a^2 {H}}\phi=0,\label{eq:theta_dm_eq}\\
    \delta_r'(a)+\frac{4}{3}\frac{ {\theta}_r}{a^2 {H}}-4\phi'=\frac{\Gamma mn_{\rm can}}{aH\rho_r}\left[\phi + \delta_c - \delta_r +\frac32\delta_c(w_c-c_s^2)\right],\label{eq:delta_r_eq}\\
     {\theta}_r'(a)-\frac{1}{4}\frac{ {k}^2}{a^2 {H}}\delta_r-\frac{ {k}^2}{a^2 {H}}\phi=\frac{ {\Gamma}mn_{\rm can}}{a {H}\rho_r}\Big(\frac{3}{4}\theta_c-\theta_r\Big),\label{eq:theta_r_eq}\\
          {k}^2\phi+3(a {H})^2\Big(a\phi'+\phi\Big)=-\frac{1}{2}a^2\frac{1}{M_{\rm Pl}^2}(\rho_{\rm can}\delta_c+\rho_r\delta_r+\rho_{\rm DM}\delta_{\rm DM})\label{eq:metric_eq}.
\end{gather}
Here the subscripts $c,r$ and $DM$ corresponds to perturbations of the cannibal, SM radiation and DM fluids respectively, and the prime denotes a derivative with respect to $a$. We have taken DM to be kinetically decoupled from both the cannibal and radiation fluids, so that its only interactions are gravitational; we will discuss the effects of adding kinetic couplings between cannibal and DM fluids below.

At $a_i$, when we begin our numerical calculations, the cannibal particles are still relativistic since $T_c(a_i)=10m$.  For adiabatic perturbations, the initial conditions for super horizon modes at $a_i$ are:
\begin{align}\label{eq:initial_condition}
\delta_r=\frac{4}{3}\delta_{\rm DM}=\delta_c=-2\phi_p && \theta_r=\theta_{\rm DM}=\theta_c=\frac{1}{2}\frac{k^2}{aH}\phi_p,
\end{align}
where $\phi_p$ is the primordial metric perturbation.  Adiabatic initial conditions for all fields are naturally obtained in the minimal cosmological scenario where the decays of a single inflaton field populate both the SM and a hidden sector containing the cannibals and DM.\footnote{Strictly speaking, these adiabatic initial conditions are applicable to $\rho_r$ as long as energy injection from cannibal decays is negligible at $a_i$.  When instead $\Gamma\rho_{\rm can}/\rho_r\gg H$ at $a_i$, $\rho_r\propto a^{-2}$, and the initial conditions for the radiation perturbations become $\delta_r=\phi_p/2$ and $\theta_r=\theta_c$.}

Our primary interest is the evolution of modes that enter the horizon prior to reheating and thus experience the epoch of modified cosmic expansion. 
Before numerically solving the perturbation equations given in eqs.~\eqref{eq:delta_can_eq}-\eqref{eq:theta_r_eq}, we first show how they simplify for modes deep inside the horizon ($k\gg aH$) during the ECDE to gain  insight into the essential physics governing the growth of DM density perturbations.
Starting with eq.~\eqref{eq:metric_eq} for the metric perturbation, we neglect the second term on the LHS of eq.~\eqref{eq:metric_eq}. We can also ignore $\rho_{\rm DM}\delta_{\rm DM}$ on the RHS of eq.~\eqref{eq:metric_eq} because $\rho_{\rm DM}$ is at least seven orders of magnitude smaller than $\rho_{\rm can}$ and $\rho_r$ prior to reheating (see figure~\ref{fig:density_evolve}), which must occur before BBN. Consequently, deep inside the horizon and prior to reheating we have
\begin{align}\label{eq:phi_subhz}
	\phi&=-\frac{3}{2}\left(\frac{aH}{k}\right)^2\frac{\rho_{\rm can}\delta_c+\rho_r\delta_r}{\rho_{\rm can}+\rho_r}.
\end{align}

Next we consider the evolution of the cannibal perturbations because they determine the evolution of DM perturbations. We use eq.~\eqref{eq:theta_can_eq} to eliminate $\theta_c$ from eq.~\eqref{eq:delta_can_eq}. In doing so, we make three approximations. First, we neglect terms proportional to $c_s^2-w_c, w_c'(a)$ and $d(c_s^2(a))/da$, as $w_c$ and $c_s$ are slowly varying before $a_{\rm fz}$ and rapidly become negligible after $a_{\rm fz}$. Second, we neglect $\phi'$ in eq.~\eqref{eq:delta_can_eq} because the variation of the metric perturbation is negligible compared to $\theta_c/(aH)$ deep inside the horizon. Third, we neglect terms proportional to $\Gamma/H$: before $a_{\rm rh}$ we have $\Gamma/H\ll1$, and after $a_{\rm rh}$, the cannibal fluid decays and becomes irrelevant. Around $a_{\rm rh}$, when $\Gamma/H\sim\mathcal{O}(1)$, the metric perturbation multiplying $\Gamma$ in eq.~\eqref{eq:delta_can_eq} is negligible compared to $\delta_c$ for modes deep within the horizon, and the sound speed term multiplying $\Gamma$ in eq.~\eqref{eq:theta_can_eq} is much smaller than one by the time of reheating (see figure~\ref{fig:c2s_fz}). 
Finally we eliminate $\phi$ using eq.~\eqref{eq:phi_subhz} to obtain
\begin{align}\label{eq:delta_can_subhz_full}
\delta_c''(a)+\Big[\frac{(a^2 {H})'}{a^2 {H}}+\frac{1}{a}(1-3w_c)\Big]\delta_c'+\frac{1}{a^2}\left(\frac{ c_s{k}}{aH}\right)^2\delta_c=
&\frac{3}{2}\frac{(1+w_c)}{a^2 }\frac{\rho_{r}\delta_r+\rho_{\rm can}\delta_c}{\rho_{\rm can}+\rho_r}.
\end{align}

Naively, eq.~\eqref{eq:delta_can_subhz_full} implies that $\delta_r$ may affect $\delta_c$ during SM radiation domination, when $\rho_r\gg \rho_{\rm can}$. 
However, subhorizon radiation perturbation oscillate, and thus their gravitational influence on $\delta_c$ is negligible. Consequently one can set $\delta_r=0$ and rewrite eq.~\eqref{eq:delta_can_subhz_full} in the form
\begin{align}\label{eq:delta_can_subhz_adom}
\frac{\textrm{d}^2\delta_c(a)}{\textrm{d}\ln^2(a)}-3w_c\frac{\textrm{d}\delta_c(a)}{\textrm{d}\ln(a)}+\left[\left(\frac{ c_s{k}}{aH}\right)^2-\frac{3}{2}(1+w_c)\frac{\rho_{\rm can}}{\rho_{\rm can}+\rho_r}\right]\delta_c=
&0.
\end{align}
The first term in the square brackets arises from thermal pressure in the cannibal fluid and induces oscillations in the cannibal density perturbation. 
The second term in the square brackets is inconsequential during SM radiation domination, but during cannibal domination, it induces growth in the cannibal perturbation due to the gravitational attraction between the cannibal particles.  When $\rho_{\rm can}\gg \rho_r$, the terms in the square brackets thus determine a Jeans wavenumber, $k_J$, for the cannibal fluid,
\begin{align}\label{eq:jeans}
    k_J\equiv\sqrt{\frac{3}{2}(1+w_c)}\ \frac{aH}{c_s}.
\end{align}
The corresponding Jeans length scale $k_J^{-1}$ determines when gravitational attraction overcomes the thermal pressure and leads to growth in $\delta_c$. 

If the cannibals dominate the universe prior to cannibal freeze-out, then the Jeans length grows until $a\sim 2a_{\rm fz}$ because $c_s^2$ and $w_c$ decay logarithmically with scale factor until $2a_{\rm fz}$ (see figure~\ref{fig:c2s_fz}) while $H\propto a^{-3/2}$ up to logarithmic factors. After $2a_{\rm fz}$, the Jeans length decays as $k_{J}^{-1}\propto 1/\sqrt{a}$ because $c_s$ now decays as $\propto1/a$, while $w_c \ll 1$ and can be neglected.   In this case, the cannibal Jeans length $k_{J}^{-1}$ is the relevant length scale separating oscillating modes from growing modes in the cannibal fluid and thereby determines 
the growth of DM modes during the ECDE.  Thus when the cannibal species is always dominant over the SM, the wave number $k_{\rm pk}$ for which $\delta_{\rm DM}$  experiences maximum growth during the ECDE can be simply related to the maximum value of the Jeans length,  $k_{\rm pk}^{-1}\sim1.4k_J^{-1}(2a_{\rm fz})$ \cite{Erickcek:2020wzd}.
This peak wavenumber $k_{\rm pk}$ determines the characteristic mass of the earliest-forming DM microhalos, as we discuss further in section~\ref{sec:params}.

We now consider scenarios in which the cannibal is initially subdominant to SM radiation.  To understand the evolution of DM perturbations in this scenario, we again need to determine the length scale that separates oscillating and growing cannibal modes. While the universe is SM dominated,  we find that this length scale is determined by the cannibal sound horizon.  After analyzing the evolution of perturbations in the next
subsection, 
we provide a new analytical estimate of $k_{\rm pk}$ for scenarios with $a_{\rm dom}>2a_{\rm fz}$ in section~\ref{sec:an_est_transfer}.

\subsection{Cannibal freeze-out during SM radiation domination}
\label{sec:perts_initraddom}

We can understand the essential behavior of the perturbations in cosmologies with an initially-subdominant cannibal density by separately considering the evolution of perturbation modes that enter the horizon prior to cannibal freeze-out and modes that enter the horizon between 
cannibal freeze-out and cannibal decay.

\subsubsection{Modes that enter the horizon prior to cannibal freeze-out}

\begin{figure}
\centering
\includegraphics[width=0.8\textwidth]{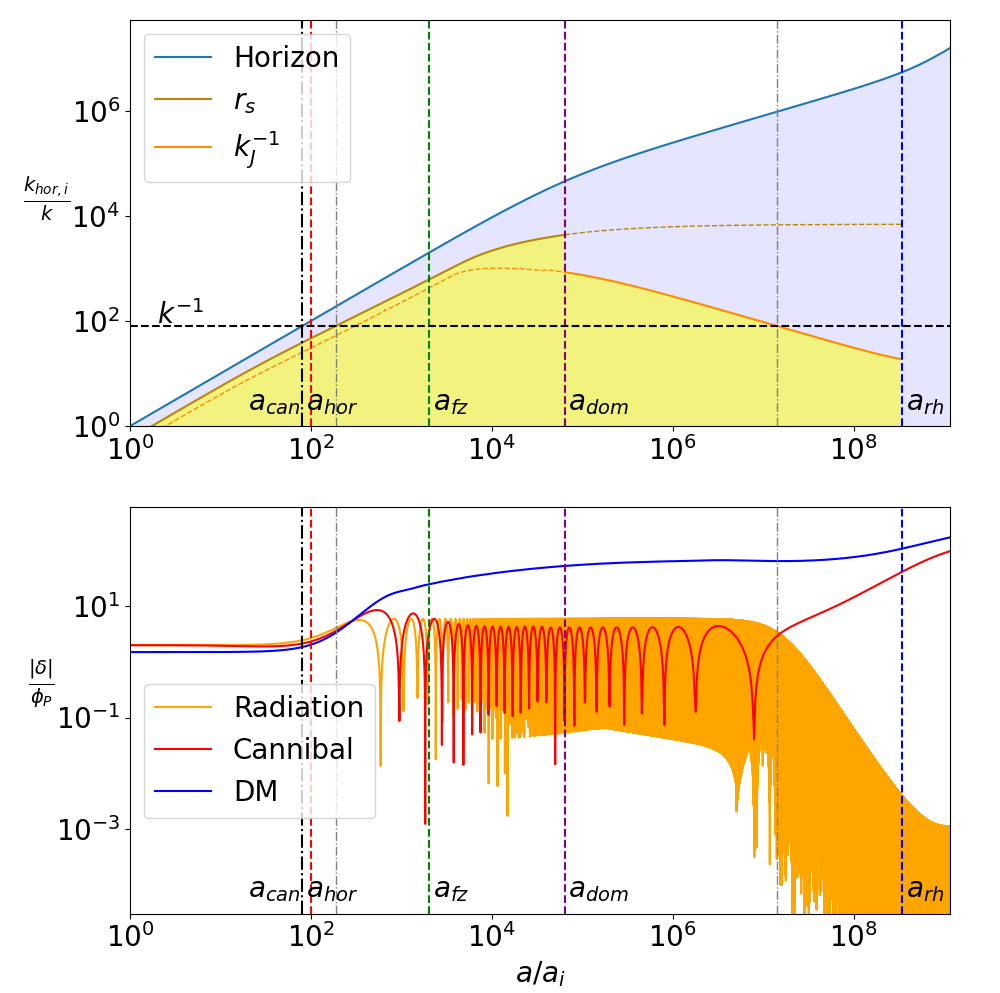}
\caption{\textbf{Top}: Comoving horizon (solid blue line), cannibal Jeans length (solid yellow line) and cannibal sound horizon (solid brown line) relative to the horizon size at $a_i$,  with $k_{\rm hor,i}^{-1}\equiv [a_iH(a_i)]^{-1}$. The dashed yellow (brown) line represents when $k_J^{-1}$ ($r_s$) is plotted in regimes outside of its validity. The horizontal black dashed line marks the Fourier mode, $k=12.5\times 10^{-3}k_{\rm hor,i}$, of the perturbations shown in the bottom panel. The shaded yellow region highlights the region within which $\delta_c$ oscillates. \textbf{Bottom:} Evolution of SM radiation, cannibal and DM density perturbations relative to the primordial metric perturbation, $\phi_p$. The vertical dashed red, green, purple, and blue lines marks the scale at $a_{\rm can}$, $a_{\rm fz}$,  $a_{\rm dom}$, and $a_{\rm rh}$ respectively. The vertical dot-dashed black line marks the scale  $a_{\rm hor}$ when the mode enters the horizon. The vertical dashed grey lines mark the scale factors when the mode enters the sound horizon and exits the Jeans horizon. The figure has been plotted for  parameters $\xi_i= 0.68$, $m= 15$~TeV, $T_{\rm rh}=10$~MeV and $\alpha_c =1$, for which we obtain $k_{\rm hor,i}^{-1}=2\times 10^{-6}$ pc.}
\label{fig:cann_dom_horz_pert}
\end{figure}

In the top panel of figure~\ref{fig:cann_dom_horz_pert} we show the comoving horizon, $(aH)^{-1}$ (solid blue), the cannibal Jeans length (solid orange), 
and the comoving cannibal sound horizon,
\begin{align}\label{eq:rs}
	r_s\equiv\int^t c_s\frac{d\tilde{t}}{\tilde{a}}=\int^{a} \frac{c_s}{\tilde{a}H}\textrm{d}\ln(\tilde{a}).
\end{align}
These scales will determine the evolution of perturbation modes that enter the horizon prior to cannibal freeze-out.  In the bottom panel of figure~\ref{fig:cann_dom_horz_pert} we show the evolution of density perturbations for one such mode, indicated in the top panel as the black-dashed line.  These results are obtained by numerically solving eqs.~\eqref{eq:delta_can_eq}-\eqref{eq:theta_r_eq} with initial conditions given by eq.~\eqref{eq:initial_condition}.

The perturbations shown in the bottom panel start to evolve once the mode enters the horizon at $a_{\rm hor}$, defined through
\begin{align}\label{eq:ahor}
a_{\rm hor}H(a_{\rm hor})\equiv k.
\end{align}
All density perturbations grow by a factor of 10 to 100 shortly after they enter the horizon. 
Inside the horizon, thermal pressure causes both the cannibal and SM radiation density perturbations to oscillate, whereas the DM density perturbation grows logarithmically during radiation domination and then approaches a constant value while the cannibal perturbation oscillates during ECDE. 

When the wavenumber $k$ is much larger than $aH/c_s$, the time scale of cannibal oscillations is much smaller than the time scale over which the instantaneous frequency and the anti-damping terms in eq.~\eqref{eq:delta_can_subhz_adom} evolve. Thus, one can use a WKB approximation  to obtain
\begin{align}\label{eq:delta_can_wkb_rs}
    \delta_c\approx\frac{C_1}{\sqrt{c_s}}\exp\left(-\int_{a_{\rm s,hor}}^{a}\frac{1-3w_c}{2}\textrm{d}\ln(\tilde{a})\right)\sin\left[kr_s+C_2\right],
\end{align}
as detailed in appendix~\ref{sec:steady-state}.
Here $a_{\rm s,hor}$ is the scale factor for which $c_sk/(aH)=1$, and $C_1$ and $C_2$ are constants determined by $\delta_c(a_{\rm s,hor})$ and $\delta_c'(a_{\rm s,hor})$. Note that the instantaneous frequency of $\delta_c$ oscillations is  set by the cannibal sound horizon.

In the bottom panel of figure~\ref{fig:cann_dom_horz_pert}, we see that the amplitude of $\delta_c$ oscillations decreases slowly for $a<2a_{\rm fz}$ and reaches a constant value for $a>2a_{\rm fz}$. The slow decay prior to $2a_{\rm fz}$ results from the logarithmic decay of $c_s$ partially compensating for the exponential in eq.~\eqref{eq:delta_can_wkb_rs}. While $a\gtrsim 2a_{\rm fz}$, $c_s$ and $w_c$ decay as $c_s=c_{s,\textrm{fz}}a_{\rm fz}/a$ and $w_c=w_{c,\textrm{fz}}a_{\rm fz}^2/a^2$ (see figure~\ref{fig:c2s_fz}). Inserting this evolution in eq.~\eqref{eq:delta_can_wkb_rs}, one can check that the amplitude of $\delta_c$ remains constant after $2a_{\rm fz}$.

Once cannibal domination begins, the Jeans length is again the scale that controls the oscillations in $\delta_c$. For instance, the $\delta_c$ oscillations in the bottom panel of figure~\ref{fig:cann_dom_horz_pert} end when the mode exits the Jeans horizon in the top panel. The linear growth of $\delta_c$ after the mode exits the Jeans horizon can be seen analytically by solving eq.~\eqref{eq:delta_can_subhz_full} while neglecting $c_s, w_c$, and $\rho_r$ and using the fact that Hubble rate evolves as $H\propto a^{-3/2}$.

The evolution of $\delta_r$ seen in the bottom panel of figure~\ref{fig:cann_dom_horz_pert}, although interesting, has no significant impact on $\delta_{\rm DM}$. Radiation perturbations have an important gravitational impact on $\delta_{\rm DM}$ only during SM radiation domination. However, during SM radiation domination $\delta_r$ oscillates, and hence its gravitational feedback on both $\delta_{\rm DM}$ and $\delta_c$ is negligible. The only time $\delta_r$ has a significant influence on the other perturbations is near horizon entry ($a\lesssim 10a_{\rm hor}$) before $\delta_r$ starts oscillating.

In the bottom right panel of figure~\ref{fig:cann_dom_horz_pert} we see that $\delta_{\rm DM}$ (blue line) grows logarithmically for $a_{\rm hor}<a<a_{\rm dom}$. This is the expected evolution for $\delta_{\rm DM}$ in a radiation-dominated universe and is given by
\begin{align}\label{eq:log_growth}
	\delta_{\rm DM}(a)=-A_s\phi_p(k)\ln\left(\frac{B_sa}{a_{\rm hor}}\right) && a_{\rm hor}<a<a_{\rm dom},
\end{align}
where $A_s=9.11$ and $B_s=0.594$ are numerical fitting factors \cite{Hu:1995en}.
After the universe becomes cannibal dominated, $\delta_{\rm DM}$ is constant until $\delta_c$ grows to be of order $\delta_{\rm DM}$, after which $\delta_{\rm DM}$ grows linearly as $\delta_{\rm DM}=\delta_c\propto a$. After reheating, $\delta_{\rm DM}$ again grows logarithmically. Consequently, the growth experienced by DM perturbations during an ECDE is determined by the growth of the cannibal perturbation.

\subsubsection{Modes that enter the horizon after cannibal freeze-out}

\begin{figure}
	\includegraphics[width=1.00\textwidth]{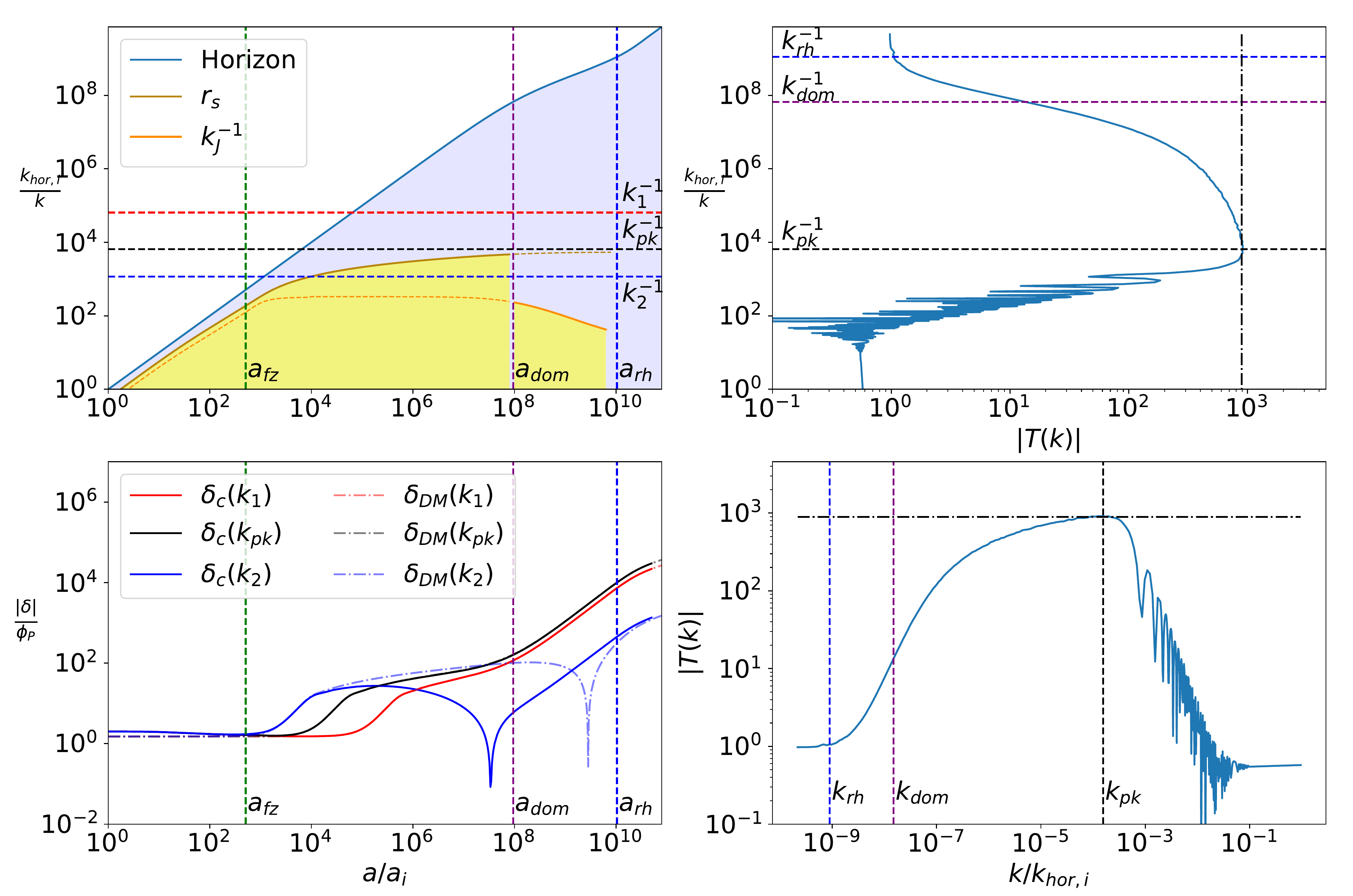}
	\caption{\textbf{Top left}: Comoving horizon (blue solid line), sound horizon (brown solid line) and Jeans length (yellow solid line) relative to $k_{\rm hor,i}^{-1}\equiv [a_iH(a_i)]^{-1}$. Horizontal dashed lines mark the wavenumbers corresponding to the perturbations in the bottom panel. \textbf{Top right}: Absolute value of the transfer function, eq.~\eqref{eq:transfer}, on the $x$-axis, for every inverse wavenumber $k^{-1}$ shown on the $y$-axis. The horizontal blue (purple) dashed line marks the wavenumber that enters the horizon at $a_{\rm rh}$ ($a_{\rm dom}$). The horizontal black dashed line marks the analytical estimate of the wavenumber $k_{\rm pk}$ that maximizes $\delta_{\rm DM}/\phi_P$, eq.~\eqref{eq:kpk_sm}. The vertical black dot-dashed line is our analytical estimate of the the transfer function at horizontal black dashed line, eq.~\eqref{eq:peak_sm}. \textbf{Bottom left}: Evolution of $\delta_c$ (solid) and $\delta_{\rm DM}$ (light dot-dashed) for $k_{\rm pk}$ and two other modes $k_1=k_{\rm pk}/10$ and $k_2=4k_{\rm pk}$. \textbf{Bottom right}: Transfer function as in the top right panel but rotated by ninety degrees. In this figure we take $\xi_i=0.1$, $m= 180$~TeV, $T_{\rm rh}=10$~MeV and $\alpha_c = 1 $.}
	\label{fig:SM_dom_horz_peak}
\end{figure}

The homogeneous cannibal fluid behaves like pressureless matter after $2a_{\rm fz}$ because $w_c\ll1$ in this regime. Thus, for modes that enter the horizon after $2a_{\rm fz}$, one might expect $\delta_c$ to simply evolve as expected for pressureless matter, i.e., with $\delta_c$ growing logarithmically between horizon entry and the end of radiation domination. However, lingering thermal pressure affects the evolution of the cannibal perturbations that enter the horizon shortly after cannibal freeze-out. The sound horizon grows logarithmically for $a_{\rm fz}<a<a_{\rm dom}$, as can be seen analytically by substituting $H\propto 1/a^2$ and $c_s=c_{s,\textrm{fz}}a_{\rm fz}/a$ in eq.~\eqref{eq:rs}:
\begin{align}\label{eq:rs_sfz}
	r_s\sim \frac{c_{s,\textrm{fz}}}{(aH)_{\rm fz}}\ln\left(\frac{a}{2a_{\rm fz}}\right) && 2a_{\rm fz}<a<a_{\rm dom}.
\end{align}
This logarithmic growth of $r_s$ is also evident in figure~\ref{fig:cann_dom_horz_pert} and in the top right panel of figure~\ref{fig:SM_dom_horz_peak}. Since the sound horizon continues to grow while the universe is radiation dominated, modes that enter the horizon after cannibal freeze-out may still oscillate. For example, in the bottom left panel of figure~\ref{fig:SM_dom_horz_peak} we can see that $\delta_c(k_2)$ undergoes oscillations once $k_2^{-1}$ enters the sound horizon in the top left panel.

To better understand the evolution of $\delta_c$ for modes entering the horizon between $2a_{\rm fz}$ and $a_{\rm dom}$, we solve eq.~\eqref{eq:delta_can_subhz_adom}. For $a>a_{\rm fz}$, the anti-damping term in eq.~\eqref{eq:delta_can_subhz_adom} rapidly decays while the frequency remains constant. Moreover, since $w_c$ is already much less than one by $a_{\rm fz}$, the anti-damping term never has a significant impact, as we have verified numerically. Consequently, for $a> a_{\rm fz}$, eq.~\eqref{eq:delta_can_subhz_adom} simplifies to a simple harmonic oscillator equation in $\ln(a)$. Using $c_s=c_{s,\textrm{fz}}a_{\rm fz}/a$ and $H=H(a_{\rm fz})a_{\rm fz}^2/a^2$ for $a>a_{\rm fz}$, we can exactly solve this simple harmonic oscillator equation to obtain
\begin{align}\label{eq:temp_pk_sol}
    \delta_c=\tilde{C}_1\sin\left(\frac{c_{s,\textrm{fz}}k}{a_{\rm fz}H(a_{\rm fz})}\ln\left(\tilde{C}_2\frac{a}{a_{\rm hor}}\right)\right) && a_{\rm fz},a_{\rm hor}<a<a_{\rm dom},
\end{align}
where $\tilde{C}_1$ and $\tilde{C}_2$ are constants.

Eq.~\eqref{eq:temp_pk_sol} suggests that the cannibal perturbation evolves logarithmically with $a$ for a short time after horizon entry. Since  $c_{s,\textrm{fz}}k/(aH)_{\rm fz} \ll 1$ for modes entering the horizon after $a_{\rm fz}$, eq.~\eqref{eq:temp_pk_sol} simplifies to
\begin{align}\label{eq:temp_pk_sol2}
    \delta_c\approx \tilde{C}_1\frac{c_{s,\textrm{fz}}k}{(aH)_{\rm fz}} \ln\left(\tilde{C}_2\frac{a}{a_{\rm hor}}\right),
\end{align}
while $a$ does not greatly exceed $a_{\rm hor}$.
Hence the naive expectation that $\delta_c$ should evolve logarithmically for modes entering the horizon after cannibal freeze-out does hold for a brief period after horizon entry. During this period of logarithmic growth, the influence of thermal pressure is initially negligible. However, the influence of thermal pressure keeps growing logarithmically until it becomes large enough that the argument of the sine becomes $\mathcal{O}$(1) and $\delta_c$ begins to oscillate. 

We can find $\tilde{C}_1$ and $\tilde{C}_2$ for modes entering the horizon after $2a_{\rm fz}$  by using the fact that the super horizon initial condition is the same for both the cannibal and DM perturbations in this regime because the cannibal particles are non-relativistic. Consequently, the early-time solution in eq.~\eqref{eq:temp_pk_sol2} should match the standard logarithmic growth of DM during radiation domination in eq.~\eqref{eq:log_growth}.
By matching eq.~\eqref{eq:temp_pk_sol2} to eq.~\eqref{eq:log_growth} we find the constants to be $\tilde{C}_1=-A_s\phi_p(aH)_{\rm fz}/(c_{s,\textrm{fz}}k)$ and $\tilde{C}_2=B_s$. It follows that the cannibal perturbation evolution within the horizon is given by
\begin{align}\label{eq:temp_pk_sol3}
    \delta_c(a)=-A_s\phi_p\frac{(aH)_{\rm fz}}{c_{s,\textrm{fz}}k}\sin\left(\frac{c_{s,\textrm{fz}}k}{(aH)_{\rm fz}}\ln\left(B_s\frac{a}{a_{\rm hor}}\right)\right) && a_{\rm hor}<a<a_{\rm dom}.
\end{align}
The fact that the cannibal and DM perturbations follow the same initial evolution after horizon entry can be seen in the bottom left panel of figure~\ref{fig:SM_dom_horz_peak}.

Note that the argument of the sine in eq.~\eqref{eq:temp_pk_sol3} is similar to $kr_s$. By using the approximate evolution of $r_s$ given by eq.~\eqref{eq:rs_sfz}, we find the difference between $kr_s$ and the argument of the sine in eq.~\eqref{eq:temp_pk_sol3}  to be $\sim c_{s,\textrm{fz}}k/(aH)_{\rm fz}\ln(a_{\rm hor}/2a_{\rm fz})$. This difference is much less than one for $a_{\rm hor}>2a_{\rm fz}$ because $c_{s,\textrm{fz}}\ll1$ and $k/(aH)_{\rm fz}=a_{\rm fz}/a_{\rm hor}<1$, which follows from the fact that $H\propto a^{-2}$ between $a_{\rm fz}$ and $a_{\rm hor}$. Thus $\delta_c$ deviates from the logarithmic growth experienced by $\delta_{\rm DM}$ approximately when the mode enters the sound horizon. This can also be seen in the left panels of figure~\ref{fig:SM_dom_horz_peak}, where the intersection of $k_2^{-1}$ and $r_s$ in the top panel coincides with $\delta_c(k_2)$ beginning to deviate from $\delta_{\rm DM}(k_2)$ in the bottom panel.

The dark matter density perturbation evolves in the same manner as in the previous subsection: it grows logarithmically between $a_{\rm hor}$ and $a_{\rm dom}$, after which it remains constant until $\delta_c$ grows to be of order $\delta_{\rm DM}$, and then it grows linearly along with the cannibal perturbation, $\delta_{\rm DM}=\delta_c\propto a$. Note that this linear growth occurs independently of whether the mode enters the horizon before or after $a_{\rm fz}$, as long as the mode is outside the cannibal Jeans horizon during cannibal domination.

So far we have discussed modes that enter the horizon prior to cannibal domination. For modes that enter the horizon after $a_{\rm dom}$, $\delta_c$ evolves as cold matter, since $k^{-1}\gg r_s(a_{\rm dom})$. Thus, the evolution of perturbations for modes that enter the horizon after $a_{\rm dom}$ is the same as those studied in early matter-dominated eras \cite{ES11, BR14, FOW14, Erickcek:2015jza, Blinov:2019jqc}.

\subsubsection{The linear matter power spectrum after an ECDE}\label{sec:transfer}

We now compare the present-day linear matter power spectrum following an ECDE to the matter power spectrum in standard cosmology. Here, by ``standard cosmology'' we mean that the universe experienced uninterrupted  radiation domination between inflationary reheating and matter-radiation equality, $a_{\rm eq}$. With $\delta_{\rm DM,s}$ denoting the DM density perturbation in the standard cosmology, we define the transfer function
\begin{align}
T(k)\equiv \frac{\delta_{\rm DM}(k,a)}{\delta_{\rm DM,s}(k,a)} ,
\end{align}
which is evaluated after matter-radiation equality.

After a perturbation mode enters the horizon, $\delta_{\rm DM,s}$ grows logarithmically with scale factor while the universe is radiation dominated, as described in eq.~\eqref{eq:log_growth}. Since an ECDE changes the scale factor at a mode's horizon entry, we must use the appropriate value of $a_{\rm hor}$ when evaluating eq.~\eqref{eq:log_growth}. With $H_s$ denoting the Hubble rate in the standard cosmology, we define $a_{\rm hor,s}$ through 
\begin{align}
	a_{\rm hor,s}H_s(a_{\rm hor,s})\equiv k .
\end{align}
The logarithmic evolution of $\delta_{\rm DM}$ continues until matter-radiation equality, after which $\delta_{\rm DM}$ grows linearly with scale factor. This evolution is described by the growing solution of the Meszaros equation \cite{1974A&A....37..225M,Hu:1995en} with initial conditions provided by eq.~\eqref{eq:log_growth},
\begin{align}\label{eq:Meszaros}
\delta_{\rm DM,s}(a)=-\frac{3A_s\phi_p(k)}{2}\ln\left(\frac{4B_se^{-3}a_{\rm eq}}{a_{\rm hor,s}}\right)\left(1+a/a_{\rm eq}\right)^{0.9} && a>a_{\rm eq}.
\end{align}
Here the exponent of $0.9$ results from the fact that the scales affected by an ECDE are much smaller than the baryon Jeans length \cite{Bertschinger:2006nq}. Consequently, $\sim$15\% of the matter density does not participate in the gravitational growth, causing the dark matter overdensities to undergo slower than linear growth. The argument of the logarithmic term in eq.~\eqref{eq:Meszaros} also obtains $\mathcal{O}(1)$ baryonic corrections as described in \cite{Hu:1995en}. However, we have ignored these  corrections because they have an insignificant effect on the final transfer function.
The value of $\phi_p$ in eq.~\eqref{eq:Meszaros} is the same as in eq.~\eqref{eq:initial_condition} because the universe is radiation dominated at $a_i$ regardless of whether the cannibal fluid or SM radiation is dominant at $a_i$.

During the radiation-dominated era that follows an ECDE, $\delta_{\rm DM}$ also grows logarithmically with scale factor for subhorizon modes. Consequently, the post-reheating evolution of $\delta_{\rm DM}$ can be described by eq.~\eqref{eq:log_growth}, but with $A_s$ and $B_s$ replaced by $k$-dependent functions $A(k)$ and $B(k)$, which encode the evolution history of $\delta_{\rm DM}$ prior to reheating. After matter-radiation equality, the evolution of $\delta_{\rm DM}$ can similarly be described using eq.~\eqref{eq:Meszaros} but with $A_s$ and $B_s$ again replaced by $A(k)$ and $B(k)$. For $k<k_{\rm rh}$, we recover $A(k)=A_s$ and $B(k)=B_s$.
It follows that
\begin{align}\label{eq:transfer}
   T(k)\equiv\frac{\delta_{\rm DM}(k,a)}{\delta_{\rm DM,s}(k,a)}
    \approx&\frac{A(k)}{A_s}\frac{\ln[4B(k)e^{-3}a_{\rm eq}/a_{\rm hor}(k)]}{\ln[4B_se^{-3}a_{\rm eq}/a_{\rm hor,s}(k)]},
\end{align}
where in the latter equality we have neglected baryonic effects in the logarithm.

In the bottom right panel of figure~\ref{fig:SM_dom_horz_peak} we plot the transfer function for a scenario with $a_{\rm dom}\gg a_{\rm fz}$. To increase computational speed while evaluating the transfer function, we ignore radiation perturbations deep inside the horizon, as the feedback of $\delta_r$ on $\delta_{\rm DM}$ and $\delta_c$ is negligible. We also neglect the heating of the cannibal fluid caused by its decay (see figure~\ref{fig:c2s_fz}), which has no noticeable impact on $\delta_{\rm DM}$ in this regime.

The transfer function is unity for modes that enter the horizon after reheating
because $\delta_{\rm DM}$ and $\delta_{\rm DM,s}$ undergo the same evolution for these modes. As we increase $k$, the transfer function increases as approximately $k^2$ until $k\sim k_{\rm dom}\equiv (aH)_{\rm dom}$. This quadratic increase results from the linear growth of $\delta_{\rm DM}$ between $a_{\rm hor}$ and $a_{\rm rh}$. Since the horizon size grows as $(aH)^{-1}\propto a^{1/2}$ between $a_{\rm dom}$ and $a_{\rm rh}$, $a_{\rm hor}\propto k^{-2}$. Consequently, $\delta_{\rm DM}(a_{\rm rh})$ scales as $\delta(a_{\rm rh})\propto a_{\rm rh}/a_{\rm hor}\propto k^2$. In the case of $\delta_{\rm DM,s}$, the modes entering the horizon earlier undergo more logarithmic growth between $a_{\rm hor,s}$ and $a_{\rm eq}$ but the same linear growth after $a_{\rm eq}$. Since the horizon size grows as $(aH_s)^{-1}\propto a$ during radiation domination, an increase in $k$ results in a linear decrease in $a_{\rm hor,s}$, which implies that $\delta_{\rm DM,s} \propto \ln[k/(8k_{\rm eq})]$ \cite{Hu:1995en}, where $k_{\rm eq}\equiv (aH)_{\rm eq}$.  Thus in this regime the transfer function goes as
$k^2/\ln[k/(8k_{\rm eq})]$.

For $k>k_{\rm dom}$, modes enter the horizon during SM radiation domination. 
Modes with larger $k$ now see $\delta_{\rm DM}$ undergo a larger logarithmic growth between $a_{\rm hor}$ and $a_{\rm dom}$ but the same linear growth between $a_{\rm dom}$ and $a_{\rm rh}$ (seen, for example, in the differing growth of $\delta_{\rm DM}(k_1)$ and $\delta_{\rm DM}(k_{\rm pk})$ in the bottom left panel of fig.~\ref{fig:SM_dom_horz_peak}). Consequently, an increase in $k$ results in a $\sim \ln[k/(8k_{\rm dom})]$ increase in the growth experienced by $\delta_{\rm DM}$.  Thus the transfer  function increases as approximately $\ln[k/(8k_{\rm dom})]/\ln[k/(8k_{\rm eq})]$, and as $k_{\rm dom}\gg k_{\rm eq}$, the $k$-dependence of the transfer function is primarily driven by the logarithm in the numerator.

As we further increase $k$ beyond $k_{\rm dom}$ in the top right panel of figure~\ref{fig:SM_dom_horz_peak}, the transfer function continues to grow until $k^{-1}$ intersects the cannibal sound horizon in the top left panel.  For these modes, the cannibal perturbations undergo oscillations while their wavelength is contained within the yellow shaded region in the top left panel, which inhibits the growth of DM perturbations. Modes with larger $k$ spend more time within the Jeans horizon, and thus have less time to grow prior to reheating.  Since the Jeans length decays as $a^{-1/2}$ for $a>a_{\rm fz}$, the envelope of the transfer function for  $k>k_{\rm pk}$ falls as $k^{-2}/\ln[k/(8k_{\rm eq})]$.
 The oscillations in $T(k)$ are caused by changes in the phase of the cannibal perturbation between horizon entry and Jeans horizon exit; the dark matter inherits these oscillations because it falls into the gravitational wells generated by the cannibal perturbations after they stop oscillating \cite{Erickcek:2020wzd}. When  $\delta_c$ keeps oscillating until $a_{\rm rh}$, however, the cannibal mode does not have a net gravitational impact on  $\delta_{\rm DM}$, and thus for  $k>k_J(a_{\rm rh})$ the oscillations in $T(k)$ stop.

Large values of the transfer function imply that the DM perturbations after an ECDE reach the nonlinear regime much earlier than they would in a standard cosmology. Once $\delta_{\rm DM}\gtrsim 1$, overdense fluctuations collapse to form halos \cite{1980lssu.book.....P}. In section~\ref{sec:params} we discuss how $k_{\rm pk}$ determines the mass of the earliest-forming microhalos, while $T(k_{\rm pk})$ determines the redshift of their formation. Due to the importance of $k_{\rm pk}$ and $T(k_{\rm pk})$ in controlling microhalo formation, in the following subsection we find analytical estimates for both quantities and highlight their connection to cannibal parameters.

\subsection{Analytical estimate of the peak of the matter power spectrum}
\label{sec:an_est_transfer}

If  cannibal reactions freeze out while the universe is SM dominated, then the matter power spectrum peaks near the smallest-scale mode that never enters the cannibal sound horizon,
i.e.,
\begin{align}
	k_{\rm pk}\approx r_s^{-1}(a_{\rm dom}).
\end{align}
We can more accurately determine $k_{\rm pk}$ by using the fact that $\delta_{\rm DM}(a_{\rm rh})=\delta_c(a_{\rm rh})$ for wavenumbers in the vicinity of $k_{\rm pk}$. For these wavenumbers, $\delta_c$ and $\delta_{\rm DM}$ undergo the same amount of linear growth between $a_{\rm dom}$ and $a_{\rm rh}$, as illustrated in the bottom left panel of figure~\ref{fig:SM_dom_horz_peak}. Consequently, we can find $k_{\rm pk}$ by finding the wavenumber that maximizes $\delta_c(a_{\rm dom})$.

As the top left panel of figure~\ref{fig:SM_dom_horz_peak} demonstrates, wavenumbers with $k\sim r_s^{-1}(a_{\rm dom})$ enter the horizon between $a_{\rm fz}$ and $a_{\rm dom}$.  We can estimate $\delta_c(a_{\rm dom})$ for these modes using the analytical solution for $\delta_c$ given in eq.~\eqref{eq:temp_pk_sol3}.  Strictly speaking, the approximations yielding eq.~\eqref{eq:temp_pk_sol3} do not include the gradual transition to cannibal domination around $a_{\rm dom}$, but as the impact of this transition is similar for all modes with $k$ between $k_{\rm dom}$ and $k_{\rm fz}$, these neglected terms will not affect the determination of $k_{\rm pk}$.

In eq.~\eqref{eq:temp_pk_sol3}, some of the $k$-dependence is hidden inside $a_{\rm hor}$. This dependence can be made explicit by using the fact that the modes near $k_{\rm pk}$ enter the horizon during radiation domination, yielding $k/(aH)_{\rm fz}=(aH)_{\rm hor}/(aH)_{\rm fz}=a_{\rm fz}/a_{\rm hor}$. Expressing $k$ in terms of $a_{\rm hor}$ and defining $\gamma\equiv a_{\rm hor}/a_{\rm fz}$, eq.~\eqref{eq:temp_pk_sol3} becomes
\begin{align}\label{eq:temp_pk_sol4}
    \delta_c(a_{\rm dom})\approx -A_s\phi_p\frac{\gamma}{c_{s,\textrm{fz}}}\sin\left(\frac{c_{s,\textrm{fz}}}{\gamma} \ln\left(\frac{B_s}{\gamma}\frac{a_{\rm dom}}{a_{\rm fz}}\right)\right).
\end{align}
Apart from the weak $k$-dependence in $\phi_p$ ($\phi_p\propto k^{-0.02}$) \cite{Aghanim:2018eyx}, the rest of the $k$ dependence is now encoded inside $\gamma$. We maximize $\delta_c(a_{\rm dom})$ by taking $\phi_p$ to be constant and
setting the derivative of $\delta_c(a_{\rm dom})$ with respect to $\gamma$ to zero. Doing so yields
\begin{align}\label{eq:temp_kpk}
    \tan\left[\frac{c_{s,\textrm{fz}}}{\gamma} \ln\left(\frac{B_s}{\gamma}\frac{a_{\rm dom}}{a_{\rm fz}}\right)\right]=\frac{c_{s,\textrm{fz}}}{\gamma}+\frac{c_{s,\textrm{fz}}}{\gamma} \ln\left(\frac{B_s}{c_{s,\textrm{fz}}\gamma}\frac{a_{\rm dom}}{a_{\rm fz}}\right).
\end{align}
This equation has multiple solutions. We want the largest $\gamma$ for which the above equation is satisfied, because the largest solution corresponds to the value of $a_{\rm hor}$ for which $\delta_c$ does not oscillate. Since oscillations suppress $\delta_c(a_{\rm dom})$, this mode is the global maximum of $\delta_c(a_{\rm dom})$. For the largest $\gamma$ satisfying the above equation, the tangent is well-approximated by a second-order Taylor expansion. After simplifying the resulting equation we obtain
\begin{align}\label{eq:gamma}
    \gamma_{\rm pk}=\frac{3c_{s,\textrm{fz}}}{2\sqrt{2}}W^{3/2}\left[2\left(\frac{B_s}{3c_{s,\textrm{fz}}}\frac{a_{\rm dom}}{a_{\rm fz}}\right)^{2/3}\right],
\end{align}
where $W$ is the Lambert function. Using $\gamma=a_{\rm hor}/a_{\rm fz}=(aH)_{\rm fz}/k$, we obtain
\begin{align}\label{eq:kpk_sm}
k_{\rm pk}^{-1}=\frac{\gamma_{\rm pk}}{(aH)_{\rm fz}}.
\end{align}
In the right panels of figure~\ref{fig:SM_dom_horz_peak}, this estimate of $k_{\rm pk}$ is shown with a black dashed line.  

Eq.~\eqref{eq:kpk_sm} holds as long as $a_{\rm dom}\gg 2a_{\rm fz}$.  If instead the universe was cannibal dominated during the freeze-out of cannibal reactions, then the peak Fourier wavenumber is \cite{Erickcek:2020wzd}
\begin{align}\label{eq:kpk_can}
	k_{\rm pk}^{-1}=1.4k_J^{-1}(2a_{\rm fz}).
\end{align}
While Ref.~\cite{Erickcek:2020wzd} considered scenarios where the cannibal density always dominated over SM radiation density prior to reheating, the derivation of eq.~\eqref{eq:kpk_can} only requires that the universe be radiation dominated when cannibal freeze-out occurs. Consequently, eq.~\eqref{eq:kpk_can} is also valid for initially subdominant cannibals as long as $a_{\rm dom}$ is smaller than $2a_{\rm fz}$. 
scenarios.

\begin{figure}
	\centering
	\includegraphics[width=0.8\textwidth]{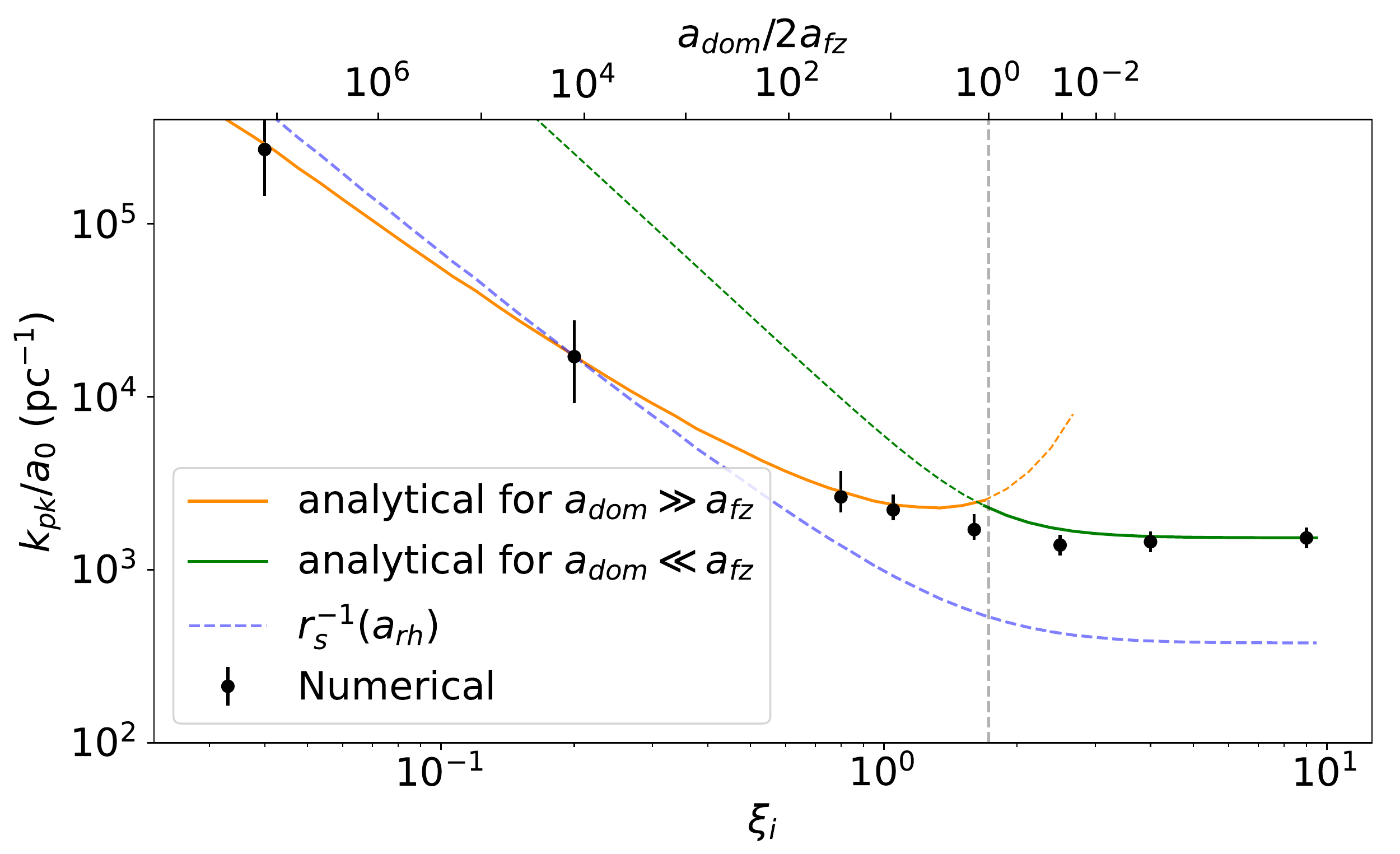}
	\caption{Wavenumber corresponding to the peak of the matter power spectrum, $k_{\rm pk}$, as a function of initial temperature ratio between cannibals and SM radiation, $\xi_i=T_c(a_i)/T_r(a_i)$. The yellow (green) solid line shows the estimate of $k_{\rm pk}$ given in eq.~\eqref{eq:kpk_sm} (eq.~\eqref{eq:kpk_can}); the dashed yellow and green dashed lines show extrapolations of these analytical estimates outside their regime of validity. Black dots are the numerically obtained values of $k_{\rm pk}$. The bars on the black dots show the range of $k$ around $k_{\rm pk}$ for which $\delta_{\rm DM}(k,a_{\rm eq})$ is within 5\% of $\delta_{\rm DM}(k_{\rm pk},a_{\rm eq})$. The blue dashed line is the inverse of the cannibal sound horizon (see eq.~\eqref{eq:rs}) at $a_{\rm rh}$. The vertical gray dashed line marks the point at which $a_{\rm dom}=2a_{\rm fz}$. In the secondary $x$-axis on top, we show the values of $a_{\rm dom}/(2a_{\rm fz})$. For $\xi_i\gtrsim3.2$, the SM radiation density is subdominant to the cannibal density until reheating, and $a_{\rm dom}$ is not defined. In this figure we take $m=300$ TeV, $\alpha_c=0.2$ and $T_{\rm rh}=8$ MeV. We have restricted the plot to $\xi_i>0.03$ because for smaller $\xi_i$ the cannibal density never exceeds SM density.}
	\label{fig:kpk_vs_xi}
\end{figure}

In figure~\ref{fig:kpk_vs_xi}, we plot the estimates of $k_{\rm pk}$ given in eqs.~\eqref{eq:kpk_sm} and~\eqref{eq:kpk_can} as functions of $\xi_i$. We also show the values of $k$ that maximize $\delta_{\rm DM}(k,a_{\rm eq})/\phi_P$ as black dots. Since $\phi_P$ is only weakly scale-dependent, the peak location in $\delta_{\rm DM}(k,a_{\rm eq})/\phi_P$ is an excellent estimate of the peak location in the matter power spectrum. The bars around the black dots indicate the range of $k$ for which $\delta_{\rm DM}(k,a_{\rm eq})/\phi_P$ is within 5\% of its maximum value. This range is larger for $a_{\rm dom}<2a_{\rm fz}$ because the logarithmic evolution of $\delta_c$ prior to $a_{\rm dom}$ causes the transfer function to be flatter near the peak (see figure~\ref{fig:SM_dom_horz_peak}). Our analytical estimates are accurate approximations to the numerical results except in the transition region where $a_{\rm dom}\sim 2a_{\rm fz}$. In figure~\ref{fig:kpk_vs_xi}, we also plot the cannibal sound horizon at reheating, $r_s(a_{h})$. We plot $r_s(a_{\rm rh})$ and not $r_s(a_{\rm dom})$ because $r_s(a_{\rm dom})$ provides an approximate estimate of $k_{\rm pk}$ only for $a_{\rm dom}>2a_{\rm fz}$, while $r_s(a_{\rm rh})$ can provide an order-of-magnitude estimate of $k_{\rm pk}$ for all values of $\xi_i$.

To understand how $k_{\rm pk}$ depends on $\xi_i$, we need to express $k_{\rm pk}$ in terms of cannibal parameters: $m,\ \alpha_c,\ T_{\rm rh},$ and $\xi_i$. We first use $\gamma_{\rm pk}\approx \ln^{3/2}(a_{\rm dom}/a_{\rm fz})/3.5$ in eq.~\eqref{eq:kpk_sm}. This approximation for $\gamma_{\rm pk}$ is an empirical relation that we found to be accurate to within $20\%$ for $10<a_{\rm dom}/a_{\rm fz}<10^5$ and $10^2<a_{\rm fz}/a_i<10^4$. Furthermore, we use $(aH)_{\rm fz}=k_{\rm hor,i}a_i/a_{\rm fz}$ because the universe is radiation dominated between $a_i$ and $a_{\rm fz}$. We can get an estimate of $k_{\rm hor,i}$ as a function of our cannibal parameters by splitting $a_i/a_0=a_i/a_{\rm rh}\times a_{\rm rh}/a_0$, estimating $a_{\rm rh}/a_0$ using eq.~\eqref{eq:entropy_cons} and $a_{\rm rh}/a_i$ using eq.~\eqref{eq:arh_est}, and using the initial densities of SM radiation and cannibals in the Hubble rate to obtain
\begin{align}\label{eq:khor_est}
	\frac{k_{\rm hor,i}}{a_0}\sim 34.5\ {\rm pc}^{-1}\times\left(\frac{T_{\rm rh}}{\rm 10\, MeV}\right)^{1/3}\left(\frac{m}{\rm GeV}\right)^{2/3}\sqrt{[1+g_*(10m/\xi_i)\xi_i^{-4}]}.
\end{align}
Finally, expressing $a_{\rm fz}$ and $a_{\rm dom}$ in terms of cannibal parameters using eq.~\eqref{eq:asfz_est_sm} and eq.~\eqref{eq:adom_est} respectively, we obtain for $a_{\rm dom}>2a_{\rm fz}$
\begin{multline}\label{eq:kpk_sm_param}
\frac{k_{\rm pk}}{a_0}\sim  3.7\times 10^{-2}\ {\rm pc}^{-1} \alpha_c^{-3/4}\xi_i^{-5/2}\left(\frac{m}{\rm GeV}\right)^{11/12}\left(\frac{T_{\rm rh}}{\rm 10\, MeV}\right)^{1/3}\left(\frac{g_*(10m/\xi_i)}{100}\right)^{5/8}\\
\times\ln^{-3/2}\left(\frac{1}{4}\alpha_c^{-3/4}\xi_i^{-9/2}\left(\frac{m}{\rm GeV}\right)^{1/4}\left(\frac{g_{*}(10m/\xi_i)}{100}\right)^{9/8}\right) {\rm pc}^{-1}.
\end{multline}
The location of the peak in the transfer function thus depends on all four cannibal parameters, but it is most sensitive to $\xi_i$.
In contrast,  $k_{\rm pk}$ can be estimated by \cite{Erickcek:2020wzd}
\begin{align}\label{eq:kpk_can_param}
\frac{k_{\rm pk}}{a_0}\sim 0.05\times \alpha_c^{-1/3}\left(\frac{m}{\rm GeV}\right)^{7/9}\left(\frac{T_{\rm rh}}{\rm 10\, MeV}\right)^{1/3} {\rm pc}^{-1},
\end{align}
for cases when $a_{\rm dom}<2a_{\rm fz}$, which is independent of the initial temperature ratio, $\xi_i$.

While our analytic estimates of $k_{\rm pk}$ in the scenarios with $a_{\rm dom}<2a_{\rm fz}$ and $a_{\rm dom}>2a_{\rm fz}$ are given by different functions, the underlying scale determining $k_{\rm pk}$ in both cases is the cannibal sound horizon, up to an order of magnitude. For $a_{\rm dom}>2a_{\rm fz}$ we have already shown that $k_{\rm pk}^{-1}\approx r_s(a_{\rm dom})$. In the case of $a_{\rm dom}<2a_{\rm fz}$ one can show that $r_s(2a_{\rm fz})$ is within an $\mathcal{O}(1)$ factor of $k_J^{-1}(2a_{\rm fz})\approx k_{\rm pk}^{-1}$. Moreover, after cannibals freeze out and dominate the universe, $r_s$ asymptotes to a constant value (see figure~\ref{fig:cann_dom_horz_pert} and \ref{fig:SM_dom_horz_peak}) as seen by inserting $H\propto a^{-3/2}$ and $c_s\propto 1/a$ in the definition of $r_s$,  eq.~\eqref{eq:rs}. Consequently, the total comoving distance traveled by the sound waves in the cannibal fluid, $r_s(a_{\rm rh})$, provides an order-of-magnitude estimate of $k_{\rm pk}$ for all scenarios. In figure~\ref{fig:kpk_vs_xi}, we compare  the dependence of $r_s^{-1}(a_{\rm rh})$ on $\xi_i$ to that of $k_{\rm pk}$. The value of $r_s^{-1}(a_{\rm rh})$ always falls within a factor of 5 from $k_{\rm pk}$ in our parameter space.

Next we find an analytic estimate of $T(k_{\rm pk})$ for scenarios with $a_{\rm dom}>2a_{\rm fz}$.\footnote{Due to the presence of $\delta_{\rm DM,s}$ in the denominator of the transfer function (eq~\eqref{eq:transfer}), the location of the peak in the transfer function is slightly different from $k_{\rm pk}$. However, this difference is negligible, as can be seen in figure~\ref{fig:SM_dom_horz_peak}.} Since the value of $\delta_{\rm DM}(k_{\rm pk},a_{\rm rh})$ is equal to that of $\delta_c(k_{\rm pk},a_{\rm rh})$, we first estimate the value of $\delta_c(a_{\rm rh})$. Since the mode with wavenumber $k_{\rm pk}$ typically remains outside of the cannibal sound horizon, $\delta_c(k_{\rm pk})$ evolves similarly to a cold matter perturbation. That is, $\delta_c$ evolves logarithmically from horizon entry at $a_{\rm hor}=\gamma_{\rm pk}a_{\rm fz}$ until $a_{\rm dom}$. The linear growth of $\delta_c(k_{\rm pk})$ after cannibal domination is well described by the growing solution of the Meszaros equation:
\begin{align}
    \delta_{c}(k_{\rm pk},a)\approx\frac{3 A_s \phi_{p}}{2} \ln \left[\frac{4 B_s e^{-3} a_{\rm dom}}{\gamma_{\rm pk} a_{\rm fz}}\right] \frac{a}{a_{\rm dom}}.
\end{align}
As $\delta_{\rm DM}(k_{\rm pk},a_{\rm rh})=\delta_c(k_{\rm pk},a_{\rm rh})$, the logarithmic growth of $\delta_{\rm DM}$ after reheating will then be of the form
\begin{align}\label{eq:delta_DM_after_arh_sm}
    \delta_{\rm DM}(k_{\rm pk},a)=\frac{3 A_s \phi_{p}}{2} \ln \left[\frac{4 B_s e^{-3} a_{\rm dom}}{\gamma_{\rm pk} a_{\rm fz}}\right]\frac{a_{\rm rh}}{a_{\rm dom}}\tilde{b}_1\ln \left(\tilde{b}_2\frac{a}{a_{\rm rh}}\right).
\end{align}
Here $\tilde{b}_{1}$ and $\tilde{b}_{2}$ parameterize the transition from linear to logarithmic growth through reheating. Numerically we find $\tilde{b}_1=1.29$ and $\tilde{b}_2=1.66$.

Comparing eq.~\eqref{eq:delta_DM_after_arh_sm} with the standard logarithmic growth of $\delta_{\rm DM}$ during radiation domination, eq.~\eqref{eq:log_growth}, we find
\begin{align}\label{eq:Akpk}
A(k_{\rm pk})=\frac{3 A_s}{2} \ln \left[\frac{4 B_s e^{-3} a_{\rm dom}}{\gamma_{\rm pk}a_{\rm fz}}\right]\frac{a_{\rm rh}}{a_{\rm dom}}\tilde{b}_1 && B(k_{\rm pk})=\tilde{b}_2\frac{a_{\rm hor}}{a_{\rm rh}}.
\end{align}
Using the above relations in the definition of the transfer function, eq.~\eqref{eq:transfer}, gives
\begin{align}\label{eq:peak_sm}
    T(k_{\rm pk})\approx&\frac{3}{2}\tilde{b}_1 \ln \left(\frac{4 B_s e^{-3} a_{\rm dom}}{\gamma_{\rm pk} a_{\rm fz}}\right)\frac{a_{\rm rh}}{a_{\rm dom}}\left[1-\frac{\ln(B_s \tilde{b}_2^{-1}a_{\rm rh}/a_{\rm hor,s})}{\ln(4B_se^{-3}a_{\rm eq}/a_{\rm hor,s})}\right].
\end{align}
This estimate is accurate to within $5\%$ as long as $a_{\rm dom}>100 a_{\rm fz}$ and $a_{\rm rh}>10a_{\rm dom}$,  and is shown as a black dot-dashed line in the right panels of figure~\ref{fig:SM_dom_horz_peak}).

We obtain a simple approximation for $T(k_{\rm pk})$ by neglecting the logarithmic factors in the square bracket in eq.~\eqref{eq:peak_sm} as they provide only an $\mathcal{O}(1)$ correction and using the fact that $\gamma_{\rm pk}$ is typically of $\mathcal{O}(1)$, yielding
\begin{align}\label{eq:peak_sm_param}
T(k_{\rm pk}) \sim 2\ln\left(\frac{a_{\rm dom}}{10a_{\rm fz}}\right)\frac{a_{\rm rh}}{a_{\rm dom}}.
\end{align}
In contrast, the peak of the transfer function for $a_{\rm dom}<2a_{\rm fz}$ is approximately \cite{Erickcek:2020wzd}
\begin{align}\label{eq:pk_can_param}
T(k_{\rm pk})\sim \frac{1}{5}\frac{a_{\rm rh}}{a_{\rm fz}}.
\end{align}
Notice that in both scenarios, the peak of the transfer function is primarily determined by the duration of cannibal domination after the freeze-out of cannibal reactions. In the scenarios with $2a_{\rm fz}<a_{\rm dom}$ one also gets an additional logarithmic enhancement due to the logarithmic growth of $\delta_{c}(k_{\rm pk})$ prior to $a_{\rm dom}$.

\subsection{The effects of DM-cannibal interactions}\label{sec:DM-can_scat}

Until now we have focused on scenarios where the DM only interacts gravitationally with the other constituents of the universe. However, if DM and the cannibal particle are part of the same hidden sector, then it is natural for the two species to have non-gravitational interactions as well. In this section we show that the presence of non-gravitational DM-cannibal interactions does not change the key features of the transfer function and does not change $k_{\rm pk}$ and $T(k_{\rm pk})$.

Scattering between the DM and cannibal particles can cause the DM to be kinetically coupled to the cannibal fluid. The scenario we consider involves an interaction between two non-relativistic particles with a mass hierarchy ($m_{\rm DM}>m$) and hence is similar to the baryon-DM interactions studied in Ref.~\cite{Dvorkin:2013cea}. Using the results of Ref.~\cite{Dvorkin:2013cea}, the momentum transfer rate ($\frac{1}{p}\frac{\textrm{d}p}{\textrm{d}t}$) experienced by a DM particle due to its $s$-wave scattering interactions with the cannibal bath is given by $n_{\rm can} (m/m_{\rm DM})\langle\sigma_{\textrm{DM,c}}v_{c}\rangle$, where $\langle\sigma_{\textrm{DM,c}}v_{c}\rangle\propto \sqrt{T_c/m}$ is the velocity averaged cross-section for cannibal-DM scattering.
When we include these interactions, eq.~\eqref{eq:theta_dm_eq}) for $\theta_{\rm DM}$ becomes 
\begin{align}\label{eq:DM_can_kinetic}
	{\theta}_{\rm DM}'(a)+\frac{1}{a} {\theta}_{\rm DM}-\frac{ {k}^2}{a^2 {H}}\phi=\frac{m}{m_{\rm DM}}\frac{n_{\rm can}\langle\sigma_{\textrm{DM,c}}v_{c}\rangle}{aH}(\theta_c-\theta_{\rm DM}).
\end{align}

The  momentum transfer rate for a cannibal particle interacting with the DM fluid is given by $n_{\rm DM}\langle\sigma_{\textrm{DM,c}}v_{c}\rangle$. Consequently, a  term similar to the RHS of eq.~\eqref{eq:DM_can_kinetic} would also appear in the $\theta_c$ equation but with an additional factor of $-\rho_{\rm DM}/\rho_{\rm can}$. As $\rho_{\rm DM}\ll \rho_{\rm can}$ prior to reheating, the effect of DM-cannibal scattering on $\theta_c$ is much smaller than its effect on $\theta_{\rm DM}$. Consequently, the DM perturbations track the cannibal perturbations while providing negligible feedback on the evolution of the cannibal perturbations. Hence we ignore the impact of DM interactions on the cannibal fluid.

\begin{figure}
\begin{subfigure}{.5\textwidth}
	\includegraphics[width=1.00\textwidth]{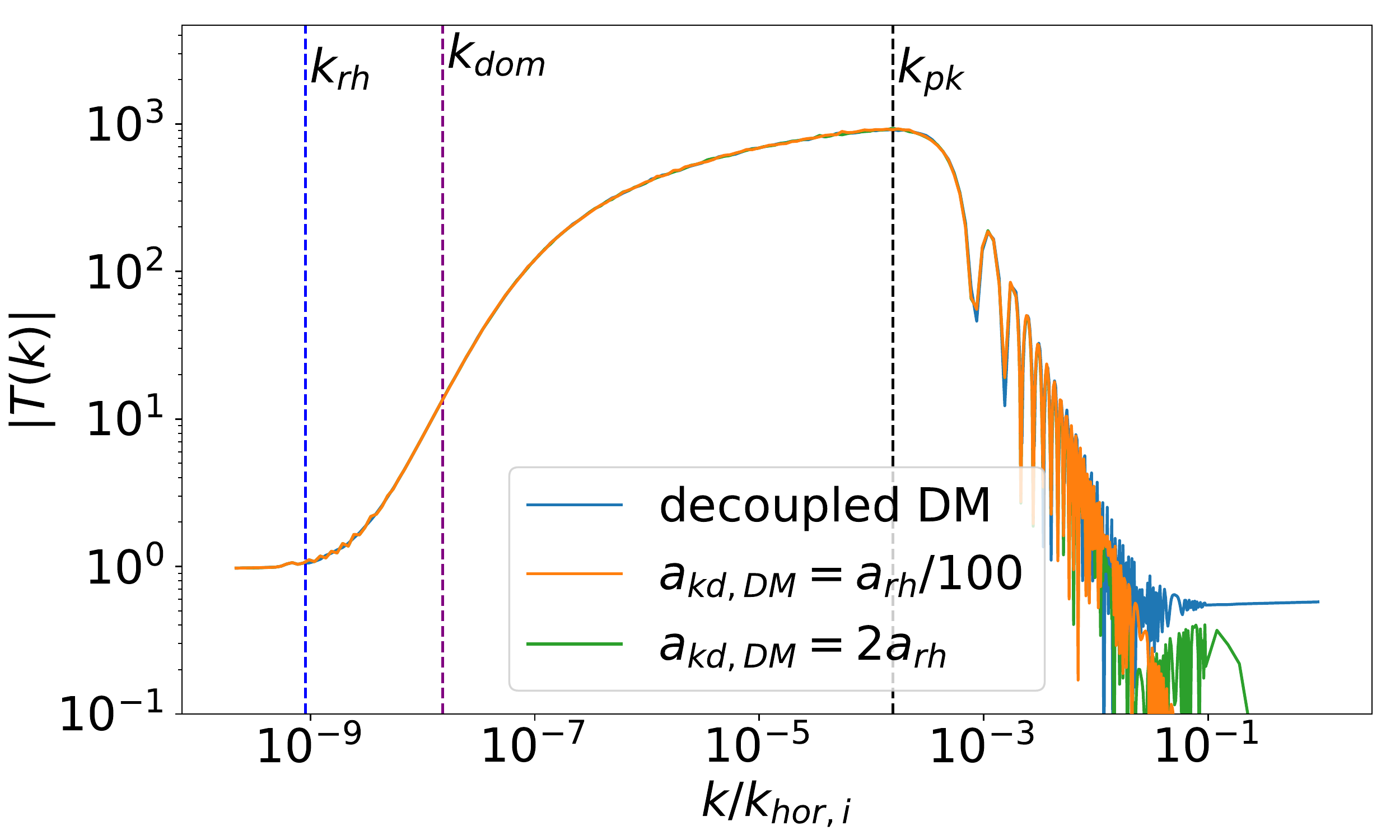}
\end{subfigure}
\begin{subfigure}{.5\textwidth}
	\includegraphics[width=1.00\textwidth]{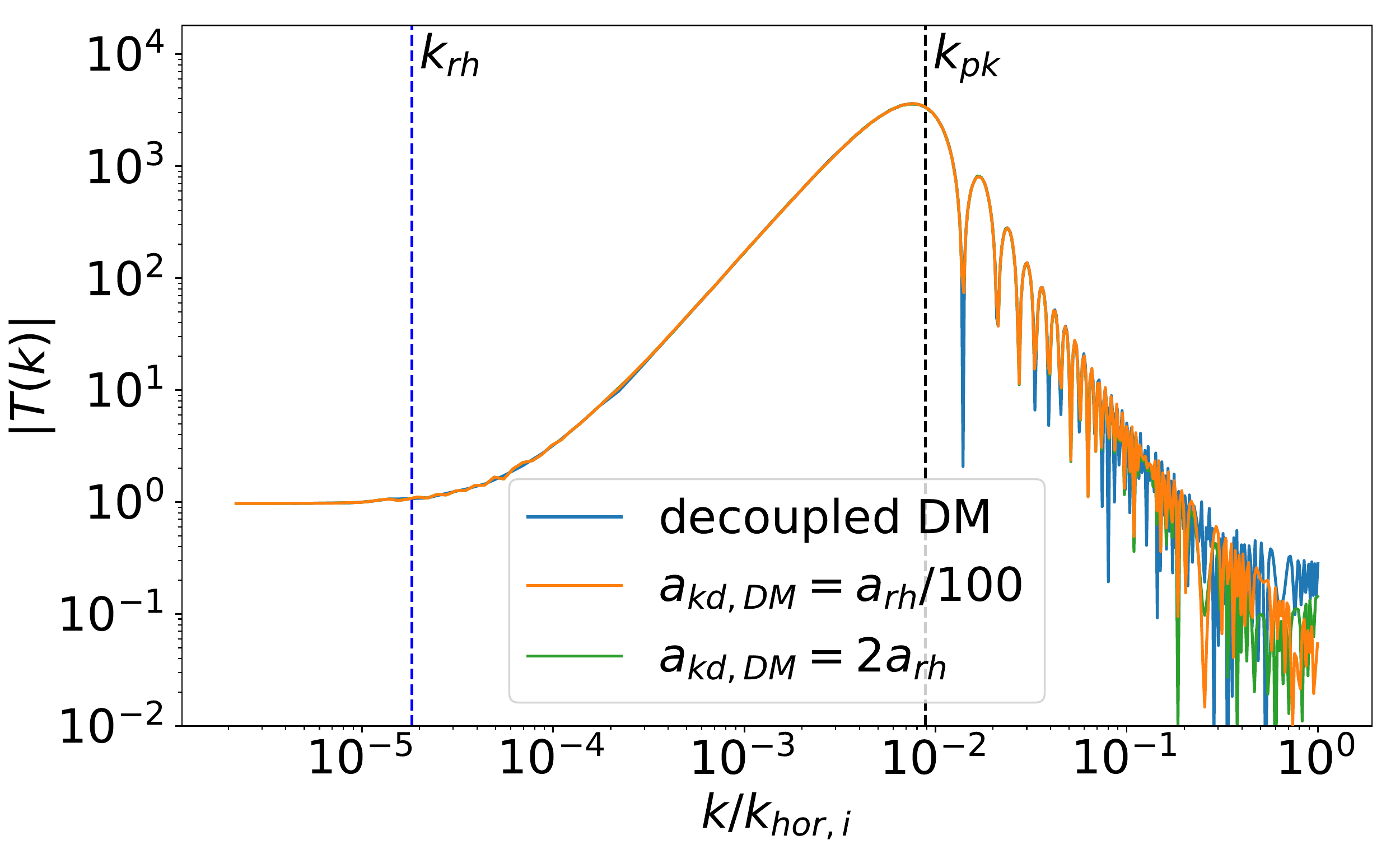}
\end{subfigure}
\caption{Transfer functions for scenarios with different cannibal-DM interaction cross-sections. The left panel shows an initially subdominant cannibal scenario with the same parameters as in figure~\ref{fig:SM_dom_horz_peak}. The right panel shows a scenario with an initially dominant cannibal density with $m= 1.8$ TeV, $T_{\rm rh} = 10$ MeV, $\alpha_c= 0.1$, and $\xi_i=10$,.  
The blue line corresponds to no kinetic coupling between DM and the cannibals. The orange line corresponds to a scenario where the DM-cannibal scattering rate falls below the Hubble rate at $a_{\rm kd,DM}=a_{\rm rh}/100$, while the green line is for a scenario with $a_{\rm kd,DM}=2a_{\rm rh}$. The transfer function is unaffected by cannibal-DM interaction for modes where $T(k)>1$.}
\label{fig:SM_dom_horz_peak_tk}
\end{figure}

In figure~\ref{fig:SM_dom_horz_peak_tk} we compare the transfer functions  resulting from scenarios with different values of $\langle\sigma_{\textrm{DM,c}}v_{c}\rangle$. We see that DM-cannibal interactions have no impact on the peak of the transfer function. In the limit of strong kinetic coupling, i.e. $n_{\rm can}\langle\sigma_{\textrm{DM,c}}v_{c}\rangle\gg H$, the DM-cannibal interactions cause $\delta_{\rm DM}$ to track $\delta_c$. However, regardless of the strength of the DM-cannibal kinetic coupling, the metric perturbation will always drive $\delta_{\rm DM}$ toward $\delta_{c}$ once the mode escapes the cannibal Jeans horizon. Consequently, the value of $\delta_{\rm DM}$ at reheating is insensitive to DM-cannibal scattering for modes that escape the cannibal Jeans horizon.

DM-cannibal interactions do affect the DM transfer function on very small scales, corresponding to modes that do not escape the cannibal Jeans horizon before reheating. For these modes, $\delta_c$ oscillates until reheating and so  never generates a coherent gravitational pull on the DM perturbation. In scenarios with only gravitational interactions, shown by the blue line in figure~\ref{fig:SM_dom_horz_peak_tk}, $\delta_{\rm DM}$ is larger than $\delta_c$ at reheating for these small-scale modes because $\delta_{c}$ oscillates while $\delta_{\rm DM}$ grows logarithmically until $a_{\rm dom}$. However, in the opposite limit where DM remains kinetically coupled to the cannibals until $2a_{\rm rh}$, shown  with the green line, $\delta_{\rm DM}$ has the same value as $\delta_c$ at $a_{\rm rh}$.

The intermediate case shown by the orange line in figure~\ref{fig:SM_dom_horz_peak_tk} is more suppressed on very small scales than the tightly coupled case shown in green, even
though the intermediate case has a smaller value of $\langle\sigma_{\textrm{DM,c}}v_{c}\rangle$. This relative suppression results from diffusion damping of the dark matter perturbations. Diffusion damping occurs when the cannibal perturbations oscillate faster than the DM-cannibal scattering rate prior to kinetic decoupling, i.e. $c_sk>n_{\rm can}\langle\sigma_{\textrm{DM,c}}v_{c}\rangle>H$. During this period, the DM perturbations oscillate with the same frequency as the oscillations in the cannibal perturbations, but the amplitude of their oscillation is highly damped.
This damping is similar to the Silk damping of baryon density perturbations \cite{1968ApJ...151..459S}. In figure~\ref{fig:SM_dom_horz_peak_tk}, the orange line is more suppressed than the green line because the diffusion damping scale, $k_{D}^{-1}\sim c_s/(n_{\rm can}\langle\sigma_{\textrm{DM,c}}v_{c}\rangle)$, is larger for smaller values of $\langle\sigma_{\textrm{DM,c}}v_{c}\rangle$. Consequently, the modes experience damping at smaller $k$ values when DM decouples shortly before reheating compared to the tightly coupled case.

\section{Beyond the perfect-fluid approximation}\label{sec:can_not_perfect}

So far we have assumed that the cannibals and the DM fluids are perfect fluids. 
The perfect-fluid approximation will break down on scales where the random motion of the particles comprising the fluid cannot be neglected, which can occur in a variety of regimes.  Even while the homogeneous cannibal fluid is in kinetic equilibrium, the cannibals still have a finite diffusion length.  For perturbations on scales smaller than the diffusion length, a perfect-fluid description is not sufficient.  Once kinetic equilibrium is lost, the random thermal motion of particles becomes important on scales quantified by either the free-streaming length or the collisionless Jeans length, depending on the gravitational forces experienced by the particles.  Again, for perturbations with wavelengths smaller than these scales, the perfect-fluid description breaks down.

Momentum exchange among cannibal particles is dominated  by elastic 2-to-2 scatterings, with a rate given by
$n_{\rm can}\langle \sigma_c v_c\rangle $.
In appendix~\ref{sec:can_kd} we derive the two-to-two scattering rate for the $\varphi^4$ theory described by eq.~\eqref{eq:L_can}. We find that the  $s-$wave contribution in the non-relativistic limit is
\begin{align}\label{eq:scat_can_rate}
    \langle\sigma_c v_c\rangle =\frac{1}{64\pi^{3/2}m^2}\left(\lambda-\frac{5}{3}\frac{g^2}{m^2}\right)^2 \times \sqrt{\frac{T_c}{m}} \equiv \sigma_{\mathrm{eff}} \sqrt{\frac{T_c}{m}} ,
\end{align}
where $\lambda$ and $g$ are the coupling constants of the cubic and quartic interactions in eq.~\eqref{eq:L_can}, respectively, and in the second equality we pulled out a factor of $v_c = \sqrt{T_c/m}$ to define an effective scattering cross-section.  Note that the parameter $\alpha_c$ that controls the cannibal number-changing interactions does not uniquely determine $\sigma_c$ because $\alpha_c$ and $\sigma_c$ depend on different combinations of $g$ and $\lambda$.  

In a Hubble time, a cannibal particle will undergo $N=(n_{\rm can}\langle\sigma_c v_c\rangle)/H$ scatterings. The average distance travelled by a cannibal particle between two collisions is $\ell_{\rm mfp}\sim 1 /(n_{\rm can}\sigma_{\mathrm{eff}})$. 
Consequently, the comoving diffusion length is given by
\begin{align}\label{eq:diff_length}
    \lambda_{\rm diff}=\frac{1}{a}\sqrt{N}\times \ell_{\rm mfp}=\frac{1}{a}\sqrt{\frac{v_c}{n_{\rm can}\sigma_{\mathrm{eff}} H}}.
\end{align}
For modes  with wavelengths shorter than the comoving diffusion length, the higher moments of the Boltzmann hierarchy can no longer be neglected, and will suppress $\delta_c$ \cite{Hu:1996mn}. The perfect-fluid approximation also breaks down for modes that oscillate faster than the 2-to-2  scattering rate, i.e. if $c_sk>n_{\rm can}\langle \sigma_{c} v_c\rangle$. Since $c_s\sim v_c$, requiring the oscillation frequency to be slower than the scattering rate is equivalent to requiring $k^{-1}>\ell_{\rm mfp}$. Since the cannibal diffusion length is larger than the mean free path prior to kinetic decoupling (as $N>1$), modes will be damped by diffusion before the scattering rate falls below the oscillation frequency.  

The diffusion length is relevant as long as the cannibal fluid maintains internal kinetic equilibrium, i.e. $n_{\rm can}\langle \sigma_cv_c\rangle>H$. We define the scale factor, $a_{\rm kd}$, at which the cannibal fluid falls out of its kinetic equilibrium through the relation
\begin{align}\label{eq:akd_def}
	n_{\rm can}(a_{\rm kd})\langle\sigma_{c} v_c(a_{\rm kd})\rangle=H(a_{\rm kd}). 
\end{align}
After kinetic decoupling, the cannibal fluid is effectively collisionless.   Cannibal number-changing interactions, which involve three particles in the initial state, freeze out substantially before the cannibal fluid loses internal kinetic equilibrium, so after kinetic decoupling the cannibal fluid evolves as pressureless matter.\footnote{The $s$-wave component of the two-to-two scattering cross-section vanishes for $\lambda = (5/3) g^2/m^2$; for couplings in the neighborhood of such values, the $p$-wave component will dominate the elastic scattering cross-section in the non-relativistic regime.  As the three-to-two cannibal interactions are phase-space suppressed as well as higher order in the couplings, they will still generically decouple earlier than the elastic scattering interactions, but to examine this specific sliver of parameter space in detail requires the retention of $p$-wave contributions beyond eq.~\ref{eq:scat_can_rate}, and is beyond the scope of this work. }

While the universe is radiation dominated, the cannibals experience no coherent gravitational force and have a comoving free-streaming length given by
\begin{align}\label{eq:can_fs}
	\lambda_{\rm fs}(a)=\lambda_{\rm diff}(a_{\rm kd})+\int_{a_{\rm kd}}^a\frac{v_c}{a^2H}da && a_{\rm kd}<a<a_{\rm dom}.
\end{align}
Here we have imposed that the cannibal diffusion length is equal to the free-streaming length at kinetic decoupling. 
When the cannibal comes to dominate the universe, metric perturbations can begin to pull particles toward overdense regions.  
In this regime, departures from perfect-fluid behavior are governed by the collisionless Jeans length. Analogous to the collisional Jeans length described in the previous section, the collisionless Jeans length determines the scale above which gravitational attraction is sufficient to overcome the random motion of particles.

To find the collisionless Jeans length we need to include the anisotropic stress, $\sigma_{\rm can}$, in eq.~\eqref{eq:theta_can_eq}, which governs the evolution of $\theta_c$. Before  kinetic decoupling, elastic cannibal scattering ensures that $\sigma_{\rm can}$ is only relevant for modes within the diffusion length. After  kinetic decoupling, the anisotropic stress is determined by the free-streaming velocity of the cannibals. Ref.~\cite{Piattella:2013cma} finds the anisotropic stress for a collisionless fluid to be given by $\sigma =-\frac{5}{3}\langle v^2\rangle \delta$, which follows from the assumption that the phase-space density of the particles remains unchanged while particles fall into gravitational potential wells. After kinetic decoupling, the sound speed term in eq.~\eqref{eq:theta_can_eq} is ill-defined and no longer appears in that equation. Consequently, the $\theta_c$ equation after cannibal kinetic decoupling is
\begin{align}\label{eq:theta_stress_eq}
	{\theta_c}'(a)&=-\frac{1}{a}(1-3w_c)\theta_c-\frac{w_c'}{1+w_c} {\theta_c}+\frac{ {k}^2}{a^2 {H}}\phi+\frac{ {k}^2}{a^2 {H}}\frac{5}{3}\langle v^2\rangle \delta.
\end{align}

We find the collisionless Jeans length from eq.~\eqref{eq:theta_stress_eq} by following the same steps as we performed for calculating the collisional Jeans length: find the simple harmonic oscillator equation for $\delta_c$ analogous to eq.~\eqref{eq:delta_can_subhz_adom} and then find the wavenumber $k$ for which the frequency becomes imaginary. Doing so, we find the collisionless Jeans length, $k_{J,c}^{-1}$, to be the same as the collisional Jeans length in eq.~\eqref{eq:jeans}, except with $c^2_s/(1+w_c)$ replaced by $5\langle v_c^2\rangle/3$:
\begin{align}\label{eq:jeans_c}
	k_{J,c}=\sqrt{\frac{9}{10\langle v_c^2\rangle}}\ aH.
\end{align}
After cannibal freeze-out and before cannibal kinetic decoupling, the sound speed is given by $c_s^2=\frac53\frac{T_c}{m}=\frac53\langle v_c^2\rangle$ and $w_c\ll 1$.
Therefore, the collisionless Jeans length has the same value as the collisional Jeans length would have had in the absence of kinetic decoupling.

\begin{figure}
	\includegraphics[width=1.00\textwidth]{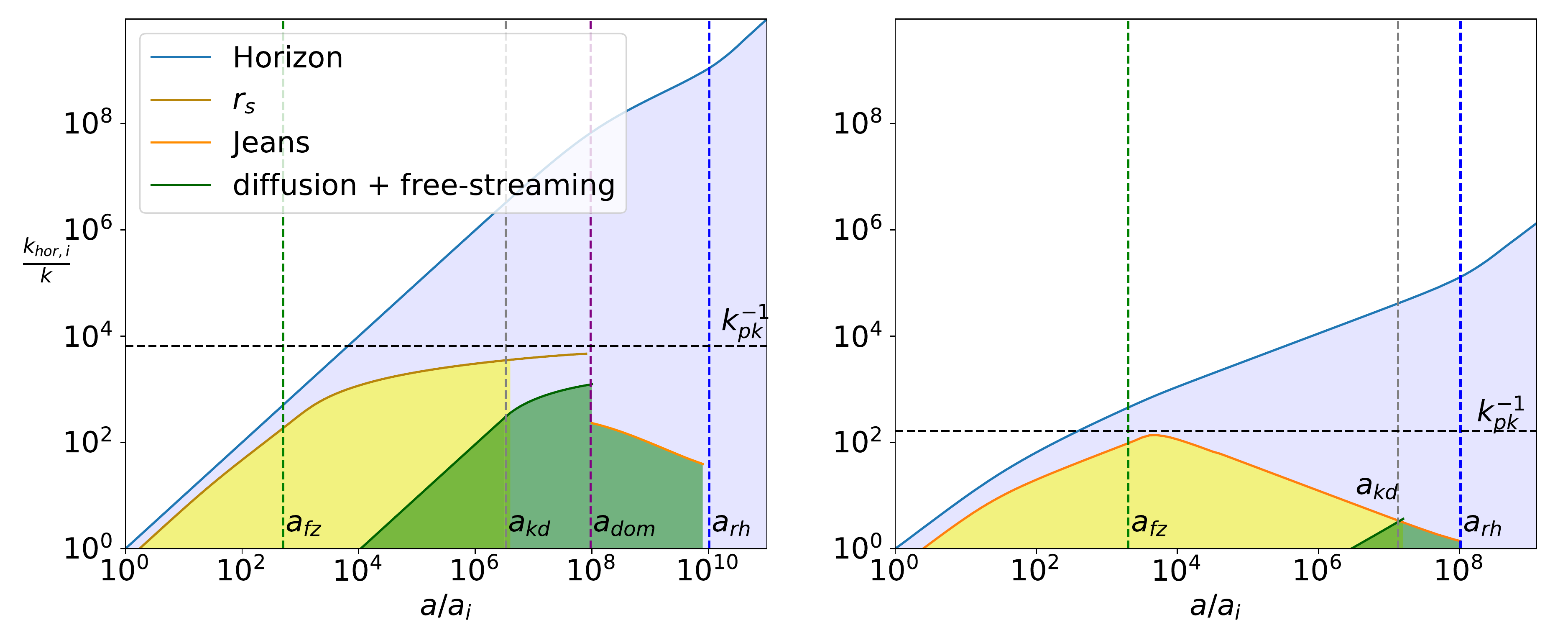}
	\caption{Length scales where the perfect-fluid approximation for the cannibals breaks down. The left panel uses the same parameters as in figure~\ref{fig:SM_dom_horz_peak} and shows a scenario where both cannibal freeze-out and kinetic decoupling occur during SM radiation domination. The right panel takes $m=6.1$ TeV, $T_{\rm rh} = 10$ MeV, $\alpha_c= 0.42$, and $\xi_i=10$ and shows a scenario where cannibal freeze-out occurs during cannibal domination. The green line prior to $a_{\rm kd}$ shows the comoving cannibal diffusion length as given in eq.~\eqref{eq:diff_length}, and in the left panel between $a_{\rm kd}$ and $a_{\rm dom}$, the comoving free-streaming length of the cannibals as given in eq.~\eqref{eq:can_fs}. The orange solid line is the collisional Jeans length, eq.~\eqref{eq:jeans}, for $a<a_{\rm kd}$ and the collisionless Jeans length, eq.~\eqref{eq:jeans_c}, for $a>a_{\rm kd}$. The brown line indicates the cannibal sound horizon, eq.~\eqref{eq:rs}. The green shaded region marks the regimes where the perfect-fluid approximation for the cannibals breaks down.  Here we show the diffusion length taking $\lambda=0$ in eq.~\eqref{eq:scat_can_rate}.}
	\label{fig:diff_horz}
\end{figure}

We show the evolution of all three length scales (diffusion, free-streaming, and collisional Jeans lengths) in fig.~\ref{fig:diff_horz}.  In the left panel we show a scenario where kinetic decoupling occurs before the cannibal density comes to dominate, and in the right panel we show a scenario where the cannibal density is always dominant.  In both panels, the solid green line shows the growth of the diffusion length up until $a_{\rm kd}$, which, in the left panel, transitions smoothly into the free-streaming length given in eq.~\eqref{eq:can_fs} in the region $a_{\rm kd}<a<a_{\rm dom}$.   In the left panel, the solid orange line after $a_{\rm dom}$  shows the collisionless Jeans length.  In the right panel, the orange line shows the collisional Jeans length before $a_{\rm kd}$ and the collisionless Jeans length after $a_{\rm kd}$.  The green shaded region indicates the scales where the cannibal particles no longer behave as a perfect fluid.  For modes that enter the green shaded region in figure~\ref{fig:diff_horz}, we expect $\delta_{c}(a_{\rm rh})$  to experience a suppression that is not captured in our suite of perturbation equations. The black dashed line shows the peak of the matter power spectrum in the perfect-fluid approximation, and thus indicates the location of the cutoff that follows from oscillations in the cannibal fluid. We see that, in the cases shown, the modes that are affected by the breakdown of the perfect-fluid approximation are already suppressed by the cannibal oscillations.
 
Cutoffs arising from imperfect-fluid behavior can be important for the transfer function when they occur on scales larger than the small-scale cutoff provided by cannibal oscillations.  
In scenarios where the cannibals freeze out during cannibal domination, the cutoff provided by cannibal interactions sets $k_{\rm pk}\approx k_J(2a_{\rm fz})/1.4$, which ensures that any deviations from perfect-fluid behavior occur at scales substantially below $k_{\rm pk}^{-1}$ when $a_{\rm fz}\ll a_{\rm kd}$.  
In scenarios where the cannibals freeze out during radiation domination, however, the situation is a little more subtle.  In this case, the perfect-fluid calculation of the cannibal cutoff gives $k_{\rm pk} \sim r_s(a_{\rm dom})$.  As both the sound horizon and the free-streaming length grow logarithmically during the period of radiation domination following $a_{\rm kd}$, the sound horizon will generically stay outside the free-streaming length, and therefore the cannibal oscillation cutoff $k_{\rm pk}^{-1}$ as given by eq.~\eqref{eq:kpk_sm} will occur at larger scales than the free-streaming length.  However, the derivation of  eq.~\eqref{eq:kpk_sm} assumes a collisional fluid.  Thus while we expect the cannibal oscillation cutoff to be the relevant small-scale cutoff for $a_{\rm kd}>a_{\rm dom}$, this conclusion does not necessarily hold if $a_{\rm kd}<a_{\rm dom}$.  The regime with $a_{\rm kd}<a_{\rm dom}$ can be realized in a small region of parameter space, as we show in section~\ref{sec:params}.
A full calculation of the small-scale cutoff in this regime would require incorporating higher moments of the Boltzmann hierarchy and is beyond the scope of this work;  see \cite{Egana-Ugrinovic:2021gnu} for related calculations in a similar model.  However in general we can expect this cutoff to lie somewhere in the vicinity of $r_s(a_{\rm kd})$ and $\lambda_{\rm fs} (a_{\rm dom})$.   These two scales are relatively similar: the sound horizon is governed by the distance traveled by sound waves in the cannibal fluid, while the free-streaming horizon is governed by the distance traveled by the non-relativistic cannibal particles in this epoch.  Both the sound speed and the cannibal particle speed are determined by the cannibal temperature, which changes only logarithmically between $a_{\rm kd}$ and $a_{\rm dom}$.  

Finally, we quantify the relationship between $a_{\rm fz}$ and $a_{\rm kd}$ in our cannibal model.  Since the $2\to 2$ and $3\to 2$ scattering cross-sections depend on different combinations of the quartic coupling $\lambda$ and cubic coupling $g/m$, we can obtain a range of possible $a_{\rm kd}$ for a fixed $a_{\rm fz}$.  To evaluate $a_{\rm kd}/a_{\rm fz}$ in terms of the Lagrangian couplings, we first divide eq.~\eqref{eq:akd_def} by eq.~\eqref{eq:afz_def}:
\begin{align}
	\frac{n_{\rm can}(a_{\rm kd})\langle\sigma_cv_c(a_{\rm kd})\rangle}{n_{\rm can}^2(a_{\rm fz})\svc}=\frac{H(a_{\rm kd})}{H(a_{\rm fz})}.
\end{align}
We then express $n_{\rm can}(a_{\rm kd})$ and $T_c(a_{\rm kd})$ in terms of their values at $a_{\rm fz}$ by using  $n_{\rm can}\propto 1/a^3$ and $T_c\propto1/a^2$ for $a>a_{\rm fz}$. Next, we approximate $mn_{\rm can}(a_{\rm fz})\approx \rho_{\rm can,eq}(a_{\rm fz})$ and $T_c(a_{\rm fz})\approx T_{c,\textrm{eq}}(a_{\rm fz})$ and  use eqs.~\eqref{eq:can_eq_evol}-\eqref{eq:can_eq_evol2} to express $\rho_{\rm can,eq}$ and $T_{c,\textrm{eq}}$ in terms of the scale factor. Finally, we set $a_{\rm fz}/a_i=10^3$ inside the logarithms to obtain
\begin{align}\label{eq:akd_mid_step}
	\frac{a_{\rm kd}^4}{a_{\rm fz}^4}\frac{H(a_{\rm kd})}{H(a_{\rm fz})}\sim 10^{-2} \left(\frac{a_{\rm fz}}{a_i}\right)^3\frac{\sigma_{\mathrm{eff}}}{m^3\svc}=10^{-2} \left(\frac{a_{\rm fz}}{a_i}\right)^3\frac{20736}{225\sqrt{5\pi}}\frac{\left[\lambda-5g^2/(3m^2)\right]^2}{(g/m)^2[(g/m)^2-3\lambda]^2}.
\end{align}
In the last equality above we used eq.~\eqref{eq:scat_can_rate} and eq.~\eqref{eq:alpha_can} for $\sigma_{\mathrm{eff}}$ and $\svc$ respectively.
When the universe is cannibal dominated between $a_{\rm fz}$ and $a_{\rm kd}$, then $H(a_{\rm fz})/H(a_{\rm kd})=(a_{\rm kd}/a_{\rm fz})^{3/2}$ and $a_{\rm fz}$ is given by eq.~\eqref{eq:asfz_est_can}. Defining $q$ as the ratio of the quartic and the cubic coupling, $q\equiv \lambda/(g/m)^2$, and expressing $(g/m)^2$ using the definition of $\alpha_c$ in eq.~\eqref{eq:akd_mid_step} yields
\begin{align}\label{eq:akd_can}
	\frac{a_{\rm kd}}{a_{\rm fz}}&\sim 10^5\alpha_c^{2/5}\left(\frac{\rm GeV}{m}\right)^{4/15}\left|\frac{q-5/3}{(3q-1)^{2/3}}\right|^{4/5}.
\end{align}
Similarly, if the universe is radiation dominated between $a_{\rm fz}$ and $a_{\rm kd}$, $H(a_{\rm fz})/H(a_{\rm kd})=(a_{\rm kd}/a_{\rm fz})^{2}$ and $a_{\rm fz}$ is given by eq.~\eqref{eq:asfz_est_sm}. Eq.~\eqref{eq:akd_mid_step} then implies that
\begin{align}\label{eq:akd_sm}
	\frac{a_{\rm kd}}{a_{\rm fz}}&\sim 2\times10^{6}\alpha_c^{5/8}\xi_i^{3/4}\left(\frac{\rm GeV}{m}\right)^{3/8}\left|\frac{q-5/3}{(3q-1)^{2/3}}\right|.
\end{align}
In most of the parameter space that realizes an ECDE, varying $\lambda$ while keeping $\alpha_c$ fixed results in a variation in $a_{\rm kd}/a_{\rm fz}$ of up to an order of magnitude. 
Figure~\ref{fig:diff_horz} shows results with $\lambda=q=0$; for the value of $\alpha_c$ shown in figure~\ref{fig:diff_horz}  increasing $\lambda$ to non-perturbative values results in a correction of less than 30\%  to the values of $a_{\rm kd}$ shown in the figure. 

Departures of the DM from perfect-fluid behavior can also be important for determining the transfer function. 
Prior to reheating, the DM free-streaming, diffusion, and collisionless Jeans lengths are always smaller than the perfect-fluid result for $k_{\rm pk}^{-1}$, as 
the DM speed $v_{\rm DM}=\sqrt{T_{\rm DM}/m_{\rm DM}}$ is always smaller than the cannibal sound speed, $c_s\sim \sqrt{T/m}$, which controls the scale of $k_{\rm pk}$.
However, DM free streaming after reheating can affect the peak of the DM transfer function in some regions of parameter space. Ref.~\cite{Erickcek:2020wzd} evaluated the post-reheating free streaming of DM in the case where DM kinetically decouples from the cannibal fluid after $a_{\rm fz}$. In this case we have $T_{\rm DM}(a_{\rm rh})=T_c(a_{\rm rh})$, which gives for the DM free-streaming length
\begin{align}\label{eq:DM_fs}
	\lambda_{\rm DM,fs} \approx\int_{t_{\rm rh}}^{t_0} v_{\rm DM}\frac{dt}{a}= \sqrt{\frac{T_{ c}(a_{\rm rh})}{m_{\rm DM}}} \frac{1}{(aH)_{\rm rh}} \int_{a_{\rm rh}}^{a_0}
	\frac{da}{a^{3}} \left[G(a) \left( \frac{1}{a}\right)^4 + G(a_{\rm eq})\left( \frac{1}{a^3 a_{\rm eq}}\right) \right]^{-1/2},
\end{align}
where we defined $G(a)\equiv g_*(a)g_*^{1/3}(a_{\rm rh})/g_{*s}^{4/3}(a)$ and dropped negligible contributions from dark energy at late times. This result is applicable regardless of whether the universe was radiation dominated or SM radiation dominated at $a_i$.  While the DM free-streaming length in any given model will depend in detail on the DM coupling to the cannibal species, eq.~\eqref{eq:DM_fs} provides an upper bound on $\lambda_{DM, fs}$: DM that decouples from the cannibals prior to $a_{\rm fz}$ will have a reduced free-streaming length as the temperature of the DM at reheating will be colder than the cannibal temperature.

If the DM free-streaming length is larger than the small-scale cutoff coming from cannibal self-interactions, then the DM transfer function will be maximized on a scale $\sim \lambda_{\rm DM,fs}^{-1}$, which depends on DM as well as cannibal microphysics. 
For $m_{\rm DM}\gtrsim 10 m$, we find DM free-streaming can provide the small-scale cutoff in the transfer function in a small portion of the parameter space, as we discuss in the following section.

\section{Implications for microhalo formation}
\label{sec:params}

In this section, we first discuss how the key features of the linear transfer function, namely $k_{\rm pk}$ and $T(k_{\rm pk})$, relate to the properties of the earliest-forming microhalos. We then express $k_{\rm pk}$ and $T(k_{\rm pk})$ as a function of the cannibal parameters $m,T_{\rm rh},\alpha_c$, and $\xi_i$.  Finally, we briefly discuss the microhalos' observational signatures and how these observations probe cannibalism in the early Universe.

After an ECDE, the DM perturbations with wavenumber $k_{\rm pk}$ have experienced the most growth. Although the stochastic nature of the primordial perturbations prevents us from knowing exactly which mode has the largest amplitude, the near scale-invariance of the primordial power spectrum implies that perturbations on scales near $k_{\rm pk}$ are the first to collapse and form gravitationally bound structures. Since perturbations that enter the horizon prior to BBN form halos that are too small to capture baryons \cite{Tseliakhovich:2010bj}, the characteristic mass of the earliest-forming halos is given by the amount of DM in a sphere of comoving radius $k_{\rm pk}^{-1}$:
\begin{align}\label{eq:Mpk_def}
	M_{\rm pk}\equiv\frac{4\pi}{3}k_{\rm pk}^{-3}\rho_{\rm DM,0},
\end{align}
where $\rho_{\rm DM,0}$ is the dark matter density today, which we take to be $\rho_{\rm DM,0}=9.7\times 10^{-48}$ GeV${}^{4}$ \cite{Aghanim:2018eyx}.
When cannibals freeze out while they dominate the energy density of the universe ($a_{\rm dom}<2a_{\rm fz}$), we calculate $M_{\rm pk}$ from the expression for $k_{\rm pk}$ given in eq.~\eqref{eq:kpk_can_param}:
\begin{align}\label{eq:Mpk_can}
M_{\rm pk}&\sim  10^{-11}M_{\odot}\left(\frac{\alpha_c}{0.1}\right)\left(\frac{\rm 10\, MeV}{T_{\rm rh}}\right)\left(\frac{\rm TeV}{m}\right)^{7/3}.
\end{align}
For $a_{\rm dom}>2a_{\rm fz}$, we calculate $M_{\rm pk}$ from the expression for $k_{\rm pk}$ given in eq.~\eqref{eq:kpk_sm_param}:
\begin{multline}\label{eq:Mpk_sm}
M_{\rm pk}\sim 3\times 10^{-13}M_{\odot}\left(\frac{\alpha_c}{0.1}\right)^{9/4}\left(\frac{\xi_i}{0.4}\right)^{15/2}\left(\frac{\rm 10\, MeV}{T_{\rm rh}}\right)\left(\frac{\rm TeV}{m}\right)^{11/4}\left(\frac{100}{g_{*}(10m/\xi_i)}\right)^{15/8}\\
\times\left(\frac{1}{6}\ln\left[500\left(\frac{0.4}{\xi_i}\right)^{9/2}\left(\frac{0.1}{\alpha_c}\right)^{3/4}\left(\frac{m}{\rm TeV}\right)^{1/4}\left(\frac{g_{*}(10m/\xi_i)}{100}\right)^{9/8}\right]\right)^{9/2}.
\end{multline}
We remind the reader that the expression for $k_{\rm pk}$ given in eq.~\eqref{eq:kpk_sm_param} is a good approximation for $10<a_{\rm dom}/a_{\rm fz}<10^5$ and $10^2<a_{\rm fz}/a_i<10^4$.
Since the peak halo mass is typically much smaller than one Earth mass, the earliest-forming halos are microhalos.

In both cases, $M_{\rm pk}$ increases as either $T_{\rm rh}$ or $m$ decreases because $M_{\rm pk}$ is determined by the sound horizon at reheating, $r_s(a_{\rm rh})$. Decreasing $T_{\rm rh}$ delays reheating and hence increases $r_s(a_{\rm rh})$. Decreasing $m$ increases $r_s(a_{\rm rh})$ by delaying the freeze-out of cannibal reactions, which increases the cannibal temperature. Since increasing $\alpha_c$ also delays the freeze-out of cannibal reactions, we see a positive correlation between $M_{\rm pk}$ and $\alpha_c$. The peak halo mass has a stronger dependence on $\alpha_c$ when the cannibals freeze out while the universe is SM radiation dominated because the Hubble rate falls faster in a radiation-dominated universe compared to a cannibal-dominated universe.

An ECDE enhances the amplitude of all perturbations with $k<k_{\rm pk}$ that enter the horizon during the ECDE. Therefore, the largest halos that are affected by the ECDE have masses equal to the amount of DM within the horizon at reheating, $M_{\rm rh}$, which is given by eq.~\eqref{eq:Mpk_def} but with $k_{\rm pk}$ replaced by $k_{\rm rh}=a_{\rm rh}H(a_{\rm rh})$.
We find $M_{\rm rh}$ in terms of cannibal parameters by taking $H(a_{\rm rh})\sim\Gamma$ and then expressing $\Gamma$ in terms of $T_{\rm rh}$.\footnote{The Hubble rate at $a_{\rm rh}$ does not equal $\Gamma$ because $a_{\rm rh}$ is defined as the scale factor when the Hubble rate equals $\Gamma$ in a standard cosmology. However, since $\rho_{\rm can}\sim \rho_r$ at $a_{\rm rh}$, $H(a_{\rm rh})$ is some $\mathcal{O}(1)$ factor times $\Gamma$.} We then express $a_{\rm rh}/a_0$ in terms of SM temperatures using entropy conservation to obtain
\begin{align}\label{eq:mrh_def}
	M_{\rm rh}\sim 10^{-4}M_{\odot}\left(\frac{\rm 10\, MeV}{T_{\rm rh}}\right)^{3}\left(\frac{10}{g_{*}(T_{\rm rh})}\right)^{1/2}.
\end{align}
While deriving the above relation we set $g_{*s}(T_{\rm rh})=g_*(T_{\rm rh})$. An ECDE increases the abundance of all halos with masses between $M_{\rm pk}$ and $M_{\rm rh}$, and these halos form earlier than they would in a standard cosmology.

Halos form when $\delta_{\rm DM}$ becomes of order unity. In a standard cosmology, the amplitude of small-scale perturbations increases only logarithmically with $k$, so microhalos with masses within several orders of magnitude of an earth mass form near a redshift of 60 \cite{Diemand:2005vz, Ishiyama:2010es}.
Since baryons do not participate in structure formation for modes that enter the horizon during an ECDE, $\delta_{\rm DM}\propto (1+z)^{-0.9}$ for $z<z_{\rm eq}$ on these scales \cite{Hu:1995en, Bertschinger:2006nq}. Consequently, the collapse redshift of the microhalos corresponding to overdensities with wavenumber $k$ increases by a factor of $\sim [T(k)]^{1.11}$ compared to that in the standard cosmology as long as the collapse occurs after matter-radiation equality, i.e. for $T(k)<30$. For $T(k)> 30$, the formation of the microhalos occurs prior to matter-radiation equality, and the exact increase in the collapse redshift depends non-trivially on $T(k)$ \cite{Blanco:2019eij}.

The central density of a dark matter halo scales with the homogeneous matter density at the time of its formation \cite{1997ApJ...490..493N, Bullock:1999he, Wechsler:2001cs}, so the microhalos that form after an ECDE have central densities that are significantly larger than those in standard cosmology \cite{Delos:2019mxl}. These central densities are large enough for the microhalos to survive within galaxies, although they experience significant mass loss \cite{2010PhRvD..81j3529B, Ishiyama:2010es, Delos:2019lik, Delos:2019tsl, Delos:2019dyh,Blinov:2021axd}.

If $T(k_{\rm pk})$ is significantly large, then the cannibals and DM particles assemble into microhalos before reheating. For modes in the vicinity of the peak in the matter power spectrum, eq.~\eqref{eq:Akpk} implies that the DM overdensity at reheating is related to the primordial metric fluctuation via  $\delta_{\rm DM}(k,a_{\rm rh})\approx A(k)\phi_P$.  For a nearly scale-invariant spectrum, we expect $\phi_P \sim 10^{-5}$ \cite{Aghanim:2018eyx}, and thus  density perturbations on all scales remain perturbative until reheating provided $A(k_{\rm pk})\lesssim10^5$. As $T(k_{\rm pk})\approx A(k_{\rm pk})/A_s$, where $A_s=9.11$, microhalos will form prior to reheating if $T(k_{\rm pk})$ exceeds $10^4$.
These microhalos are destroyed once reheating occurs because they are primarily composed of cannibal particles. When the cannibal particles decay, DM particles free stream out of the microhalos with typical speeds given by the virial speed within the microhalos, which is of order $10^{-2}$ \cite{Blanco:2019eij}. The subsequent free streaming of DM particles acts to erase the structure within the comoving horizon at the time of reheating, thus 
washing out much of the enhanced structure resulting from the ECDE. 

The peak amplitude of the transfer function, and thus the formation time of the first microhalos, can be directly related to the properties of the cannibal field. In the case where cannibal freeze-out occurs during cannibal domination ($a_{\rm dom}<2a_{\rm fz}$), we use the expression for $T(k_{\rm pk})$ given in eq.~\eqref{eq:pk_can_param} and express $a_{\rm fz}$ and $a_{\rm rh}$ using eq.~\eqref{eq:asfz_est_can} and eq.~\eqref{eq:arh_est} respectively to obtain
\begin{align}\label{eq:Tpk_can}
T(k_{\rm pk})&\sim 2\times 10^{3}\left(\frac{0.1}{\alpha_c}\right)^{2/3}\Big(\frac{m}{\rm TeV}\Big)^{14/9}\left(\frac{\rm 10\, MeV}{T_{\rm rh}}\right)^{4/3}\left(\frac{10}{g_*(T_{\rm rh})}\right)^{1/3}.
\end{align}
For $a_{\rm dom}>2a_{\rm fz}$, we use eq.~\eqref{eq:peak_sm_param} for $T(k_{\rm pk})$  and express $a_{\rm fz}$, $a_{\rm rh}$, and $a_{\rm dom}$ using eq.~\eqref{eq:asfz_est_sm}, eq.~\eqref{eq:arh_est}, and eq.~\eqref{eq:adom_est} respectively to obtain
\begin{multline}\label{eq:Tpk_sm}
T(k_{\rm pk})\sim 2\times 10^2\left(\frac{\xi_i}{0.4}\right)^4\Big(\frac{m}{\rm TeV}\Big)^{4/3}\left(\frac{\rm 10\, MeV}{T_{\rm rh}}\right)^{4/3}\left(\frac{10}{g_*(T_{\rm rh})}\right)^{1/3}\left(\frac{100}{g_{*}(10m/\xi_i)}\right)\\
\times \frac{1}{6}\ln\left[50\left(\frac{0.4}{\xi_i}\right)^{9/2}\left(\frac{0.1}{\alpha_c}\right)^{3/4}\left(\frac{m}{\rm TeV}\right)^{1/4}\left(\frac{g_{*}(10m/\xi_i)}{100}\right)^{9/8}\right].
\end{multline}
In both cases, $T(k_{\rm pk})$ is approximately proportional to $m/T_{\rm rh}$ because for a given $\alpha_c$ and $\xi_i$ this ratio determines the post-freeze-out duration of the ECDE. Since $\delta_{\rm DM}(k_{\rm pk})$ grows faster during this period than at any other time prior to matter-radiation equality, increasing this duration increases $T(k_{\rm pk})$. The amplitude of the transfer function at $k_{\rm pk}$ has a power-law dependence on $\alpha_c$ when the cannibals freeze out in a cannibal-dominated universe, while it only depends logarithmically on $\alpha_c$ when the cannibals freeze out in a SM radiation-dominated universe. This difference in sensitivity to $\alpha_c$ reflects the linear growth of $\delta_{\rm DM}(k_{\rm pk})$ between $a_{\rm fz}$ and $a_{\rm rh}$ for $a_{\rm dom}<2a_{\rm fz}$, as opposed to its logarithmic growth between $a_{\rm fz}$ and $a_{\rm dom}$ for $a_{\rm dom}>2a_{\rm fz}$.   These two growth histories for $\delta_{\rm DM}(k_{\rm pk})$ also explain why $T(k_{\rm pk})$ is independent of $\xi_i$ when $a_{\rm dom}<2a_{\rm fz}$, but is strongly dependent on $\xi_i$ when $a_{\rm dom}>2a_{\rm fz}$: since $\xi_i$ determines $a_{\rm dom}/a_i$, it sets the transition from logarithmic to linear growth when cannibals freeze out prior to the start of the ECDE.

\begin{figure}
	\begin{subfigure}{.5\textwidth}
		\includegraphics[trim=0 50 0 0,width=1.00\textwidth]{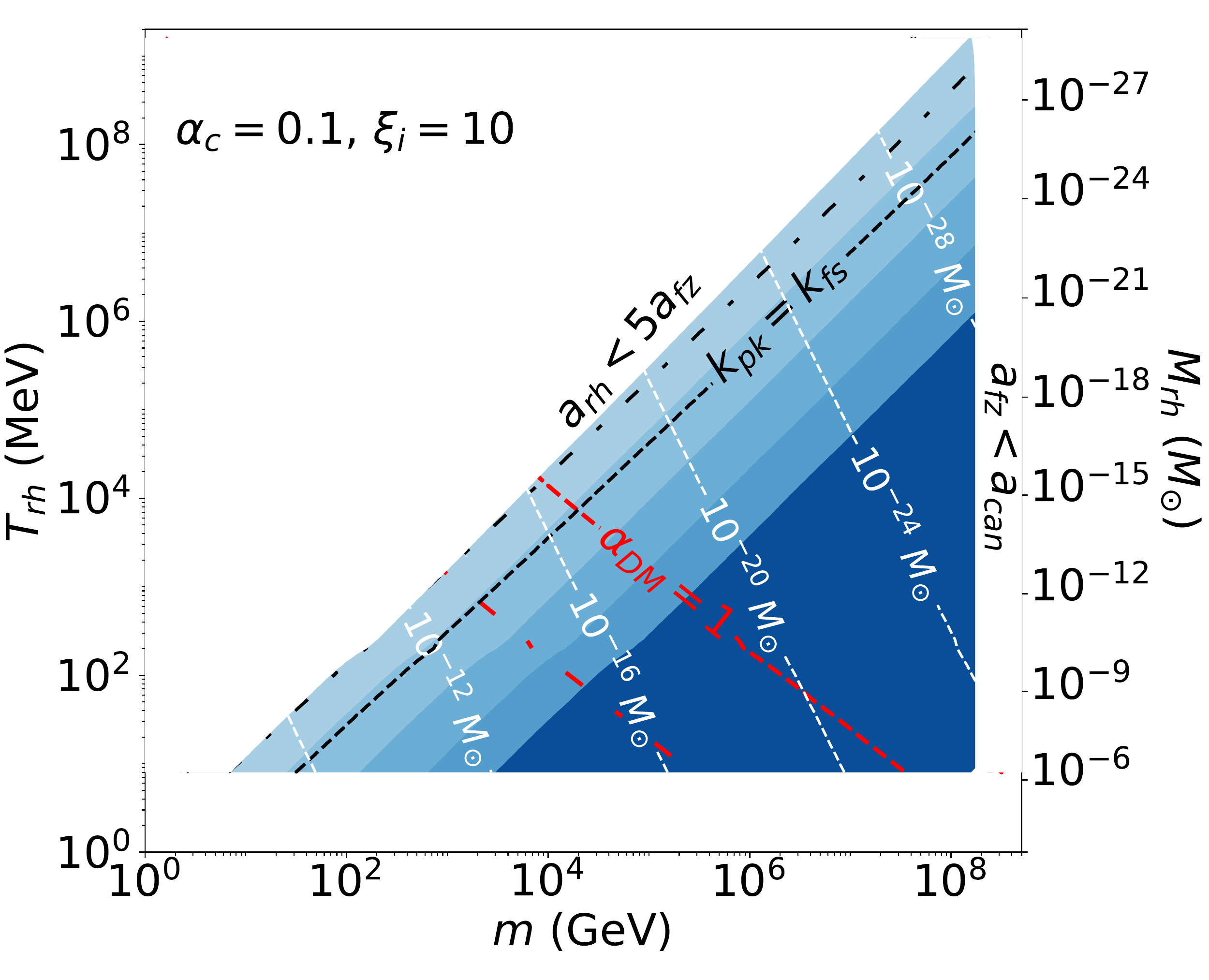}
	\end{subfigure}
	\begin{subfigure}{.5\textwidth}
		\includegraphics[trim=0 50 0 0,width=1.00\textwidth]{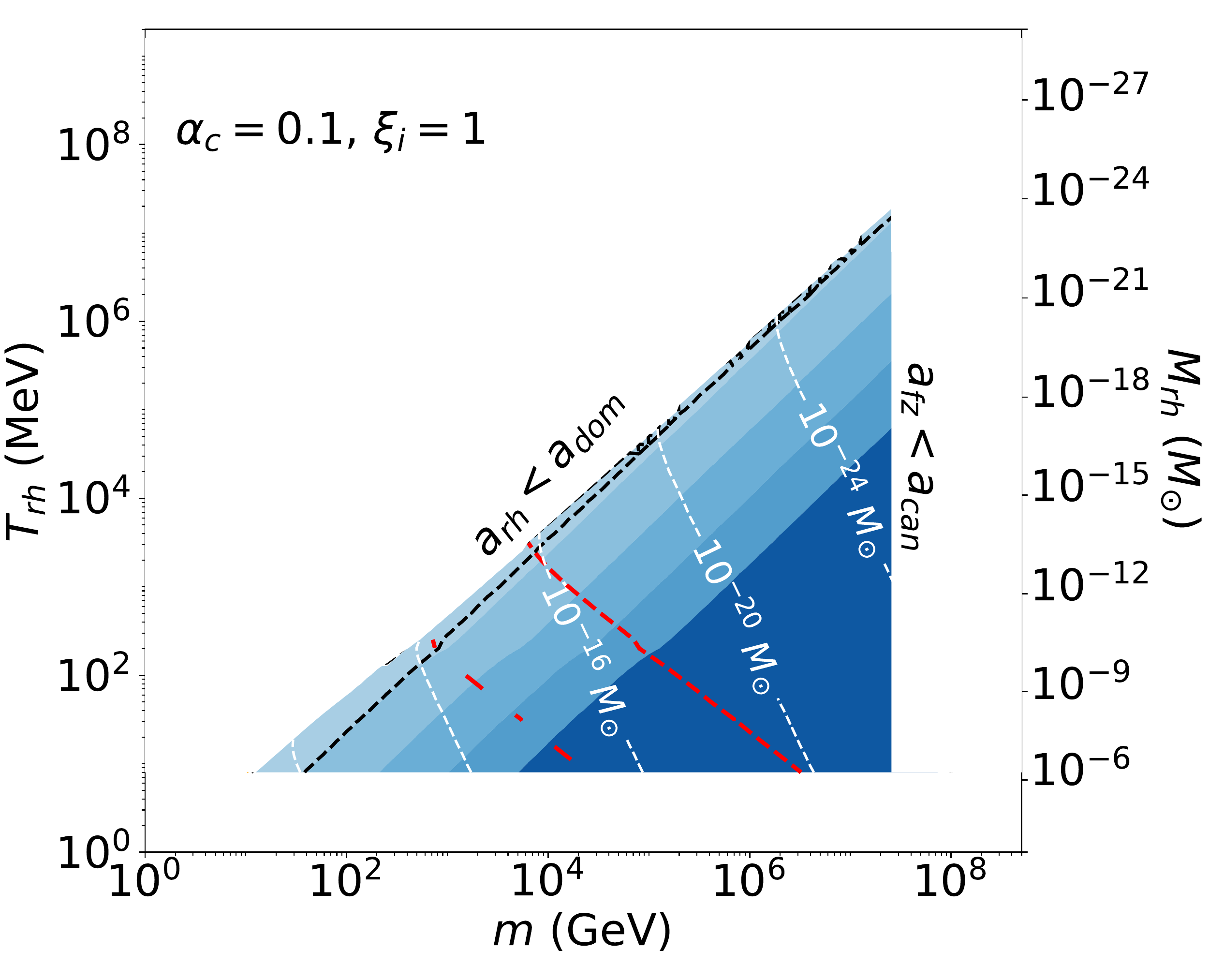}
	\end{subfigure}\\
	\begin{subfigure}{.5\textwidth}
	\includegraphics[trim=0 50 0 0,width=1.00\textwidth]{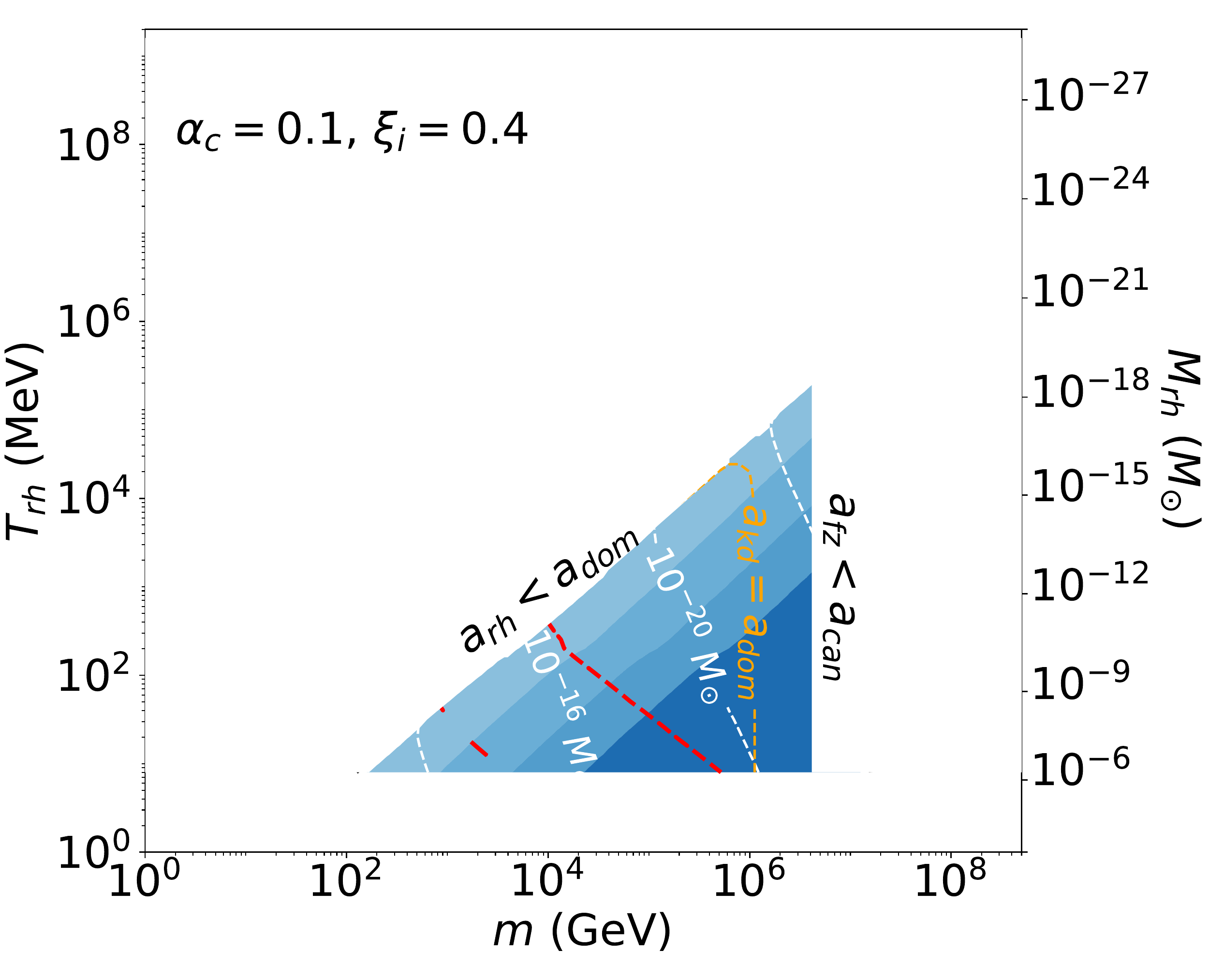}
	\end{subfigure}
	\begin{subfigure}{.5\textwidth}
		\includegraphics[trim=0 50 0 0,width=1.00\textwidth]{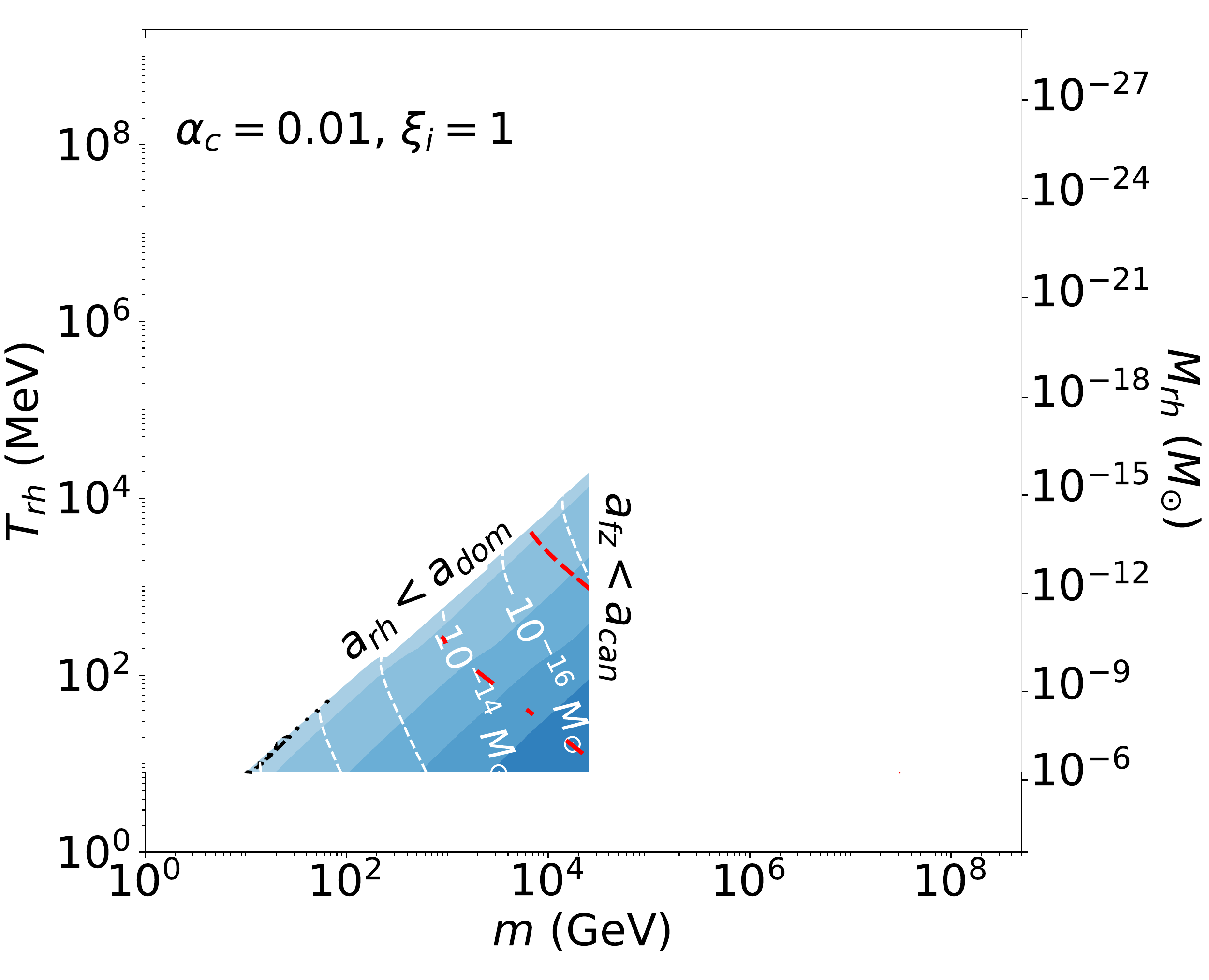}
	\end{subfigure}\\
\begin{center}
		\begin{subfigure}{.5\textwidth}
		\includegraphics[trim=0 20 0 600,clip,width=\textwidth]{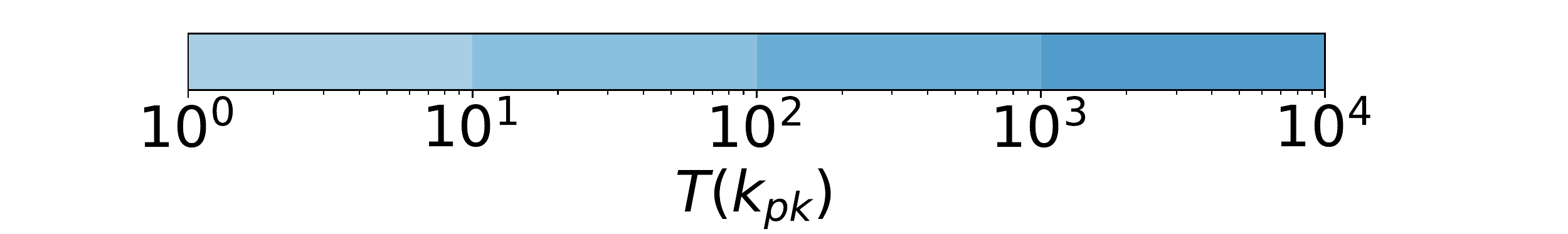}
	\end{subfigure}
\end{center}
	\caption{Colored contours show the value of the DM transfer function at the scale that maximizes the matter power spectrum, $T(k_{\rm pk})$, as a function of $m$ and $T_{\rm rh}$ for fixed $\alpha_c$ and $\xi_i$. In the top left panel, the cannibal density exceeds the SM radiation density up until reheating. In all other panels, the cannibals freeze out in a SM radiation-dominated universe. The white dashed contours show the mass scale of the first microhalos, eq.~\eqref{eq:Mpk_def}. The secondary $y$-axis shows the microhalo mass scale corresponding to modes entering the horizon at reheating. In the white space on the top left, reheating either occurs prior to cannibal freeze-out (top left panel) or the cannibal density does not dominate the universe prior to the decay of the cannibals (all other panels). In the white space on the right, cannibalism does not occur.
	Above the red dashed (dot-dashed) line, thermal freezeout cannot generate the observed DM abundance for $m_{\rm DM}\geq10m$  ($m_{\rm DM}\geq100m$). To the left of the black dashed (dot-dashed) line, the post-reheating free streaming of DM modifies $k_{\rm pk}$ if DM kinetically decouples from the cannibals after $a_{\rm fz}$ and $m_{\rm DM}=10m$ ($m_{\rm DM}=100m$). In the bottom left panel, the cannibal fluid becomes collisionless prior to $a_{\rm dom}$ to the right of the orange-dashed line, and consequently cannibal free streaming modifies $k_{\rm pk}$.}
	\label{fig:param_space}
\end{figure}

In figure~\ref{fig:param_space}, white-dashed contours show $M_{\rm pk}$ as a function of $m$ and $T_{\rm rh}$ for fixed values of $\xi_i$ and $\alpha_c$. The $M_{\rm pk}$ contours were calculated from the expression for $k_{\rm pk}$ given in eq.~\eqref{eq:kpk_can} if the cannibals freeze out in a cannibal-dominated universe and eq.~\eqref{eq:kpk_sm} if the cannibals freeze out in a SM radiation-dominated universe. The colored contours show $T(k_{\rm pk})$, which is evaluated by numerically solving the cosmological perturbation equations for $k_{\rm pk}$.  The secondary $y$-axis on the right shows the values of $M_{\rm rh}$ obtained from eq.~\eqref{eq:mrh_def}.
The parameter space with $5\lesssim T(k_{\rm pk})\lesssim 10^4$ is the region that generates a significantly enhanced abundance of microhalos with masses between $M_{\rm pk}$ and $M_{\rm rh}$. As $T(k_{\rm pk})$ is roughly proportional to $m/T_{\rm rh}$, there is an enhanced abundance of microhalos for a band of $m/T_{\rm rh}$ values. 
 
The parameter space shown in figure~\ref{fig:param_space} is bounded on all sides by three conditions. First, the reheat temperature defined by eq.~\eqref{eq:Trh_def} must exceed 8.1 MeV to be compatible with the constraints from BBN and the CMB \cite{deSalas:2015glj,Hasegawa:2019jsa}. Second, significantly increasing the microhalo abundance requires a period of cannibal domination following cannibal freeze-out, i.e. ${\rm max}(a_{\rm fz},a_{\rm dom})<a_{\rm rh}$.  Finally, as here we are  specifically interested in the impact of cannibal interactions on perturbation growth, we require an epoch of cannibalism to occur,  i.e., $a_{\rm can}<a_{\rm fz}$.

A period of cannibalism only occurs if the cannibals remain in chemical equilibrium after they become non-relativistic.  For a fixed value of $\alpha_c$, this condition imposes an upper bound on $m$ because $a_{\rm fz}/a_i$ decreases as $m$ increases.
Using eq.~\eqref{eq:asfz_est_can} and eq.~\eqref{eq:asfz_est_sm} for $a_{\rm fz}/a_i$ and the fact that $a_{\rm can}\approx 100a_i$, the $a_{\rm fz}>a_{\rm can}$ condition can be rewritten as:
\begin{align}\label{eq:m_fz_can_bound}
\Big(\frac{m}{\rm GeV}\Big)<\begin{cases}10^{8}\left(\dfrac{\alpha_c}{0.1}\right)^3 & a_{\rm dom}<2a_{\rm fz}, \\
	\\
2\times 10^{6}\left(\dfrac{\xi_i}{0.4}\right)^2\left(\dfrac{\alpha_c}{0.1}\right)^3 \left(\dfrac{100}{g_*(10m/\xi)}\right)^{1/2} & a_{\rm dom}>2a_{\rm fz}.
\end{cases}
\end{align}
If $m$ exceeds the bound in eq.~\eqref{eq:m_fz_can_bound}, then the number-changing self-interactions decouple while the cannibal particles are relativistic, and any subsequent ECDE is simply a matter-dominated era. The evolution of perturbations in such scenarios (without cannibal interactions) have already been studied in the context of decoupled hidden sector theories  \cite{Blanco:2019eij,Zhang:2015era}.  

To obtain a substantial amount of growth, reheating must occur well after the cannibal reactions freeze out ($a_{\rm rh}>5a_{\rm fz}$) and in a cannibal-dominated universe ($a_{\rm rh}>a_{\rm dom}$).\footnote{Our numerical calculations neglect  cannibal decays during cannibal freezeout. For the parameter space of interest for enhanced structure formation, this is an excellent approximation, but for  $a_{\rm rh}<5a_{\rm fz} $, the impact of cannibal decays can be nonnegligible during freezeout. Thus we only show numerical results for $ a_{\rm rh}>5a_{\rm fz} $.}  This requirement imposes a lower bound on $m$ for a given $T_{\rm rh}$, following from the expressions for $a_{\rm rh}$, $a_{\rm fz}$, and $a_{\rm dom}$ given in eq.~\eqref{eq:arh_est}, eq.~\eqref{eq:asfz_est_can}, and eq.~\eqref{eq:adom_est}:
\begin{align}\label{eq:m_rh_bound}
\Big(\frac{m}{\rm GeV}\Big)>\begin{cases}10\left(\dfrac{\alpha_c}{0.1}\right)^{3/7}\Big(\dfrac{T_{\rm rh}}{\rm 10\, MeV}\Big)^{6/7}\left(\dfrac{g_*(T_{\rm rh})}{10}\right)^{3/14} & a_{\rm dom}<2a_{\rm fz},\\ \\
70\left(\dfrac{0.4}{\xi_i}\right)^{3}\left(\dfrac{T_{\rm rh}}{\rm 10\, MeV}\right)\left(\dfrac{g_{*}(10m/\xi_i)}{100}\right)^{3/4}\left(\dfrac{g_*(T_{\rm rh})}{10}\right)^{1/4} & a_{\rm dom}>2a_{\rm fz}.
\end{cases}
\end{align}
If $a_{\rm rh}<5a_{\rm fz}$, the modes that enter the horizon during the ECDE do not escape the cannibal Jeans horizon much prior to reheating.  Consequently, there is no significant enhancement of DM perturbations, and the $a_{\rm rh}<5a_{\rm fz}$ section of parameter space does not provide interesting cosmological signatures.  

As the initial density of the cannibals decreases relative to the SM radiation density, the upper and lower bounds on $m$ given by eqs.~\eqref{eq:m_fz_can_bound} and \eqref{eq:m_rh_bound} become more restrictive, as seen in figure~\ref{fig:param_space}. For smaller $\xi_i$, larger values of $m/T_{\rm rh}$ are needed to give sufficient time for the cannibal density to overcome the SM radiation density prior to reheating. Decreasing $\xi_i$ also increases the Hubble rate at a given cannibal temperature, so smaller values of $m/\alpha_c^3$ are required to keep the cannibals in equilibrium after they become non-relativistic.

Figure~\ref{fig:param_space} also illustrates how decreasing $\alpha_c$ shrinks the region of cannibal parameter space that yields a substantially enhanced  microhalo population.
If cannibal freeze-out occurs while the universe is cannibal dominated, then decreasing $\alpha_c$ makes it possible for lighter cannibals to freeze out before reheating (for fixed $T_{\rm rh}$). However, decreasing $\alpha_c$ also reduces the parameter space where the cannibal particles will freeze out while non-relativistic, so the net effect of lowering $\alpha_c$ is to reduce the range of $m$ values that can realize $a_{\rm can}<a_{\rm fz}<5a_{\rm rh}$.
If cannibal freeze-out occurs during SM radiation domination, the lower bound on $m$ is set by the requirement that $a_{\rm dom}<a_{\rm rh}$, which is independent of $\alpha_c$. Consequently, in the right panels in figure~\ref{fig:param_space}, only the upper bound on $m$ moves as $\alpha_c$ is changed.

If either $\xi_i$ or $\alpha_c$ becomes too small, then it is not possible for particles that undergo an epoch of cannibalism to significantly affect the growth of structure because they do not dominate the universe after freezing out. Since decreasing $T_{\rm rh}$ reduces the lower bound on $m$ given by eq.~\eqref{eq:m_rh_bound}, we compare the lower and upper bounds on $m$ at the smallest reheat temperatures allowed by BBN and the CMB constraints. With $T_{\rm rh}= 10$ MeV, it is possible to satisfy both upper and lower bounds on $m$ if
\begin{align}\label{eq:alpha_bound}
	 \alpha_c>\begin{cases}
	 	2\times 10^{-4}\left(\dfrac{g_*(T_{\rm rh})}{10}\right)^{1/12} & a_{\rm dom}\ll 2a_{\rm fz}\\
	 	\\
	 	10^{-3}\xi_i^{-5/3}	  \left(\dfrac{g_*(10m/\xi_{i})}{100}\right)^{5/12}\left(\dfrac{g_*(T_{\rm rh})}{10}\right)^{1/12}& a_{\rm dom}\gg 2a_{\rm fz}.
	 \end{cases}
\end{align}

Not all regions of cannibal parameter space that realize an ECDE are compatible with DM production through thermal freeze-out.   If we suppose that the DM relic abundance is fixed by annihilations to hidden sector species (e.g., the cannibal itself), then for a given set of cannibal parameters, $\{m,T_{\rm rh}, \alpha_c , \xi_i\}$, we can solve for  the value of the DM annihilation cross-section, $\langle \sigma v\rangle_{\rm DM}$, that generates the observed DM abundance. 
If the DM annihilation cross-section takes the form
\begin{align}\label{eq:alpha_DM}
\langle \sigma v\rangle_{\rm DM}\equiv \frac{\pi\alpha_{\rm DM}^2}{m_{\rm DM}^2},
\end{align}
requiring $\alpha_{\rm DM}<1$ for perturbativity then implies an upper bound on $m_{\rm DM}$.

We can estimate this upper bound using a sudden freezeout approximation for the DM abundance,
\begin{align}
\label{eq:sudden_fzout_DM}
	n_{\rm DM}(a_{\rm f,DM})\equiv \frac{H(a_{\rm f,DM})}{\langle\sigma v\rangle_{\rm DM}},
\end{align}
which defines the scale factor at DM freeze-out, $a_{\rm f,DM}$. 
The DM number density today is
\begin{align}
	n_{\rm DM}(a_0)\approx n_{\rm DM}(a_{\rm f,DM})\left(\frac{a_{\rm f,DM}}{a_0}\right)^3=\frac{H(a_{\rm f,DM})}{\langle\sigma v\rangle_{\rm DM}}\left(\frac{a_{\rm f,DM}}{a_i}\times \frac{a_{i}}{a_{\rm rh}}\times \frac{a_{\rm rh}}{a_0}\right)^3.
	\label{nDMtoday}
\end{align}
As the cannibals evolve like radiation prior to $a_i$, we have
\begin{align}
	\frac{a_{\rm f,DM}}{a_i} = \frac{T_c(a_i)}{T_c(a_{\rm f,DM})}=\frac{10m}{m_{\rm DM}}x_{\rm DM},
\end{align}
where $x_{\rm DM}\equiv m_{\rm DM}/T_c(a_{\rm f,DM})$. We use eq.~\eqref{eq:entropy_cons} and eq.~\eqref{eq:arh_est} to express $a_{\rm rh}/a_0$ and $a_{\rm rh}/a_i$ in terms of $T_{\rm rh}$ and $m$. Given that
\begin{align}
H(a_{\rm f,DM})=H(a_i)\frac{a_i^2}{a_{\rm f,DM}^2}=H(a_i)\frac{m_{\rm DM}^2}{(10m)^2}x_{\rm DM}^{-2},
\end{align} 
eq.~\eqref{nDMtoday} implies that
\begin{align}\label{eq:svdm_est}
	\langle \sigma v\rangle_{\rm DM}\approx \frac{10^{-12}}{{\rm GeV}^2} \sqrt{[1+g_*(10m/\xi_i)/\xi_i^4]} \left(\frac{x_{\rm DM}}{10}\right)\left(\frac{\Omega_{\rm dm}h^2}{0.12}\right)^{-1}\left(\frac{T_{\rm rh}}{\rm 10\, MeV}\right)\left(\frac{\rm TeV}{m}\right),
\end{align}
where we assume $g_{*s}(T_{\rm rh})=g_*(T_{\rm rh})$. 
Keeping $\alpha_{\rm DM}<1$ then demands that
\begin{align}\label{eq:m_DM_fzout}
\left(\frac{x_{\rm DM}}{10}\right)\left(\frac{m_{\rm DM}/m}{10}\right)^2\left(\frac{\Omega_{\rm dm}h^2}{0.12}\right)^{-1}\left(\frac{T_{\rm rh}}{\rm 10\, MeV}\right)\left(\frac{m}{\rm TeV}\right)\sqrt{[1+g_*(10m/\xi_i)/\xi_i^4]}<10^4.
\end{align}
The above bound still depends on $x_{\rm DM}$. While $x_{\rm DM}$ is typically determined by inserting the equilibrium number density into Eq.~\eqref{eq:sudden_fzout_DM}, this procedure makes $x_{\rm DM}$ logarithmically dependent on $\langle \sigma v\rangle_{\rm DM}$.  To avoid this dependency, we instead determine $x_{\rm DM}$ through 
\begin{align}
	n_{\rm DM,eq}(x_{\rm DM})\frac{a_{\rm f,DM}^3}{a_0^3}=n_{\rm DM}(a_0).   
\end{align}
Thus, given $m_{\rm DM}/m$, eq.~\eqref{eq:m_DM_fzout} provides an upper bound on $m$. The red dashed line in figure~\ref{fig:param_space} shows the upper bound on $m$ resulting from the condition $\alpha_{\rm DM}<1$ for $m_{\rm DM}=10m$.\footnote{While decreasing $m_{\rm DM}$ relative to $m$ relaxes the upper bound on $m$,  our analysis assumes the DM and cannibal species to be chemically decoupled by $T_c(a_i)=10m$, and thus we consider $m_{\rm DM}\gtrsim 10m$.} For larger values of $m$, alternative production mechanisms such as freeze-in can still generate the observed DM density \cite{Hall:2009bx}.

As discussed in section~\ref{sec:can_not_perfect}, departures from perfect-fluid behavior for either the cannibals or DM can be important in some regions of parameter space.  The impact of DM free streaming depends on its kinetic coupling to the cannibal fluid and is model-dependent. In figure~\ref{fig:param_space} the black dashed lines show where the free-streaming horizon $\lambda_{\rm DM,fs}$, given in eq.~\eqref{eq:DM_fs}, equals $k_{\rm pk}$ for $m_{\rm DM}=10m$, in the case where DM kinetically decouples from the cannibal fluid after cannibal freeze-out.
Above and to the left of this line, DM free-streaming, rather than cannibal interactions, can determine the peak of the transfer function.  
 
 To better illustrate when DM free-streaming can be relevant, we simplify eq.~\eqref{eq:DM_fs} by neglecting the temperature dependence of $g_*$ and the DM density inside the Hubble rate. This yields
\begin{align}
	\lambda_{\rm DM,fs}\sim \sqrt{\frac{T_{c}(a_{\rm rh})}{m_{\rm DM}}}(aH)_{\rm rh}^{-1}\log(a_{\rm eq}/a_{\rm rh})\sim\sqrt{\frac{m}{m_{\rm DM}}}k_{J}^{-1}(a_{\rm rh})\log(a_{\rm eq}/a_{\rm rh}).
\end{align}
In the last relation we have used the definition of the Jeans length, eq.~\eqref{eq:jeans}, and the fact that $c_s^2=5T_c/(3m)$ and $w_c=T_c/m\ll1$ for $a\gg a_{\rm fz}$. For scenarios where cannibals freeze out in a cannibal-dominated universe, we have $k_J^{-1}(a_{\rm rh})/k_{\rm pk}^{-1}\approx  k_J(2a_{\rm fz})/k_J(a_{\rm rh})$. Consequently, DM free streaming becomes relevant when $a_{\rm rh}$ is close to $a_{\rm fz}$ and the ratio $m/m_{\rm DM}$ is not too small. In contrast, when cannibals freeze out in a SM-dominated universe, we have $k_J^{-1}(a_{\rm rh})/k_{\rm pk}^{-1}\approx k_J^{-1}(a_{\rm rh})/r_s(a_{\rm dom})$. Here, the logarithmic growth of the sound horizon until $a_{\rm dom}$ increases the gap between $k_J^{-1}(a_{\rm rh})$ and $k_{\rm pk}^{-1}$ (as seen in the top left panel of figure~\ref{fig:SM_dom_horz_peak}). In the bottom left panel of figure~\ref{fig:param_space}, this gap is large enough that the DM free-streaming horizon remains less than $k_{\rm pk}^{-1}$ for $m_{\rm DM}\geq 10m$.

In the bottom left panel of figure~\ref{fig:param_space},  the cannibal fluid becomes collisionless prior to $a_{\rm dom}$, i.e. $a_{\rm kd}<a_{\rm dom}$, in the region right of the orange dashed line. In this regions, the peak of the transfer function is determined by the cannibal free-streaming horizon instead of the cannibal sound horizon.  In computing the $a_{\rm kd}=a_{\rm dom}$ boundary shown in figure~\ref{fig:param_space} we set $\lambda=0$ when evaluating $\langle \sigma_cv_c\rangle$ through eq.~\eqref{eq:scat_can_rate}.
To see how the timing of cannibal kinetic decoupling depends on cannibal parameters more generally, we begin with the inequality $a_{\rm kd}>a_{\rm dom}$, write $a_{\rm kd}=a_{\rm kd}/a_{\rm fz}\times a_{\rm fz}$, and subsequently use eqs.~\eqref{eq:akd_sm}, ~\eqref{eq:asfz_est_sm}, and~\eqref{eq:adom_est} for $a_{\rm kd}/a_{\rm fz}$, $a_{\rm fz}$, and $a_{\rm dom}$, respectively. The condition $a_{\rm kd}>a_{\rm dom}$  then becomes
\begin{align}\label{m_kd_bound}
	 m<6\times10^{5}\left(\frac{\alpha_c}{0.1}\right)^{11/5}\left(\frac{\xi_i}{0.4}\right)^{42/5} \left|\frac{q-5/3}{(3q-1)^{2/3}}\right|^{8/5}{\rm GeV},
\end{align}
where $q\equiv \lambda/(g/m)^2$.
This restriction on $m$ is only relevant if it is more constraining than eq.~\eqref{eq:m_fz_can_bound}. Consequently, the restriction on $m$ in eq.~\eqref{m_kd_bound} becomes relevant when
\begin{align}\label{eq:xi_kd}
\xi_i<0.68\left(\frac{\alpha_c}{0.1}\right)^{1/8}\left|\frac{q-5/3}{(3q-1)^{2/3}}\right|^{-1/4}.
\end{align}
For values of $\xi_i$ larger than the RHS of eq.~\eqref{eq:xi_kd}, $a_{\rm kd}$ is always greater than $a_{\rm dom}$, and cannibal free streaming does not affect $k_{\rm pk}$ in the parameter space where the cannibals significantly enhance microhalo abundance and also undergo cannibalism. Thus, there is no $a_{\rm kd}=a_{\rm dom}$ boundary in the panels with $\xi_i=1$ or 10 in figure~\ref{fig:param_space}. As we decrease $\xi_i$ below the RHS of eq.~\eqref{eq:xi_kd}, a larger fraction of the parameter space has $a_{\rm kd}<a_{\rm dom}$. However, the parameter space where the cannibals significantly enhance microhalo abundance also shrinks as we decrease $\xi_i$, until there is no allowed parameter space for $\xi_i<0.07(\alpha_c/0.1)^{-3/5}$ (see eq.~\eqref{eq:alpha_bound}). Consequently, cannibal free streaming may affect $k_{\rm pk}$ only in the narrow parameter space between $0.07(\alpha_c/0.1)^{-3/5}<\xi_i<0.68(\alpha_c/0.1)^{1/8}\left|\frac{q-5/3}{(3q-1)^{2/3}}\right|^{-1/4}$. Furthermore, as discussed in section~\ref{sec:can_not_perfect}, we expect only a marginal correction to the perfect-fluid result for $k_{\rm pk}$  in the parameter space where $a_{\rm kd}<a_{\rm dom}$.

The early-forming microhalos generated by an ECDE have large enough central densities to survive accretion into galaxies \cite{Delos:2019lik, Delos:2019tsl, Delos:2019dyh}. While sub-Earth-mass halos are too diffuse to be detected by photometric microlensing searches \cite{Li:2012qha} and too small to be detected via astrometric microlensing \cite{Erickcek:2010fc,Li:2012qha,VanTilburg:2018ykj}, they can be detected by pulsar timing arrays \cite{dror2019pta, Ramani:2020hdo} and by their impact on stellar microlensing  within galaxy clusters \cite{Dai:2019lud, Blinov:2019jqc, Blinov:2021axd}. Furthermore, if the dark matter is a thermal relic, early-forming halos significantly boost the dark matter annihilation rate regardless of their masses, and the isotropic gamma-ray background places powerful constraints on the microhalo population \cite{Blanco:2019eij, Delos:2019dyh}.  If dark matter annihilation is eventually detected in dwarf spheroidal galaxies, the emission profile could distinguish annihilation within microhalos from both decaying dark matter and dark matter annihilation outside of microhalos \cite{Delos:2019dyh}.

A full analysis of the observational constraints on cannibalism within a hidden sector lies beyond the scope of this article, but we can use constraints on EMDE cosmologies to forecast which regions of cannibal parameter space are likely to be probed by current and future observations.  Constraints on EMDE cosmologies are often expressed in terms of a generic exponential cutoff scale: $P(k) \propto \exp[-k^2/k_\mathrm{cut}^2]$.  Ref.~\cite{Delos:2021rqs} showed that weekly observations of 500 pulsars over 20 years with an rms timing residual of 10 ns would detect microhalos arising from an EMDE with $k_\mathrm{cut}/k_\mathrm{rh} > 20$ and $T_\mathrm{rh} \leq 30$ MeV.  Increasing the observational period to 40 years extends the reach of pulsar timing arrays to reheat temperatures up to 100 MeV with 200 pulsars; see also Ref.~\cite{Lee:2021zqw}.  The EMDE transfer function \cite{ES11} implies that $T(k_{\rm pk}) \simeq 25$ for $k_\mathrm{cut}/k_\mathrm{rh} = 20$, nearly independently of 
the reheat temperature.  
If cannibal reactions freeze out during the ECDE, then the power spectrum on scales $k \lesssim k_\mathrm{pk}$ is the same in EMDE and ECDE cosmologies, and ECDE scenarios with $25 \lesssim T(k_{\rm pk}) \lesssim 10^4$ would generate pulsar timing signals that are at least as strong as those produced by an EMDE with $k_\mathrm{cut}/k_\mathrm{rh} \simeq 20$.  If cannibal reactions freeze out prior to cannibal domination, then the ECDE power spectrum differs more substantially from the EMDE power spectrum analyzed by Ref.~\cite{Delos:2021rqs} for $k \gtrsim k_\mathrm{dom}$, 
but we can still predict which ECDE scenarios are likely to be accessible by pulsar timing arrays.  If $k_\mathrm{cut}/k_\mathrm{rh} = 20$, then the EMDE power spectrum peaks at $k_\mathrm{pk} \simeq 27 k_\mathrm{rh}$.  Therefore, ECDE scenarios with $T( 27 k_\mathrm{rh} ) \gtrsim 25$ and $T(k_{\rm pk}) \lesssim 10^4$ will generate a microhalo population that is at least as detectable as the microhalos that result from an EMDE with $k_\mathrm{cut}/k_\mathrm{rh} = 20$.  For $\alpha_c = 0.1$, 
obtaining $T( 27 k_\mathrm{rh} ) \gtrsim 25$ requires 
$T(k_\mathrm{pk}) \gtrsim 100$ for $\xi_i = 1$ and $T(k_\mathrm{pk}) \gtrsim 500$ for $\xi_i = 0.4$.  Estimates of potential sensitivity from observations of cluster caustic microlensing are at a far more preliminary stage, but suggest broadly similar reach for $T_\mathrm{rh}$  and $T(k_\mathrm{pk})$ individually \cite{Blinov:2021axd}.

The best current constraints on EMDE cosmologies with thermal relic dark matter come from the isotropic gamma-ray background \cite{Blanco:2019eij, Delos:2019dyh}.  Since the dark matter annihilation rate within microhalos does not change after the microhalos form and the microhalos track the dark matter density, dark matter annihilations within early-forming microhalos generate the same constant emission per dark matter mass as decaying dark matter.  It is therefore possible to define an effective dark matter decay lifetime for these scenarios \cite{Blanco:2019eij}:
\begin{equation} 
\tau_\mathrm{eff} = \left(\frac{10^{-10} \,\mathrm{GeV}^{-2}}{\langle \sigma v \rangle_\mathrm{DM}}\right)\left(\frac{m_\mathrm{DM}}{10^6 \,\mathrm{GeV}}\right) \left(\frac{7 \times 10^{38} \,\mathrm{seconds}}{B_0}\right),
\end{equation}
where $B_0 \equiv \langle \rho_\mathrm{DM}^2 \rangle/ \bar{\rho}_\mathrm{DM}^2$ is the structure boost factor generated by the microhalos.  This effective lifetime should be compared to bounds on dark matter lifetime for particles with twice the mass.  When accounting for emission from astrophysical sources, Ref. \cite{Blanco:2018esa} found that Fermi-LAT observations of the IGRB \cite{Fermi-LAT:2014ryh} demand that $\tau_\mathrm{eff} \gtrsim 10^{28}$ seconds for $m_\mathrm{DM}$ between 10 GeV and $10^9$ GeV and a wide range of annihilation channels.  

The microhalo boost factor that arises from an EMDE has been calculated for scenarios in which all modes with $k_\mathrm{rh} < k < k_\mathrm{cut}$ enter the horizon during the EMDE \cite{Erickcek:2015jza, Delos:2019dyh} and for scenarios that include a radiation-dominated era prior to the EMDE with $k_\mathrm{cut} > k_\mathrm{dom}$ \cite{Blanco:2019eij}.  The former case generates a sharp peak in the matter power spectrum that is qualitatively similar to the peak generated when cannibal reactions freeze out during the ECDE, while the latter generates the same plateau feature as an ECDE that starts after cannibals freeze out.  However, for the limited range of $T(k_\mathrm{pk})$ values that were considered in both analyses, the two scenarios have values of $B_0$ that differ by less than a factor of 10, and much of that variation can be attributed to differing assumptions regarding the microhalo density profiles \cite{Delos:2019dyh}.  The fact that $B_0$ is largely insensitive to changes in reheat temperature for fixed $T(k_\mathrm{pk})$ further supports the conclusion that the shape of the peak in the power spectrum does not significantly affect the dark matter annihilation rate: it does not matter how the microhalos are distributed in mass as long as they have the same formation time and contain the same fraction of the dark matter, both of which are determined by $T(k_\mathrm{pk})$.

The ECDE scenarios shown in Figure~\ref{fig:param_space} generally require $\langle \sigma v \rangle_\mathrm{DM} \gtrsim 10^{-12}$ GeV$^{-2}$ to generate the observed DM abundance through thermal freeze-out, which implies that $m_\mathrm{DM} \lesssim 2\times 10^6$ GeV is required to satisfy the unitarity bound.  For these parameters, $B_0 \lesssim 10^{13}$ is required to keep $\tau_\mathrm{eff} > 10^{28}$ seconds if the annihilations are predominantly $s$-wave so that $\langle \sigma v \rangle_\mathrm{DM}$ is independent of the DM velocity.  Refs. \cite{Erickcek:2015jza, Delos:2019dyh} did not consider boost factors this large because they restricted their analyses to microhalos that form after matter-radiation equality, but Ref. \cite{Blanco:2019eij} included microhalos that form during radiation domination and found that $B_0 \gtrsim 10^{13}$ for $T(k_\mathrm{pk}) \gtrsim 80$.  However, if we only consider ECDE scenarios with $T(k_\mathrm{pk}) \lesssim 80$, then $\langle \sigma v \rangle_\mathrm{DM} \gtrsim 10^{-10}$ GeV$^{-2}$ and $m_\mathrm{DM} \lesssim 2\times 10^5$ GeV.  For these parameters, the IGRB bound on $\tau_\mathrm{eff}$ demands that $B_0 \lesssim 10^{10}$, which corresponds to $T(k_\mathrm{pk}) \lesssim 20$.  It therefore seems likely that all of the ECDE parameter space in Figure~\ref{fig:param_space} that contains dark matter that thermally froze out (via $s$-wave annihilations) prior to the ECDE is already ruled out by observations of the IGRB.

\section{Summary and conclusions}
\label{sec:conclusions}
We have shown that an early cannibal-dominated era (ECDE) leaves a distinctive peak in the matter power spectrum. Perturbation modes that enter the horizon after the freeze-out of cannibal reactions but before the end of the ECDE are enhanced. On smaller scales, the pressure  generated by the self-heating of the cannibal particles suppresses the growth of dark matter perturbations. Consequently, the properties of the cannibal field generally establish the minimum halo mass following an ECDE. In Ref.~\cite{Erickcek:2020wzd}, we calculated this minimum halo mass in scenarios where the cannibals freeze out during the ECDE. In this paper, we show how the properties of the cannibal field establish the minimum halo mass even if the cannibal reactions freeze out prior to cannibal domination.	

Cannibals are generically predicted in theories with thermally decoupled hidden sectors that have a mass gap and a number-changing self-interactions.
If the lightest particle in such a hidden sector remains in chemical equilibrium after it becomes non-relativistic, it undergoes a period of cannibalism. During cannibalism, the cannibal number-changing self-interactions convert the particles' rest-mass energy into kinetic energy to maintain chemical equilibrium while conserving entropy within the cannibal fluid. The period of cannibalism ends when the rate of number-changing self-interaction falls below the Hubble rate. Such a cannibal fluid can easily come to dominate the energy density of the universe even if the hidden sector was initially colder than the SM bath. The ECDE ends when the cannibal particles decay into relativistic SM particles prior to the onset of BBN.

During the ECDE, we find that sub-horizon cannibal density perturbations grow linearly with the scale factor on scales that are larger than the cannibal Jeans length. The DM perturbations follow the cannibal density perturbations because the DM particles fall into the gravitational potential wells formed by the cannibals. Consequently, the enhancement of the DM perturbations after an ECDE relative to those in the standard cosmology reflects the cannibal perturbation spectrum and contains information about the cannibal self-interactions. This enhancement of the DM perturbations due to an ECDE is unaffected by possible scattering between the DM and the cannibals.

Enhanced small-scale DM perturbations collapse earlier than they otherwise would and hence lead to an enhanced population of halos at high redshift. Since an ECDE only affects perturbations on scales that enter the horizon during the ECDE, perturbations on these scales form microhalos with masses far less than the mass of the Sun. The characteristic mass of the earliest-forming microhalos, $M_{\rm pk}$, is determined by the scale with the largest enhancement in DM perturbations ($k_{\rm pk}$) whereas the formation time of these microhalos is determined by the amplitude of the enhancement, which is given by $T(k_{\rm pk})$.

The location of the peak of the DM power spectrum is determined by the process that counteracts gravitationally induced growth and prevents structure formation on small scales. In earlier works that have studied microhalo formation due to an early matter-dominated era (EMDE), this cutoff in the matter power spectrum was assumed to be generated by DM free streaming \cite{ES11, BR14, FOW14, Erickcek:2015jza} or axion DM oscillations \cite{Blinov:2019jqc}. Consequently, the peak scale is determined by DM microphysics. If the DM belongs to a hidden sector whose lightest particle causes the EMDE, then the DM particle may be cold enough that the relativistic pressure of the lightest hidden-sector particle sets the small-scale cutoff \cite{Blanco:2019eij,Zhang:2015era}. We showed in Ref.~\cite{Erickcek:2020wzd} that the cutoff in the matter power spectrum following an ECDE is typically generated by the thermal pressure in the cannibal fluid and is independent of DM microphysics when there is no period of SM radiation domination prior to the ECDE. In this work, we extended our analysis to scenarios in which the cannibals freeze-out while cannibal density is subdominant to SM density and showed that the cannibal thermal pressure still determines the cutoff. We find the cutoff scale to be given by the cannibal sound horizon at reheating, up to an order of magnitude, irrespective of the initial temperature ratio between the cannibal fluid and SM plasma and the properties of the DM particles. The only exceptions occur in narrow bands of parameter space where the DM free-streaming horizon overcomes the cannibal sound horizon or where the cannibal fluid becomes collisionless prior to cannibal domination. 

While the cannibal sound horizon provides a rough estimate of the wavenumber at which the power spectrum is maximized, $k_{\rm pk}$, we have also found a more accurate (within $\sim10$\%) expression for $k_{\rm pk}$. In the case where the freeze-out of cannibal reactions occurs during cannibal domination we find $k_{\rm pk}\approx k_J(2a_{\rm fz})/1.4$, where $k_J^{-1}$ is the cannibal Jeans length, and $a_{\rm fz}$ is the scale factor when cannibal reactions freeze out \cite{Erickcek:2020wzd}. If the freeze-out of cannibal reactions occurs during SM radiation domination, then we find $k_{\rm pk}$ to be given by eq.~\eqref{eq:kpk_sm}. These analytical estimates allow us to provide a map between the key microhalo properties, $M_{\rm pk}$ and $T(k_{\rm pk})$, and the cannibal particle properties. 

The peak amplitude of enhancement in dark matter perturbations due to an ECDE, $T(k_{\rm pk})$, is determined by how long cannibals dominate the universe after they freeze out, which depends on the cannibal mass, $m$, its $3\rightarrow2$ reaction coupling, $\alpha_c$, its initial temperature relative to the SM, $\xi_i$, and the reheat temperature, $T_{\rm rh}$. A longer period of post-freeze-out cannibal domination leads to larger values of $T(k_{\rm pk})$ and earlier microhalo formation.  If $T(k_{\rm pk})\gtrsim10^4$, the cannibals and DM particles assemble into halos prior to reheating. After the cannibals decay, the DM particles are released from these halos with sufficient velocity that their subsequent motion erases nearly all the perturbations that are within the horizon at reheating \cite{Blanco:2019eij}. Consequently, an ECDE will generate a significantly enhanced abundance of microhalos for $5\lesssim T(k_{\rm pk})\lesssim10^4$. Since $T(k_{\rm pk})$ is roughly proportional to the ratio $m/T_{\rm rh}$, a band of $m/T_{\rm rh}$ values is expected to yield an enhanced microhalo population. The upper and lower limits of this band are fixed by $\alpha_c$ and $\xi_i$. The range of possible values for $m,\ T_{\rm rh},\ \alpha_c,$ and $\xi_i$, is further constrained by the requirement that cannibals undergo cannibalism and that reheating occurs early enough to avoid altering the neutrino abundance, which would spoil the success of BBN \cite{Kawasaki:1999na,Kawasaki:2000en,Hannestad:2004px,Ichikawa:2005vw} and alter the anisotropies in the CMB \cite{deSalas:2015glj,Hasegawa:2019jsa}. Thus, we have identified a bounded region in the cannibal parameter space that produces an enhanced abundance of microhalos due to an ECDE. Within this parameter space, we provide estimates for the masses of the earliest-forming halos and their formation times in terms of the properties of the cannibal field. 

Finally, we briefly discussed potential observational sensitivity to this enhanced microhalo population.  
We expect the microhalos generated by ECDEs with reheat temperatures up to $T_\mathrm{rh} \simeq 100$ MeV with $T( 27 k_\mathrm{rh} ) \gtrsim 25$ and $T(k_{\rm pk}) \lesssim 10^4$ to be detectable in the future pulsar timing arrays analyzed in Refs. \cite{Delos:2021rqs, Lee:2021zqw}, while the results of Refs. \cite{Fermi-LAT:2014ryh,Blanco:2018esa,Blanco:2019eij, Delos:2019dyh} imply that the observed IGRB likely excludes $s$-wave thermal relic DM in almost all ECDE scenarios.  Cluster caustic microlensing is a promising alternative gravitational means of detecting the ECDE-enhanced population of microhalos in the low-redshift universe, but projections for such observations are not developed enough to allow for similarly definitive statements.   

It is important to remember, however, that all of these observational probes are sensitive to the internal structure of the microhalos. While it is possible to predict the density profiles of the first microhalos from the matter power spectrum \cite{Delos:2019mxl}, it is unknown how subsequent mergers between microhalos and their further evolution within galactic halos affect their internal structure.  Analyses that employ different assumptions regarding the microhalos' density profiles, substructure, and survival rate give similar but not identical bounds on EMDE cosmologies.  There is also a great deal of uncertainty regarding how the gravitational heating of the dark matter following structure formation during the EMDE or ECDE affects the subsequent formation of microhalos \cite{Blanco:2019eij}, and it has been suggested that microhalo remnants could persist through reheating \cite{Barenboim:2021swl}.
Therefore, we cannot yet establish robust observational constraints on cannibalism within a hidden sector.  
Nevertheless, we have identified which regions of cannibal parameter space enhance the microhalo abundance, which demonstrates how observations of small-scale structure provide a window into the evolution and particle content of the early Universe.

\paragraph{Acknowledgements.} We thank Marc Kamionkowski and Josh Ruderman for useful conversations.  The work of ALE is supported in part by NSF CAREER grant PHY-1752752. The work of PR and JS is supported in part by DOE CAREER grant DE-SC0017840.  Portions of this work were performed at the Kavli Institute for Theoretical Physics, which is supported in part by the National Science Foundation under Grant No. NSF PHY-1748958.

\appendix


\section{Homogeneous cannibal evolution}
\label{sec:A}

In this appendix, we determine the cosmological evolution of the cannibal fluid as it becomes non-relativistic as well as when the cannibal reactions freeze out. We consider the case where the cannibal particles become non-relativistic before both freeze-out and decay.

\subsection{Equilibrium evolution}
\label{sec:anl_cannibalism}
\begin{figure}
	\begin{subfigure}{.5\textwidth}
		\includegraphics[width=1.00\textwidth]{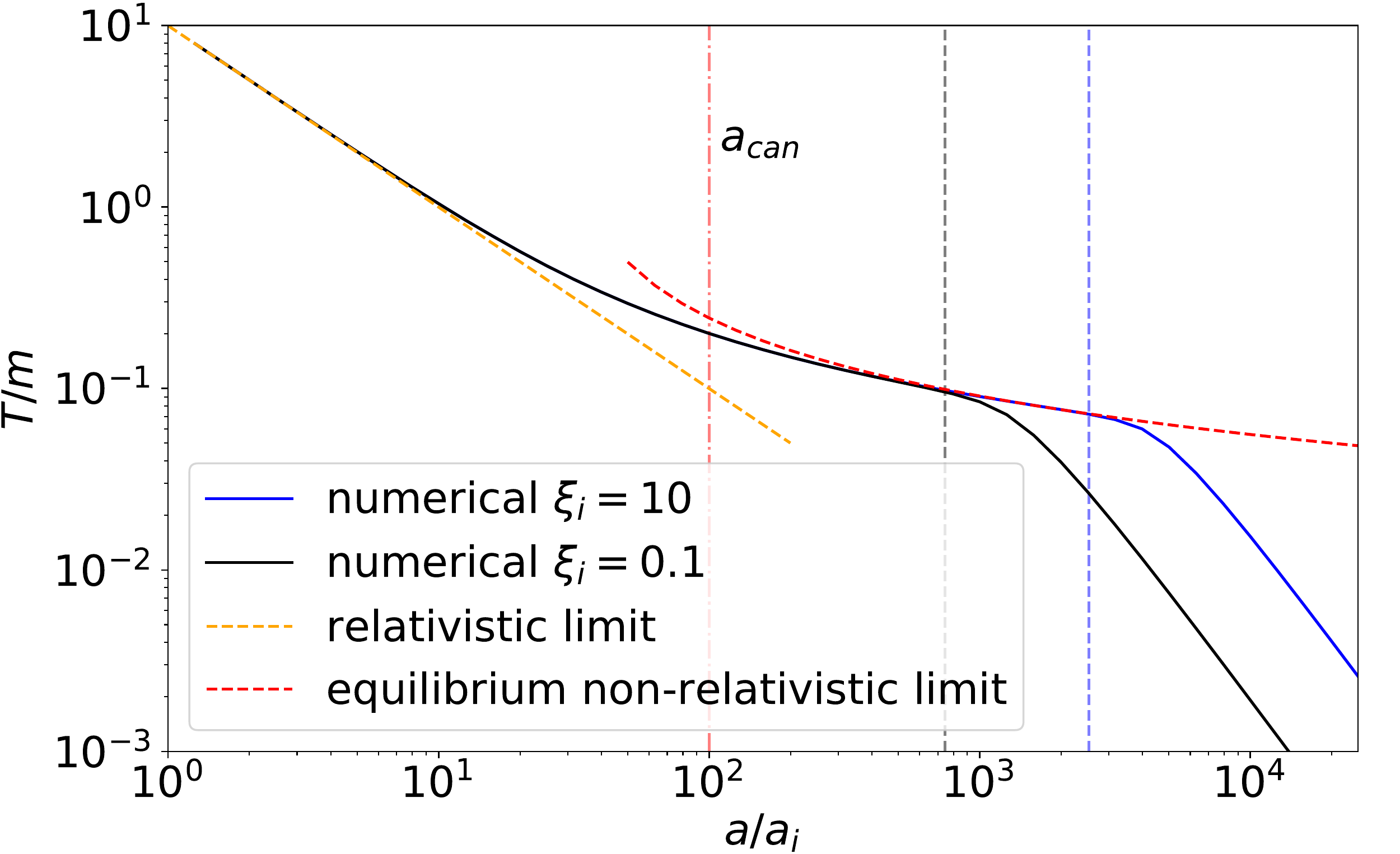}
	\end{subfigure}
	\begin{subfigure}{.5\textwidth}
		\includegraphics[width=1.00\textwidth]{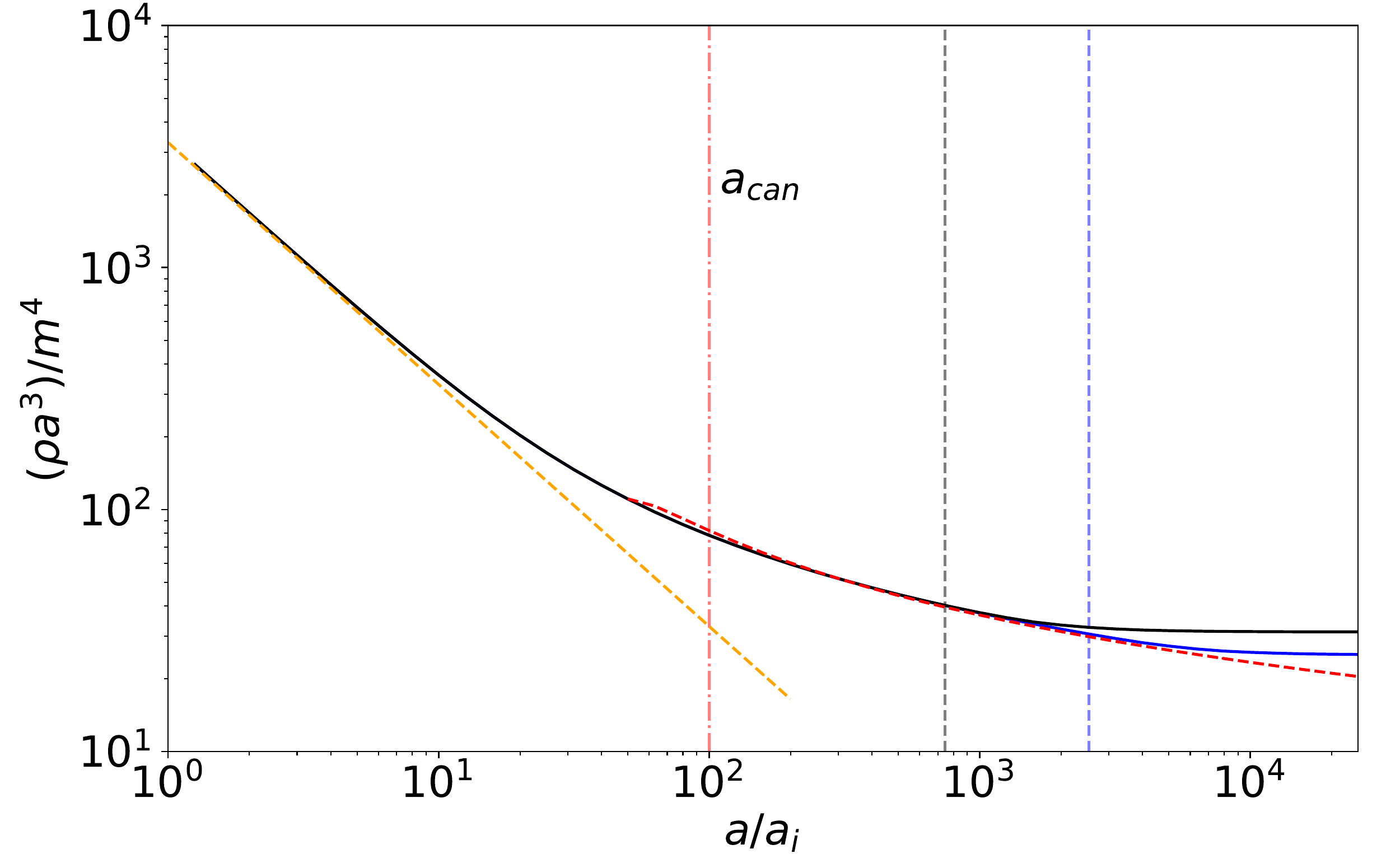}
	\end{subfigure}
	\caption{Temperature (left) and density (right) of the cannibal fluid as a function of scale factor.  Solid blue (black) lines show numerical results for the cannibal when the universe is cannibal- (SM radiation-) dominated and correspond to $m=35$ TeV, $\alpha_c=1$, $T_{\rm rh}=10$ MeV, and $\xi_i=10$ ($\xi_i=0.1$).
	The orange and red dashed lines show analytical results for the equilibrium cannibal fluid applicable in the relativistic and non-relativistic limits, respectively (eqs.~\eqref{eq:x_semi_tru} and~\eqref{eq:rhocaneq}). The vertical red dot-dashed line marks the onset of cannibalism. The remaining vertical dot-dashed lines indicate $a_{\rm fz}$ for the parameter point corresponding to the solid line of the same color.}
	\label{fig:can_intro}
\end{figure}

We are interested in finding the equilibrium cannibal density and temperature as a function of scale factor. We begin by writing the Boltzmann equation for the cannibal energy density:
\begin{align}\label{eq:can_eq_evolution_z}
a\frac{d\rho_{\rm can,eq}}{da}+3(1+w_{c,\textrm{eq}}(a))\rho_{\rm can,eq}=0.
\end{align}
While the cannibal fluid is in equilibrium, its energy density and equation of state can be written in terms of the cannibal temperature, $T_c$, using the relations
\begin{align}\label{eq:can_eq_def3}
\rho_{\rm can,eq}=&\frac{m^4}{2\pi^2}\int_1^{\infty}d\tilde{E} \tilde{E}^2 \sqrt{\tilde{E}^2-1}f_{\rm eq}(\tilde{E}x)\equiv \frac{m^4}{2\pi^2}h(x),\\\label{eq:can_w_def3}
w_{c,\textrm{eq}}=&\frac{\int_1^{\infty} d\tilde{E} (\tilde{E}^2-1)^{3/2}f_{\rm eq}(\tilde{E}x)}{3\int_1^{\infty}d\tilde{E} \tilde{E}^2 \sqrt{\tilde{E}^2-1}f_{\rm eq}(\tilde{E}x)}\equiv \frac{g(x)}{h(x)},
\end{align}
where $\tilde{E}\equiv E/m$, $x\equiv m/T_c$, and $f_{\rm eq}$ is the Bose-Einstein distribution at zero chemical potential. Consequently, eq.~\eqref{eq:can_eq_evolution_z} can be integrated to find $x$ as a function of scale factor through
\begin{align}\label{eq:F_def}
-\ln(a/a_i)=\int_{0.1}^{x} \frac{h'(\tilde{x})}{3[h(\tilde{x})+g(\tilde{x})]}d\tilde{x}\equiv F(x),
\end{align}
where we have used $x(a_i)=0.1$ and primes denotes derivatives with respect to $x$. We evaluate $F(x)$ at several values of $x$ and use the resulting table to define an interpolating function for $x$ as a function of $F=\ln(a/a_i)$. We find $\rho_{\rm can,eq}(a)$ by inserting the resulting $x(a)$ into eq.~\eqref{eq:can_eq_def3}. For $a<a_{\rm fz}/3$ in figure~\ref{fig:can_intro}, the blue and black curves shows the evolution of the equilibrium cannibal density and temperature obtained using this procedure.

In the limits $x\gg 1 $ and $x\ll 1 $, we find simple analytical expressions for $F(x)$ using the asymptotic expansions
\begin{align}
h(x)&\approx\begin{cases}\dfrac{\pi^4}{15}\dfrac{1}{x^4} & x\ll 1\\ \\ \dfrac{1}{x^{3/2}}\Big(1+\dfrac{27}{8x} +\dfrac{705}{128x^2}+O(x^{-3})\Big)\sqrt{\dfrac{\pi}{2}}e^{-x} & x\gg 1\end{cases}\label{eq:h_x}\\
g(x)&\approx\begin{cases}\dfrac{\pi^4}{45}\dfrac{1}{x^4} & x\ll 1\\ \\ \dfrac{1}{x^{3/2}}\Big(\dfrac{1}{x}+\dfrac{15}{8x^2}+O(x^{-3})\Big)\sqrt{\dfrac{\pi}{2}}e^{-x} & x\gg 1.\end{cases}
\end{align}
Using the $x\ll 1$ limits for $g(x)$ and $h(x)$ in eq.~\eqref{eq:F_def} gives the expected $T\propto 1/a$ scaling for relativistic particles. The $x\gg 1$ limits give us the evolution of the cannibal fluid during cannibalism. To connect the non-relativistic evolution of the cannibal fluid to its early relativistic evolution we need to integrate in the semi-relativistic regime ($x\sim 1$) where no simple analytical expressions are available. To handle the integration in the semi-relativistic regime, we break up the integral in eq.~\eqref{eq:F_def} into two integrals: one in the region $0.1<\tilde{x}<10$, and one in the region $10<\tilde{x}<x$. Then we use the large-$x$ approximations for $h(x)$ and $g(x)$ in the second integral to obtain
\begin{align}\label{eq:temp_A}
F(10)-\frac{1}{3}\int_{10}^{x} \left(1+\frac{1}{2\tilde{x}}+\frac{35}{8\tilde{x}{}^2}+\mathcal{O}(\tilde{x}{}^{-3})\right)d\tilde{x}\approx-\ln(a/a_i).
\end{align}
Taking $F(10)=-6.5$ in eq.~\eqref{eq:temp_A} implies
\begin{align}\label{eq:x_tru}
x=3\ln\left(\frac{a/a_i}{17.5x^{1/6}}\right)+\frac{35}{8x}+\mathcal{O}\left(x^{-2}\right).
\end{align}
To obtain a simpler relation between $x$ and $a$, we neglect the $1/x$ term and set $x= 10$ in the logarithm, which is approximately true during cannibalism, as seen in figure~\ref{fig:can_intro}. With these simplifications,
\begin{align}
x\approx 3\ln\left(\frac{a}{25.6a_i}\right)\label{eq:x_semi_tru}.
\end{align}
In the left panel of figure~\ref{fig:can_intro}  the red dashed curve shows this result for the temperature evolution, which accurately describes the evolution of the cannibal fluid once it becomes non-relativistic.

To determine the evolution of the cannibal density in the non-relativistic regime we first write its density in the large $x$ limit by using eq.~\eqref{eq:h_x} in eq.~\eqref{eq:can_w_def3}.
Since $h(x)$ has an exponential dependence on $x$, we use eq.~\eqref{eq:x_tru} instead of eq.~\eqref{eq:x_semi_tru} in the exponential term. Expanding the resulting equation to order $1/x^2$ gives
\begin{align}
\rho_{\rm can,eq}\approx m^4\left[\sqrt{\frac{\pi}{2}}\frac{(17.5)^3}{2\pi^2}\right]\frac{1}{(a/a_i)^3x}\Big[1-\frac{1}{x}+ \mathcal{O}(x^{-2})\Big].
\end{align}
Since the temperature of cannibal particles remains of order $ 0.1m$ during cannibalism, as seen in left panel of figure~\ref{fig:can_intro}, the next-to-leading order term in $1/x$ above provides a $\sim 10$\% correction. As the above relation no longer depends exponentially on $x$, we express $x$ in terms of $a$ using eq.~\eqref{eq:x_semi_tru} to obtain
\begin{align}\label{eq:rhocaneq}
\rho_{\rm can,eq}\approx \frac{148m^4}{(a/a_i)^3\ln(a/[25.6a_i])}\Big[1-\frac{1}{3\ln(a/[25.6a_i])}+ \mathcal{O}(x^{-2})\Big].
\end{align}
In the right panel of figure~\ref{fig:can_intro} we show the above estimate of $\rho_{\rm can,eq}$ as the red-dashed line. We can see that eq.~\eqref{eq:rhocaneq} accurately matches the blue and black lines for $a\gtrsim a_{\rm can}$.
\\

\subsection{Cannibal freeze-out}
\label{sec:can_fz}
As discussed in section~\ref{sec:bkgd}, we define $a_{\rm fz}$ through 
\begin{align}\label{eq:afz_def2}
	\svc  n_{\rm can}^2(a_{\rm fz})=H(a_{\rm fz}).
\end{align}
To accurately evaluate $a_{\rm fz}$ we need to solve for the evolution of $n_{\rm can}$ near freeze-out.
 
For $a<a_{\rm fz}/3$, the three-to-two interaction rate is strong enough to maintain chemical equilibrium and the cannibal density and temperature are accurately described by the equilibrium evolution discussed in the previous subsection. However, for $a>a_{\rm fz}/3$, the cannibal fluid starts to depart from its chemical equilibrium and we need to solve the Boltzmann equation for its number density:
\begin{align}\label{eq:app_ncan_fz}
	a\frac{dn_{\rm can}}{da}+3n_{\rm can}&=\frac{\svc  n_{\rm can}^2}{H}(n_{\rm can,eq}(x)-n_{\rm can})-\frac{\Gamma}{H} n_{\rm can}.
\end{align}
Since the cannibal particles are non-relativistic by the time freeze-out occurs, we use Maxwell-Boltzmann statistics in eq.~\eqref{eq:can_density_eq} to obtain
\begin{align}
    n_{\rm can}=e^{\mu/T}n_{\rm can,eq}&= e^{\mu x/m}\frac{m^3}{2\pi^2}\frac{K_2(x)}{x};\label{eq:appA_n_K}\\
    \rho_{\rm can}=e^{\mu/T}\rho_{\rm can,eq}&= e^{\mu x/m}\frac{m^4}{2\pi^2}\Big(\frac{x K_1(x)+3K_2(x)}{x^2}\Big);\\
    w_c=\frac{\mathcal{P}_{\rm can}}{\rho_{\rm can}}=\frac{\mathcal{P}_{\rm can,eq}}{\rho_{\rm can,eq}}&=\frac{K_2(x)}{x K_1(x)+3K_2(x)};\label{eq:w_nr}
\end{align}
where $K_i(x)$ is the modified Bessel function of $i^{th}$ order.

To find the evolution equation for $x$ we begin by expressing $\rho_{\rm can}$ in terms of $n_{\rm can}$ and $x$ using eqs.~\eqref{eq:appA_n_K}-\eqref{eq:w_nr},
\begin{align}\label{eq:app_rho_n}
    \rho_{\rm can}= mn_{\rm can}\frac{1}{xw_c(x)}.
\end{align}
Using the above relation to express $\rho_{\rm can}$ in terms of $n_{\rm can}$ in the energy conservation equation, eq.~\eqref{eq:num_can_density}, yields
\begin{align}
    \frac{m}{xw_c}\left(a\frac{dn_{\rm can}}{da}+3n_{\rm can}+\Gamma nxw_c(x)\right)-\frac{mn_{\rm can}}{x^2}\left(\frac{w_c+xw_c'(x)}{w_c^2}a\frac{dx}{da}-3x\right)=0.
\end{align}
We then simplify the first bracket using eq.~\eqref{eq:app_ncan_fz} to obtain the evolution equation for $x$,
\begin{align}\label{eq:fzout_T}
    a\frac{dx}{da}-3\frac{xw_c^2}{w_c+xw_c'(x)}=\frac{xw_c}{w_c+xw_c'(x)}\frac{\langle\sigma v^2\rangle}{H} n_{\rm can}(n_{\rm can,eq}-n_{\rm can})+\frac{xw_c(xw_c-1)}{w_c+xw_c'(x)}\frac{\Gamma}{H}.
\end{align}

Eq.~\eqref{eq:app_ncan_fz} and eq.~\eqref{eq:fzout_T} form coupled differential equations which are evaluated together to solve for $n_{\rm can}(a)$ and $x(a)$. In the left panel of figure~\ref{fig:can_intro} we plot the temperature (blue and black lines) for $a>a_{\rm fz}/3$ by numerically solving eq.~\eqref{eq:app_ncan_fz} and eq.~\eqref{eq:fzout_T}. In the right panel we plot the cannibal density for $a>a_{\rm fz}/3$ using the numerically evaluated $x$ and $n_{\rm can}$ in eq.~\eqref{eq:app_rho_n}. Notice that the evolution of the cannibal thermodynamic quantities is very similar in the two cases where the universe is cannibal dominated (blue line) or SM radiation dominated (black line), with the only difference being the specific value of $a_{\rm fz}$. Here $a_{\rm fz}$ is evaluated by finding where equality in eq.~\eqref{eq:afz_def2} is satisfied for numerically obtained $n_{\rm can}$.

\subsection{Cannibal decay}
\label{sec:decayderivation}

The collision term describing the energy transferred from the cannibal bath to SM-radiation due to a cannibal particle decaying into two SM particles is given by
\begin{equation}
	\frac{\textrm{d}\rho_{\rm can}}{\textrm{d} t}+3(1+w_c)H\rho_{\rm can}=\hat{C}_E=  -\int d\Pi  d\Pi_1 d\Pi_2 \,(2\pi)^4\delta^4
	(p-p_1-p_2) \, S|{\cal M}_{\Gamma}|^2 f_{c}(p) E,
\end{equation} 
where $f_c$ is the distribution function for the cannibals, $d\Pi_k = d^3p_k/[(2\pi)^3 2E_k]$, $|{\cal M}_{\Gamma}|^2$ is the matrix element corresponding to cannibal decays into radiation, $S$ is the symmetry factor, and variables with subscripts 1 and 2 correspond to the daughter particles while those with no subscripts correspond to the cannibal particle. We have neglected the contribution from final state effects as well as those from inverse decays. These approximations are valid as we consider the cannibal decays to become important ($\Gamma\sim H$) when the cannibal particles are non-relativistic.

Next, we perform the phase space integration of the daughter particles in the rest frame of the cannibal particle by using the definition of the rest frame decay width,
\begin{align}\label{eq:gamma_def}
	2m\Gamma\equiv \int d\Pi_1 d\Pi_2 \,(2\pi)^4\delta^4
	(p-p_1-p_2) \,S |{\cal M}_{\Gamma}|^2.
\end{align}
Doing so simplifies the collision term to
\begin{equation}
	\hat{C}_E=  -m\Gamma \int \frac{d^3p}{(2\pi)^3} f_{c}(p)=-mn_{\rm can}\Gamma.
\end{equation} 
The net energy transfer depends on the product $m n_{\rm can}$ and not on $\rho_{\rm can}$ because the longer cosmic rest-frame lifetime experienced by higher-energy particles exactly compensates for the increased energy released in their decays.

\section{WKB approximation for cannibal density perturbations}\label{sec:steady-state}
In this appendix we find an analytical expression for the evolution of cannibal density perturbations while the perturbations are rapidly oscillating, i.e. when $c_sk\gg aH$.

In the limit $c_sk\gg aH$, the sub-horizon evolution of cannibal density perturbations given in eq.~\eqref{eq:delta_can_subhz_adom} simplifies to 
\begin{align}\label{eq:generic_pert}
	\delta_c''(a)+\Big[\frac{(a^2 {H})'}{a^2 {H}}+\frac{1}{a}(1-3w_c)\Big]\delta_c'+\frac{1}{a^2}\left(\frac{ c_s{k}}{aH}\right)^2\delta_c=0,
\end{align}
where prime denotes derivative with respect to $a$.
Without any loss of generality, we can express the solution to the above equations as
\begin{align}
	\delta=D(a)e^{i\int \Omega(a)/(a^2H)da},
\end{align}
where $D $ and $\Omega$ determine the amplitude and frequency of oscillations, respectively.
Substituting this ansatz into eq.~\eqref{eq:generic_pert} gives
\begin{align}\label{eq:full_B_eq}
	\left[D ''+\left(\frac{(a^2 H)'}{a^2 H}+\frac{1-3w_c}{a}\right)D '+\frac{D }{{(a^2 H)^2}}\left(c_s^2k^2-\Omega^2\right)\right]\nonumber\\
	+i\left[2D '\frac{\Omega}{a^2H}+D \frac{\Omega'}{a^2H}+\left(\frac{1-3w_c}{a}\right)D \frac{\Omega}{a^2H}\right]=0.
\end{align}
Both the imaginary part and the real part above have to equal zero. Setting the imaginary part to zero implies
\begin{align}
	D ^2\ \Omega e^{\int_{a_*}^a [1-3w_c(\tilde{a})]d\ln(\tilde{a})}={\rm constant},\label{eq:A2omega_WKBfree}
\end{align}
where $a_*$ is some initial scale factor.

We solve the real part of eq.~\eqref{eq:full_B_eq} by assuming that the oscillation frequency is much greater than the rate at which the amplitude is changing: $c_s^2k^2/(a^2H)^2\gg D ''/D$ and $D '/(aD )$. This WKB approximation allows us to neglect the terms containing derivatives of $D$ when setting the real part of eq.~\eqref{eq:full_B_eq} to zero, which implies that
\begin{align}
	\Omega(a)\approx\pm c_sk.
\end{align}
Calculating the amplitude $D $ by substituting the above relation in eq.~\eqref{eq:A2omega_WKBfree} gives the full solution as
\begin{align}\label{eq:wkbfree}
	\delta (a)=\frac{1}{\sqrt{c_sk}} e^{-\int_{a_*}^a [1-3w_c(\tilde{a})]d\ln(\tilde{a})/2}\left(Ae^{i\int_{a_*}^ac_sk/(\tilde{a}^2H)d\tilde{a}}+Be^{-i\int_{a_*}^ac_sk/(\tilde{a}^2H)d\tilde{a}}\right).
\end{align}
Here $A$ and $B$ are constants determined by the initial conditions at $a_*$. We can equivalently express the above equation in terms of the sine function:
\begin{align}
	\delta (a)=C_1\frac{1}{\sqrt{c_sk}} e^{-\int_{a_*}^a [1-3w_c(\tilde{a})]d\ln(\tilde{a})/2}\sin\left(\int_{a_*}^a\frac{c_sk}{\tilde{a}^2H}d\tilde{a}+C_2\right),
\end{align}
where the initial conditions now determine the constants $C_1$ and $C_2$.

Note that the damping due to the Hubble term present in the coefficient of $\delta'$ in eq.~\eqref{eq:generic_pert} is exactly compensated by the Hubble term appearing in the frequency. Hence the expansion of the universe does not lead to damping of perturbations as one might naively think by looking at eq.~\eqref{eq:generic_pert}. In fact by rewriting eq.~\eqref{eq:generic_pert} in terms of conformal time, $d\eta=da/(a^2H)$, one can see that the Hubble damping term does not appear.
 
\section{Cannibal 2-to-2 scattering rate}
\label{sec:can_kd}

The 2-to-2 scattering rate can be computed from the forward piece of the collision term, $\Gamma_{\rm sc}=\hat{C}/n_{\rm can}$, where $\hat{C}$ is given by
\begin{align}
    \hat{C}=\int \frac{d^3p_1}{2E_1(2\pi)^3}\frac{d^3p_2}{2E_2(2\pi)^3}\frac{d^3p_3}{2E_3(2\pi)^3}\frac{d^3p_4}{2E_4(2\pi)^3} (2\pi)^4\delta^4(p_1+p_2-p_3-p_4)S|\mathcal{M}|^2 f(p_1)f(p_2).
\end{align}
Here $|\mathcal{M}|^2$ is the matrix element, $S=1/4$ includes the identical particle factors of initial and final states and $f$ is the phase-space distribution. Since we are primarily concerned with the scattering rate when the cannibal particles are non-relativistic, we have dropped the final state phase space distributions.

When the cannibal fluid is in kinetic equilibrium, $f$ is given by the Maxwell-Boltzmann distribution,
\begin{align}
	f(p)=e^{(\mu-E)/T_c}.
\end{align}
In equilibrium, the collision term can be written as 
\begin{align}
\label{eq:appc_C}
\hat{C}=&e^{2\mu/T_c}\frac{T_c}{64\pi^4} \frac{1}{2} \int_{4m^2}^{\infty} ds  \mathcal{A}(s) \sqrt{s-4m^2}  K_1(\sqrt{s}/T_c)
\end{align}
where $K_1$ is the modified Bessel function of the second kind, $s$ is the Mandelstam variable, and $ \mathcal{A}(s) $ is the integral of the squared matrix element over final state phase space,
\begin{align}
\label{eq:aofs}
\mathcal{A}(s) &=\frac{1}{2}\frac{1}{8\pi} \frac{\sqrt{s-4m^2}}{\sqrt{s}} \int \frac{d\Omega}{4\pi} |\mathcal{M}|^2 .
\end{align}
We have included factors of $1/2$ in both eq.~\eqref{eq:appc_C} and eq.~\eqref{eq:aofs} to account for  identical particles appearing in both the final and initial states.

For the cannibal Lagrangian given in eq.~\eqref{eq:L_can}, we find the matrix element describing scattering is, in the non-relativistic limit,
\begin{align}
	|\mathcal{M}|^2=& \left(\lambda-\frac{5}{3}\frac{g^2}{m^2}\right)^2.
\end{align}
Inserting the above matrix element in eq.~\eqref{eq:appc_C} and expanding the integrand in $T/m$
gives the leading contribution to the collision term in the non-relativistic limit:
\begin{align}
	\hat{C}\approx&e^{2(\mu-m)/T_c} \left(\frac{mT_c}{2\pi}\right)^3 \, \sqrt{\frac{T_c}{m}} \, \frac{\left(\lambda-\frac{5}{3}\frac{g^2}{m^2}\right)^2}{64 \pi^{3/2}m^2}	.
\end{align}

Expressing the chemical potential in terms of number density and the temperature of the cannibal fluid using the relation
\begin{align}
    n_{\rm can}=e^{\mu/T_c}n_{\rm can,eq}=e^{(\mu-m)/T_c}\left(\frac{mT_c}{2\pi}\right)^{3/2} 
\end{align}
yields our desired result
\begin{align}\label{eq:scat_can}
    \Gamma_{\rm sc}=\frac{\hat{C}}{n_{\rm can}}=&n_{\rm can}\langle\sigma_c v_c\rangle,
\end{align}
where 
\begin{align}
	\langle \sigma_c v_c\rangle = \frac{1}{64\pi^{3/2}m^2}\left(\lambda-\frac{5}{3}\frac{g^2}{m^2}\right)^2 \sqrt{\frac{T_c}{m}}.
\end{align}

\section{Perturbed collision operators for a decaying semi-relativistic particle }
\label{sec:der_pert}
In this section we derive the contribution from cannibal decays to the cosmological perturbation equations. We include the decay terms up to first order in $w_c$ and $c_s^2$, or equivalently up to first order in $T_c/m$.

We begin by writing the Boltzmann equations for a generic particle in a perturbed FRW universe, whose metric given by
\beq
ds^2 = -[1+2\psi]dt^2 + a^2(t)[1-2\phi](dx^2+dy^2+dz^2).
\eeq
Expressing the particle's phase space distribution in the form, $f(\vec{p},\vec{x},t)=\bar{f}(p,t)+\delta\!f(\vec{p},\vec{x},t)$, where $\bar{f}$ and $\delta\! f$ are unperturbed homogeneous and perturbed inhomogeneous pieces, respectively, the Fourier transform of the Boltzmann equation is given, to first order in perturbations, by
\begin{align}\label{eq:Liouville}
	\frac{df}{dt} = \frac{\partial f}{\partial t} + i\frac{\vec{k}\cdot\vec{p}}{aE} \delta f - \left[H-\frac{d\phi}{dt}\right]\frac{p^2}{E}\frac{\partial f}{\partial E} - i\frac{\vec{k}\cdot\vec{p}}{a}\psi \frac{\partial \bar{f}}{\partial E}=\frac{1+\psi}{E}\hat{C}[f].
\end{align}
Here $\vec{k}$ is the comoving Fourier wavenumber and $\hat{C}$ is the collision operator.

We are interested in the collision operator that describes the cannibal particle decaying into pairs of relativistic SM particles. The corresponding collision operators for the cannibal and radiation distributions are then given by
\begin{align}
	\hat{C}_{\Gamma}[f_c(p)] &= -\frac12 \int d\Pi_1 d\Pi_2 (2\pi)^4  \delta(E-E_1-E_2)\delta^3(\vec{p} - \vec{p}_1-\vec{p}_2)S|{\cal M}_{\Gamma}|^2f_c(p)\label{eq:C0}\\
	\hat{C}_{\Gamma}[f_r(p_1)] &= \int d\Pi d\Pi_2 (2\pi)^4  \delta(E-E_1-E_2)\delta^3(\vec{p} - \vec{p}_1-\vec{p}_2)|{\cal M}_{\Gamma}|^2f_c(p) \label{C1},
\end{align}
where $f_c$ and $f_r$ are distribution functions for the cannibals and relativistic SM particles, respectively, $d\Pi_k = d^3k/[(2\pi)^3 2E_k]$, $|{\cal M}_{\Gamma}|^2$ is the matrix element corresponding to cannibal decays into radiation, and $S$ is the identical particle factor. The collision term for SM radiation does not include a factor of $1/2$ because two SM particles are produced in the decay. We have neglected the contribution from final state effects as well as those from inverse decays because the cannibal decays become important ($\Gamma\sim H$) when $T_c\ll m$.

To obtain the evolution equations for density and velocity perturbations we take the energy-weighted phase space integral ($\int \frac{d^3p}{(2\pi)^3}E$) and the phase space integral of the first moment ($\intp [\vec{k}\cdot\vec{p}]$) of eq.~\eqref{eq:Liouville}.

\subsection{Cannibal Perturbation Equations}
First we use the definition of the rest-frame decay width, $\Gamma$, in eq.~\eqref{eq:gamma_def} to evaluate eq.~\eqref{eq:C0} for the cannibal collision operator:
\begin{align}
	\hat{C}_{\Gamma}[f_c] = -m \Gamma f_c,
\end{align}
where $m$ is the mass of the cannibal particle.
The Boltzmann equation for cannibals (eq.~\eqref{eq:Liouville}) will also include collision operators originating from cannibal self-interactions. However, these collision operators do not contribute to the perturbation equations for energy density or velocity as the self-interactions do not affect the energy and momentum of the fluid. Consequently, only the contribution from the decay collision operator remains after integrating the Boltzmann equation for cannibals over $\int \frac{d^3p}{(2\pi)^3}E$:
\begin{align}
	\frac{\partial \rho_{\rm can}}{\partial t} + \frac1a (\bar{\rho}_{\rm can}+\bar{\mathcal{P}}_{\rm can})\theta_{\rm can} + 3\left[H-\frac{d\phi}{dt}\right](\rho_{\rm can}+\mathcal{P}_{\rm can})&= -m \Gamma(1+\psi)n_{\rm can}
	\label{rhophi}.
\end{align}
To obtain the above result we used the definitions of energy density ($\rho$), number density ($n$), and pressure ($\mathcal{P}$) in terms of $f$. We also used the definition of the divergence of fluid velocity:
\begin{align}
	 \theta=\frac{i}{\bar{\rho}(1+w)} \intp (\vec{k}\cdot \vec{p}) \delta\!f.\label{eq:id2}
\end{align}
Writing $\rho$ and $\mathcal{P}$ in terms of homogeneous and perturbed pieces,
\begin{align}\label{eq:id3}
	\rho=\bar{\rho}(1+\delta) && \mathcal{P}=w\bar{\rho}+c_s^2\bar{\rho}\delta,
\end{align}
in eq.~\eqref{rhophi} and using the energy conservation equation of the cannibal fluid (eq.~\eqref{eq:num_can_density}) to evaluate $\textrm{d}\bar{\rho}_{\rm can}/\textrm{d} t$, we obtain
\begin{align}
	\dot{\delta}_c +\frac1a(1+w_c )\theta_c - 3\dot{\phi}(1+w_c ) + 3H(c_s^2-w_c) \delta_c = -\Gamma \frac{ m \bar{n}_{\rm can}}{\bar{\rho}_{\rm can}} \left[\psi + \frac{\delta\!n_{\rm can}}{\bar{n}_{\rm can}} - \delta_c\right],
	\label{fulldeltaphi}
\end{align}
where dot denotes differentiation with respect to $t$.

While $T_c\ll m$, we can further simplify the RHS by relating $n_{\rm can}$ to $\rho_{\rm can}$ and $\mathcal{P}_{\rm can}$ using
\begin{align}\label{eq:ncan_approx}
	{\rho}_{\rm can} \approx  m \intp \left(1+\frac{p^2}{2 m^2}\right) f_{c} = m{n}_{\rm can} +\frac32 \mathcal{P}_{\rm can}.
\end{align}
It follows that
\begin{align}
	\frac{\delta\!n_{\rm can}}{\bar{n}_{\rm can}} &\approx \delta_c \frac{1-\frac32 c_s^2}{1-\frac32 w_c}\approx \delta_c \left[1+\frac32(w_c - c_s^2)+  {\cal O}(w_c^2) \right]. \label{deltanovern}
\end{align}
Using the above result to evaluate $\delta\!n_{\rm can}/n_{\rm can} $ in eq.~\eqref{fulldeltaphi}, we obtain the perturbation equation for $\delta_c$ by expanding the terms proportional to $\Gamma$ to first order in $w_c$ and $c_s^2$,
\begin{align}
	\dot{\delta}_c +\frac1a(1+w_c )\theta_c - 3\dot{\phi}(1+w_c ) + 3\left(H-\frac{\Gamma}2\right)(c_s^2-w_c) \delta_c = -\Gamma \psi \left(1-\frac32 w_c\right)
	\label{approxdeltaphi}.
\end{align}

Next we calculate the perturbation equation for the divergence of fluid velocity, $\theta$, by evaluating the phase-space integration of the first moment ($\intp \left(\vec{k}\cdot\vec{p}\right)$) of the Boltzmann equation for cannibals (eq.~\eqref{eq:Liouville}). Note that all integrals that are odd in $\hat{p}$ will be proportional to $\delta\!f$, so any products of such integrals with metric perturbations can be neglected. Most of the remaining integrals can be evaluated using the definitions of $\rho$, $n$, $\mathcal{P}$, and $\theta$. The only integral not covered by these definitions contains $(\hat{k}\cdot\hat{p})^2$ in the integrand, which is contained in the definition of anisotropic stress:
\begin{align}
\sigma \equiv - \frac{1}{\bar{\rho}(1+w)}\intp \frac{p^2}{E} \left[\left(\hat{k}\cdot\hat{p}\right)^2 - \frac13\right] \delta\!f.
\end{align}
Finally, we use $\dot{\bar{\mathcal{P}}}=c_s^2\dot{\bar{\rho}}_{\rm can}$ and evaluate $\dot{\bar{\rho}}_{\rm can}$ using the energy conservation equation (eq.~\eqref{eq:num_can_density}) to obtain
\begin{multline}
	\dot{\theta}_c + H(1-3c_s^2)\theta_c - \frac{k^2}{a}\psi + \frac{k^2}{a} \sigma_{\rm can} - \frac{c_s^2k^2}{a(1+w_c)} \delta_c \\=m\Gamma\left[\frac{\bar{n}_{\rm can}(1+c_s^2)	}{\bar{\rho}_{\rm can} (1+w_c)}\theta_c- \frac{i}{\bar{\rho}_{\rm can}(1+w_c)}\intp \left(\vec{k}\cdot\vec{p}\right) \frac{\delta\!f}{E}\right].
\end{multline}
We further simplify the square bracket on the RHS by approximating $E\approx m+p^2/(2m)$ and using eq.~\eqref{eq:ncan_approx}. Simplifying the resulting expression by keeping only leading order terms in $w_c$ and $c_s^2$ and then using the definition of $\theta$, we obtain
\begin{multline}\label{eq:thetac_eq_midstep}
	\dot{\theta}_c + H(1-3c_s^2)\theta_c - \frac{k^2}{a}\psi + \frac{k^2}{a} \sigma_{\rm can} - \frac{c_s^2k^2}{a(1+w_c)} \delta_c \\
	=\Gamma\left[\left(-\frac52 w_c + c_s^2\right)\theta_c+\frac{i}{\bar{\rho}_{\rm can}(1+w_c)}\intp \frac{p^2}{2m^2}\left(\vec{k}\cdot\vec{p}\right)\delta\!f\right].
\end{multline}

To simplify the integral on the RHS, we note that the distribution function for a non-relativistic perfect fluid can be written as
\begin{align}
	f=e^{(\mu+\delta\mu)/T}e^{-(E-\vec{p}\cdot\vec{v})/T}\approx \bar{f}(E)-\bar{f}'(E)(\delta\mu+\vec{p}\cdot\vec{v}),
\end{align}
where $\delta\mu$ and $\vec{v}$ encode the density and velocity perturbations in the fluid. Using the above expression of $f$ in the $\theta$ definition (eq.~\eqref{eq:id2}) we obtain $\theta=i\vec{k}\cdot\vec{v}$. Consequently, the integral on the RHS of eq.~\eqref{eq:thetac_eq_midstep} simplifies to yield
\begin{align}
	\intp \frac{p^2}{2m^2}\left(\vec{k}\cdot\vec{p}\right) \delta f\approx -\intp \frac{p^2}{2m^2}\left(\vec{k}\cdot\vec{p}\right) \bar{f}'(E)(\vec{p}\cdot\vec{v}) = -i\bar{\rho}_{\rm can}\theta_c\left[ \frac52 w_c + {\cal O}(w_c^2) \right]\label{thetaRHS}.
\end{align}
Therefore, the perturbation equation for $\theta_c$ is given by
\begin{align}
	\dot{\theta}_{c} + H(1-3c_s^2)\theta_c - \frac{k^2}{a}\psi + \frac{k^2}{a} \sigma_{\rm can} - \frac{c_s^2k^2}{a(1+w_c)} \delta_c &=\Gamma c_s^2\theta_c
\end{align}
to leading order in $w_{c}$ and $c_s^2$ for terms proportional to $\Gamma$.

\subsection{Radiation Perturbation Equations}

We find the equation for radiation density perturbations by taking the energy-weighted phase space integral ($\int \frac{d^3p}{(2\pi)^3}E$) of the Boltzmann equation for radiation (eq.~\eqref{eq:Liouville}). The resulting integral of the collision term on the RHS is same as that encountered for the cannibal except with an opposite sign,
\begin{align}
	(1+\psi)\int \frac{d^3p_1}{(2\pi)^3E_1}SE_1\hat{C}_{\Gamma}[f_r(p_1)]=-(1+\psi)\int \frac{d^3p}{(2\pi)^3E}E\hat{C}_{\Gamma}[f_c(p)]=(1+\psi)mn_{\rm can}\Gamma,
\end{align}
where $\hat{C}_{\Gamma}[f_c(p)]$ and $\hat{C}_{\Gamma}[f_r(p_1)]$ are given in eq.~\eqref{eq:C0} and eq.~\eqref{C1}. The above equality is a direct consequence of energy conservation, which sets the energy of the daughter particle equal to half of the energy of the cannibal particle, $E_1=E/2$. Since the expression on the LHS now features integration over the phase space of both radiation particles, the symmetry factor $S$ appears.

Similar to the cannibal case, we simplify the LHS of the energy-weighted phase space integral of the Boltzmann equation by using the definitions in eq.~\eqref{eq:id3} and using the energy conservation equation for $\rho_r$ (eq.~\eqref{eq:num_r_density}) to yield
\begin{align}
	\dot{\delta}_r +\frac43 \frac{\theta_r}{a} - 4 \dot{\phi} &=  \frac{m \Gamma \bar{n}_{\rm can}}{\bar{\rho}_r} \left[\psi + \frac{\delta\!n_{\rm can}}{\bar{n}_{\rm can}} - \delta_r \right].
\end{align}
Above we have made use of the fact that $w=c_s^2=1/3$ for radiation. The $\delta n_{\rm can}$ in the RHS can be further simplified using eq.~\eqref{deltanovern} to give
\begin{align}
	\dot{\delta}_r +\frac43 \frac{\theta_r}{a} - 4 \dot{\phi}= \frac{m \Gamma \bar{n}_{\rm can}}{\bar{\rho}_r}\left[\psi + \delta_c - \delta_r +\frac32\delta_c(w_c-c_{s}^2)\right].
\end{align}

Next, we find the perturbation equations for the divergence of the radiation fluid velocity by evaluating the phase-space integration of the first moment ($\intp [\vec{k}\cdot\vec{p}]$) of the Boltzmann equation for radiation (eq.~\eqref{eq:Liouville}). The RHS of the resulting equation is of the form
\begin{multline}
	(1+\psi)\int \frac{d^3p_1}{(2\pi)^3E_1}(\vec{p}_1\cdot \vec{k})\hat{C}_{\Gamma}[f_r(p_1)]\\
	=2(1+\psi)\int d\Pi d\Pi_1 d\Pi_2 (2\pi)^4 (\vec{p}_1\cdot \vec{k}) \delta(E-E_1-E_2)\delta^3(\vec{p} - \vec{p}_1-\vec{p}_2)S|{\cal M}_{\Gamma}|^2f_c(p).
\end{multline}
In the above integral we replace $\vec{p}_1\cdot \vec{k}\rightarrow (\vec{p}_1+\vec{p}_2)\cdot \vec{k}/2$ as the labels $1$ and $2$ are interchangeable. Moreover, by momentum conservation we have $\vec{p}_1+\vec{p}_2=\vec{p}$, which yields
\begin{align}
	(1+\psi)\int &\frac{d^3p_1}{(2\pi)^3E_1}(\vec{p}_1\cdot \vec{k})\hat{C}_{\Gamma}[f_r(p_1)]\nonumber\\
	=&(1+\psi)\int d\Pi \left[\int d\Pi_1 d\Pi_2 (2\pi)^4\delta^4(p - p_1-p_2)S|{\cal M}_{\Gamma}|^2\right](\vec{p}\cdot \vec{k})f_c(p)\\
	=&\Gamma\intp \left(\vec{k}\cdot\vec{p}\right) \frac{\delta f}{\sqrt{1+p^2/m^2}}\approx-i\Gamma\bar{\rho}_{\rm can}\theta_c \left[1 -\frac32 w_c\right].
\end{align}
Here in the second line we first expanded the denominator to first order in $p^2/m^2$ and then used the definition of $\theta$ (eq.~\eqref{eq:id2}) along with the result given in eq.~\eqref{thetaRHS} to obtain the final answer.

We simplify the phase space integration of the first moment of the LHS of Boltzmann equation in the same way as we did for cannibal perturbations. Expressing the cannibal energy density in terms of the cannibal number density using eq.~\eqref{eq:ncan_approx} gives
\begin{align}
	\dot{\theta}_r-\frac{k^2}{4a} \delta_r - \frac{k^2}a \psi + \frac{k^2}a \sigma_r= \Gamma \frac{m\bar{n}_{\rm can}}{\bar{\rho}_r}\left[\frac34\theta_c -\theta_r \right].
\end{align}

\bibliographystyle{utphys}
\bibliography{cannibal}
\end{document}